\documentclass[aps,twocolumn,showpacs,amsmath,amssymb,prx,superscriptaddress,floatfix,nobalancelastpage]{revtex4-1}
\usepackage{graphicx}% Include figure files
\usepackage{bm}% bold math
\usepackage{bbold}% bold math

\usepackage{inputenc}
\usepackage{verbatim}
\usepackage{amsmath,amsfonts,amssymb,dsfont}
\usepackage{color}

\usepackage[hidelinks]{hyperref}

\renewcommand{\L}{{\mathcal L}}
\newcommand{\N}{{\mathcal N}}

\newcommand{\E}{{\mathcal E}}
\newcommand{\D}{{\mathcal D}}
\newcommand{\A}{{\mathcal A}}

\newcommand{\C}{{\mathcal C}}

\renewcommand{\l}{{{L }}}
\newcommand{\Tr}{{\mathrm{Tr}}}
\newcommand{\var}{{\mathrm{Var}}}

\definecolor{green}{rgb}{0.0, 0.5, 0.0}

\newcommand{\er}[1]{Eq.~\eqref{#1}}

\newcommand{\era}[2]{Eqs.~(\ref{#1}) and (\ref{#2})}

\newcommand{\eraaa}[4]{Eqs.~(\ref{#1}), (\ref{#2}), (\ref{#3}) and (\ref{#4})}

\usepackage[toc,page]{appendix}
\newcommand{\nocontentsline}[3]{}
\newcommand{\tocless}[2]{\bgroup\let\addcontentsline=\nocontentsline#1{#2}\egroup}
\newcommand{\Appendix}[1]{
	\refstepcounter{section}
	\vspace*{3ex}
	\section*{Appendix \thesection:\hspace*{1.5ex} #1 }
}
\newcommand{\SubAppendix}[1]{\vspace*{1.5ex}\tocless\subsection{#1}}
\newcommand{\SubSubAppendix}[1]{\tocless\subsubsection{#1}}
\begin{document}

\title{Coherence, entanglement, and quantumness in closed and open systems with conserved charge, with an application to many-body localization}

\author{Katarzyna Macieszczak}
\author{Emanuele Levi}
\affiliation{School of Physics and Astronomy, University of Nottingham, University Park, Nottingham NG7 2RD, United Kingdom}
\affiliation{Centre for the Mathematics and Theoretical Physics of Quantum Non-Equilibrium Systems, University of Nottingham, University Park, Nottingham NG7 2RD, United Kingdom}
\author{Tommaso Macr\`i}
\affiliation{International Institute of Physics, 59078-400 Natal, Rio Grande do Norte, Brazil}
\affiliation{Departamento de F\'isica Te\'orica e Experimental, Universidade Federal do Rio Grande do Norte, 59072-970 Natal, Rio Grande do Norte, Brazil}
\author{Igor Lesanovsky}
\author{Juan P. Garrahan}
\affiliation{School of Physics and Astronomy, University of Nottingham, University Park, Nottingham NG7 2RD, United Kingdom}
\affiliation{Centre for the Mathematics and Theoretical Physics of Quantum Non-Equilibrium Systems, University of Nottingham, University Park, Nottingham NG7 2RD, United Kingdom}


\date{\today}

\begin{abstract}
 While the scaling of entanglement in a quantum system can be used to distinguish many-body quantum phases, it is usually hard to quantify the amount of entanglement in mixed states of open quantum systems, while measuring entanglement experimentally, even for the closed systems, requires in general quantum state tomography. In this work we show how to remedy this situation in system  with a \emph{fixed} or \emph{conserved charge},  e.g., density or magnetization,  due to an emerging relation between quantum correlations and coherence. First, we show how, in these cases, the presence of multipartite entanglement or quantumness can be faithfully witnessed simply by detecting coherence in the quantum system, while bipartite entanglement or bipartite quantum discord are implied by asymmetry (block coherence) in the system. Second, we prove that the relation between quantum correlations and coherence is also quantitative. Namely, we establish upper and lower bounds on the amount of multipartite and bipartite entanglement in a many-body system with a fixed local charge, in terms of the amount of coherence and asymmetry present in the system. Importantly, both for pure and mixed quantum states, these bounds are expressed as closed formulas, and furthermore, for bipartite entanglement, are experimentally accessible by means of the multiple quantum coherence spectra. In particular, in one-dimensional systems, our bounds may detect breaking of the area law  of entanglement entropy. We illustrate our results on the example of a many-body localized system, also in the presence of dephasing. 
\end{abstract}

\maketitle

\tableofcontents

\section{Introduction} 

Classical phases of matter are often distinguished by observable order parameters, such as densities or magnetizations, and by the properties of their fluctuations \cite{Chandler1987}.  For quantum matter, it has been shown that the entanglement properties of quantum states are reliable indicators of quantum phase behavior, both for quantum phase transitions in the ground state \cite{Osterloh2002,Calabrese2004} and for excited-state phase transitions, such as the one that leads to many-body localization (for reviews see \cite{Nandkishore2015b,Altman2015,Abanin2017}; for other kinds
of quantum phase transitions at the excitation level see, e.g.,~\cite{Caprio2008}).  While entanglement encodes the properties of quantum fluctuations, in contrast to classical order parameters, measures of entanglement~\cite{Bennett1996,Vedral1997,Vedral1998,Henderson2000,Plenio2007} are difficult to calculate for mixed states, and even for pure states are generally not directly observable, except through full quantum tomography~\cite{Haffner2005}.  Entanglement is known, however, to be a necessary resource for quantum information protocols~\cite{Horodecki2009}. This means that the success of performing a given quantum protocol can be used to determine the presence of entanglement  in the corresponding quantum state, that is, , it can be used as a {\em witness} of entanglement.
\begin{figure}[ht]
	\begin{center}
		\vspace*{-7mm}
		\includegraphics[width=0.9\columnwidth]{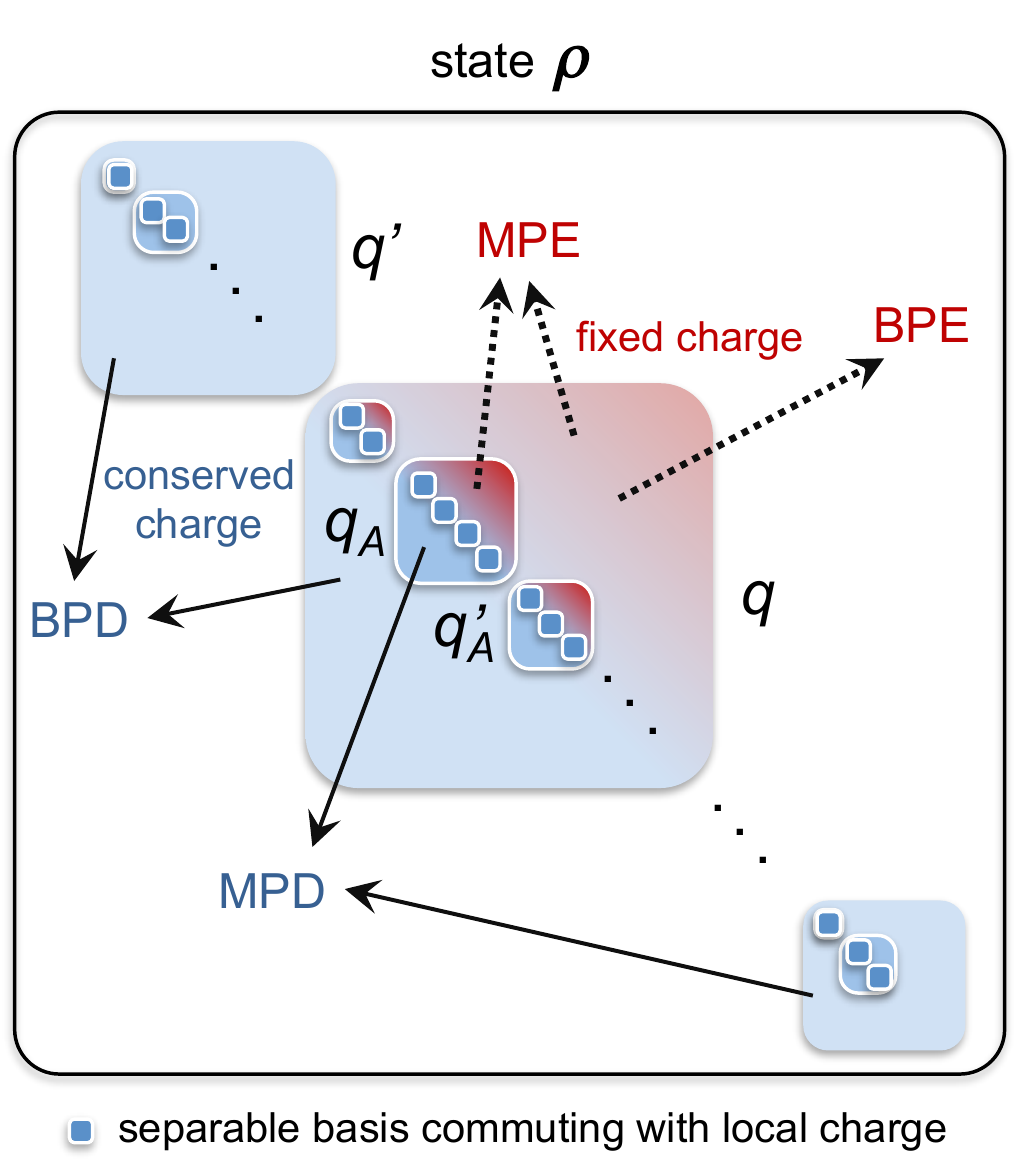} 
		\caption{
			{\bf Coherence and asymmetry implies multipartite and bipartite quantum correlations in the presence of fixed or conserved charge}. A state $\rho$ of the conserved charge $Q$, $[\rho,Q]=0$, is block diagonal [light blue (light gray) squares] with respect to the charge eigenspaces (with values of $Q$ denoted by $q$), while when a charge value is fixed [(middle) light blue square shaded into red], $Q\rho=q\rho$, it is supported within only a single block corresponding to $q$ eigenspace. When the charge is local, $Q=\sum_{k=1}^N Q^{(k)}$, and $Q^{(k)}$ is nondegenerate for each subsystem, it uniquely defines the separable basis without coherence [dark blue (dark gray) squares]. As we show in Secs.~\ref{subsec:witnessMPE} and~\ref{subsec:witnessMPD}, when the charge $Q$ is fixed or conserved, any coherence in this basis (faithfully) implies MPE or MPD, respectively. Furthermore, when the system is divided into two parts $A$ and $B$, the local charges $Q^{(A)}\equiv\sum_{k\in A} Q^{(k)}$ are in general degenerate. Nevertheless, the block coherence (asymmetry) with respect to $Q^{(A)}$ eigenspaces [blue (gray)] still implies BPE or BPD when the charge $Q$ is fixed or conserved, respectively, as we show in Sec.~\ref{subsec:witnessBPE}.
			%what is red what is blue
			%
		}\vspace*{-7mm}
		\label{fig:matrix}
	\end{center}
\end{figure}

Quantum metrology refers to the possibility of decreasing parameter estimation errors beyond those set by the classical central limit theorem~\cite{Giovannetti2004,Giovannetti2006,Giovannetti2011}.  This enhanced scaling is only due to the presence of quantum correlations associated with multipartite entanglement (MPE) of the quantum state being probed, and thus can be used as an entanglement witness. In particular, the sensitivity of a quantum state to perturbations of the parameter being estimated, and thus the usefulness of the state as a quantum sensor, is quantified by the quantum Fisher information (QFI)~\cite{Helstrom1967,Helstrom1968,Braunstein1994}.  The QFI guarantees the presence of MPE whenever its value exceeds a {\em separability threshold}, which, however, in general scales with the system size.  This is especially a problem for mixed states, where the noise that leads to mixedness of the state contributes to estimation errors, possibly decreasing the QFI below the threshold even when entanglement is present.

%This establishes a connection between enhanced parameter estimation and MPE, and therefore QFI is often used as an entanglement witness.
%A complication that arises in the most general case is that

In this paper we show that for systems with a fixed local charge, for example, when the number of particles is fixed, witnessing entanglement simplifies as follows. We prove that the state is coherent in local charge basis if and only if it is multipartite entangled, which is illustrated in Fig.~\ref{fig:matrix}. Therefore, for estimation of parameters encoded by diagonal observables, the separability threshold of the QFI vanishes, and MPE can be witnessed efficiently. Moreover, for nondegenerate diagonal observables, the QFI becomes a {\em faithful witness} of MPE, i.e., the presence of MPE is always manifested in a nonzero QFI. Similarly, for a given bipartition, the block coherence, i.e.,\ the asymmetry, of the charge difference in the partition implies the presence of bipartite entanglement (BPE). Therefore, in systems with a fixed local charge, also BPE can be witnessed at a zero threshold, by the QFI for block-diagonal observables (commuting with charges of the two system parts). We also demonstrate that for systems with a conserved, rather than fixed, local charge, i.e.,\ when the state is block diagonal with respect to different charge values, the coherence and the asymmetry imply a weaker type of quantum correlations instead of entanglement: multipartite quantum discord (MPD)~\cite{Modi2010,Okrasa2011} and bipartite quantum discord (BPD)~\cite{Ollivier2001,Zurek2000,Henderson2001}, respectively, cf. Fig.~\ref{fig:matrix}.

Importantly, we further show that the emerging relation of entanglement to coherence and asymmetry is also quantitative. The amount of multipartite entanglement~\cite{Vedral1997,Vedral1998}  in a many-body quantum system with a fixed charge can be  faithfully bounded from above by the amount of coherence~\cite{Aberg2006,Vaccaro2008,Gour2009,Baumgratz2014,Levi2014} when quantified by the relative entropy (cf.~\cite{Bromley2015,Yao2015,Adesso2016}).  Furthermore, also bipartite entanglement of formation~\cite{Bennett1996,Hill1997,Wootters1998,Henderson2000} is bounded from below by the asymmetry~\cite{Aberg2006,Vaccaro2008,Gour2009} quantified by the relative entropy  with respect to charge difference.  In particular, for a pure quantum state, the asymmetry is a lower bound for the entanglement entropy, i.e.,\ the von Neumann entropy of a reduced subsystem state. Our bounds are expressed as closed formulas and thus can be easily calculated for both  pure and mixed states. Furthermore, in one-dimensional systems, our bounds may detect breaking of the area law of entanglement entropy. We also derive a lower bound on bipartite entanglement  quantified by the convex roof of negativity~\cite{Vidal2002,Lee2003}), which can be accessed experimentally by the multiple quantum coherence spectra~\cite{Baum1985,Macri2016,Garttner2017,Garttner2018}.

The implications of the connection between entanglement and coherence in systems with particle conservation are important in practice. Take, for example, the case of disordered quantum many-body systems that display a thermal to many-body localized (MBL) transition or cross-over, which is driven by the strength of the quenched disorder \cite{Nandkishore2015b,Altman2015,Abanin2017}. Here the entanglement characteristic of many-body eigenstates serves to distinguish between thermal and MBL phases: In the bulk of the  energy spectrum thermal eigenstates have bipartite entanglement, as measured by the entanglement entropy, that scales with the size of the partition (volume law), as they are believed to obey the eigenstate thermalisation hypothesis \cite{DAlessio2016};  for MBL eigenstates, instead it scales with the size of the boundary of the bipartition (area law).  Nevertheless, in the MBL phase, the entanglement entropy of an initially separable state grows slowly (logarithmically in time) towards an asymptotic value that scales with volume, a feature that distinguishes MBL from the noninteracting case of Anderson localization. 
Since the entanglement entropy is not directly observable there have been attempts to connect it to observable quantities for closed (i.e., nondissipative) systems.  These observable proxies have included fluctuations in the number of particles within the partition~\cite{Bardarson2012}, diagonal entropies~\cite{Serbyn2013}, and QFI itself~\cite{Smith2016}.  There are a number of problems in connecting these observables to bipartite entanglement: In Ref.~\cite{Bardarson2012} logarithmic growth is observed on a much shorter time regime than that of the entanglement entropy; in Ref.~\cite{Serbyn2013} the logarithmic growth is absent, as diagonal entropy corresponds to the asymptotic value of the entanglement entropy; and in Ref.~\cite{Smith2016} the QFI, while growing logarithmically, does not usually exceed the separability threshold. 

For systems with a conservation of density the relation between these approaches is clarified by our results here from the emerging connection between coherence and entanglement. We find that the behavior of fluctuations in the number of particles is connected to the charge asymmetry in the bipartition into half chains, while fluctuations of the imbalance are related to the charge asymmetry in the staggered bipartition. The amount of asymmetry further bounds the entanglement from below. In particular, in a many-body localized phase, for the bipartition into half chains, we find numerically that the asymmetry breaks the area law present in the Anderson-localized phase, although it saturates at earlier times than the entanglement entropy. Finally, the diagonal entropy measures the coherence in closed systems, which here bounds from above both the bipartite and multipartite entanglement, and thus necessarily follows the volume law.

The paper is organized as follows.  In Sec.~\ref{sec:coherence} we review approaches for witnessing MPE by measuring the QFI above a separability threshold, as well as quantifying the amount of coherence.  Section~\ref{sec:witness} onwards contains the results of the paper. In Sec.~\ref{sec:witness}
we show that when conservation laws are present, coherence and asymmetry can be connected to multipartite and bipartite entanglement (for a fixed charge) or discord (for a conserved charge). As a consequence, entanglement or discord can be witnessed by the QFI for appropriately chosen observables with a zero separability threshold. In Sec.~\ref{sec:measures} we make these relations quantitative by showing how coherence and asymmetry monotones can serve as bounds on the amount of bipartite and multipartite entanglement present in a quantum system with a fixed charge.   Throughout the paper we illustrate our results with the example of a many-body localized system, which is discussed in detail in Sec.~\ref{sec:MBL}. We finish with a brief conclusion and outlook in Sec.~\ref{sec:conclusions}.

\section{Coherence and entanglement} \label{sec:coherence}

Enhanced quantum protocols rely on two properties: the possibility of creating superpositions between states in the computational basis and the entanglement between subsystems. Therefore, the success rate of performing a given quantum protocol  above a certain threshold can be used to certify the presence of a resource required by the protocol. 

In this section we first review how the QFI, which bounds errors in quantum phase estimation,  is used as a witness of both coherence and entanglement. This method will be used later in Sec.~\ref{sec:witness}, where we will show how, in the presence of a fixed or conserved local charge, the relation between quantum correlation and coherence is strengthened, with the coherence implying the presence of entanglement or quantum discord. Second, we will also review how to quantify the amount of coherence present in a system state by coherence monotones, which will prove relevant in Sec.~\ref{sec:measures}, where we will derive bounds on the amount of entanglement or quantum discord in systems with a fixed or conserved local charge, respectively.

%Sth about measures of entanglement!!! 

\subsection{Witnessing coherence} 

Let  $\{ |i\rangle : i = 1,\ldots,D \}$ be an orthonormal basis, with $D$ being the dimension of the system Hilbert space. Let $\rho$ be a density matrix describing a state of the system. The state is \emph{coherent} if $\rho$ features nonvanishing off-diagonal terms $\rho$ in the given basis. The state is \emph{incoherent} if $\rho$ is diagonal. Therefore, the notion of coherence is related to the possibility of creating superpositions (see e.g.,~\cite{Aberg2006,Vaccaro2008,Gour2009,Baumgratz2014,Levi2014}). \\

 One of the quantum protocols for which coherence is a resource is quantum estimation of a parameter unitarily encoded with an observable $M$ diagonal in the computational basis, 
 $M=\sum_{i} m_i |i\rangle\!\langle i| $. In this case, the phases are encoded in the off-diagonal terms in a density matrix $\rho$ describing the system state, 
 \begin{equation}
 e^{-i\phi M }\rho\, e^{-i\phi M }= \sum_{ij} e^{-i\phi(m_i-m_j)} \rho_{ij}\,|i\rangle\!\langle j|, 
 \end{equation}
 where $\rho_{ij}= \langle i|\rho|j\rangle$.
 The errors of unbiased estimation of $\phi$ are bounded from below by the inverse of the QFI~\cite{Helstrom1967,Helstrom1968,Braunstein1994,Giovannetti2004,Giovannetti2006,Giovannetti2011},
\begin{equation}
{\rm QFI}(M,\rho) \equiv \sum_{ij} \frac{2 ( \lambda_{i} -\lambda_{j})^{2}}{\lambda_{i} + \lambda_{j}} |\langle \lambda_{i} | M | \lambda_{j} \rangle|^{2} ,
\label{qfi}
\end{equation}
where $| \lambda_{i} \rangle$ is an (orthonormal) eigenstate of $\rho$ that corresponds to an eigenvalue $\lambda_i$, i.e., $\rho=\sum_i \lambda_i | \lambda_{i} \rangle\!\langle\lambda_{i} |$. In particular, if the state $\rho$ is diagonal in the computational basis, i.e., incoherent, the QFI also equals zero, as there are no phases encoded in the state.  Furthermore, when $M$ is nondegenerate the QFI can only vanish when $\rho$ is diagonal, and in this case the QFI becomes a faithful witness of coherence, meaning that any nonzero QFI guarantees the presence of coherence.  

Consider for example the Werner state~\cite{Werner1989} of two qubits (two spins $\frac{1}{2}$), 
\begin{eqnarray} \label{Werner}
\rho_W&\equiv&\frac{1-p}{4}\mathds{1}+ p \,|\Psi^-\rangle\! \langle \Psi^-|\\\nonumber
&=&\frac{1}{4}\left( \begin{array}{c c c c} 1-p&&&\\&1+p&-2p&\\&-2p&1+p&\\&&&1-p \end{array}\right),
\end{eqnarray}
where the Bell state  $|\Psi^-\rangle\equiv(|\!\!\downarrow \!\uparrow \rangle-|\!\!\uparrow \!\downarrow \rangle)/\sqrt{2}$ (with $|\!\!\uparrow \rangle$ denoting an spin-up state and $|\!\!\downarrow \rangle$ denoting the spin-down state in the $z$ direction) and $0\leq p\leq 1$. The density matrix in the second line of Eq.~\eqref{Werner} is shown in the computational basis $|\!\!\downarrow \!\downarrow \rangle$, $|\!\!\downarrow \!\uparrow \rangle$, $|\!\!\uparrow \!\downarrow \rangle$, and $|\!\!\uparrow \!\uparrow \rangle$. In this basis, the Werner state is coherent, i.e., non-diagonal, for $p>0$. However, for $M$ chosen as the $z$ magnetization, $M_z=\sum_{k=1}^2 S^z_k$ (with $S^z_k$ denoting the $z$ magnetization of the $k=1,2$ spin),  we have ${\rm QFI}(M_z,\rho_W)=0$ for all $p$ due to degeneracy of $M_z$ in the subspace of $|\!\downarrow \!\uparrow \rangle$, $|\!\uparrow \!\downarrow \rangle$. For the imbalance in the $z$ direction (the staggered $z$ magnetization), $I_z=\sum_{k=1}^2 (-1)^k S^z_k$,  however, we have ${\rm QFI}(I_z,\rho_W)=8p^2/(1+p)$, so the coherence is faithfully witnessed for all $p>0$ (due to $I_z$ being nondegenerate in the subspace of $|\!\downarrow \!\uparrow \rangle$, and $|\!\uparrow \!\downarrow \rangle$, where coherence is present).

When the state is pure  $\rho=|\psi\rangle\!\langle\psi|$, the QFI is simply proportional to the quadratic fluctuations ${\rm QFI}(M,\rho) =  4\var(M,\rho)  $ , i.e., the variance, 
\begin{equation}
 \var(M,\rho) \equiv  \Tr{(\rho M^2)} - [\Tr{(\rho M)}]^2. 
\end{equation}
For a mixed state $\rho$, however, the variance is higher,
\begin{equation}
{\rm QFI}(M,\rho)\leq 4 \, \var(M,\rho).
\label{qvc1}
\end{equation}
 Therefore, the variance cannot be used as a witness of coherence, as it captures also classical fluctuations due to mixedness, and can be nonzero even in incoherent diagonal states (cf. Fig~\ref{fig:QFI_open}). For the example of the Werner state~\eqref{Werner}, we have $\var(M_z,\rho_W)=(1-p)/2$ and  $\var(I_z,\rho_W)=(1+p)/2$, which are both nonzero at $p=0$, although $\rho_W$ is diagonal and thus incoherent. 

We thus conclude that for open systems described by mixed density matrices, it is necessary to calculate and measure the QFI rather than the variance in order to witness coherence. Although the QFI can be measured experimentally only for certain families of states, e.g., thermal states~\cite{Hauke2016}, lower bounds are accessible~\cite{Pezze2009,Hyllus2010,Hyllus2012,Toth2012,Strobel2014,Toth2014,Pezze2016,Frerot2016,Malpetti2016,Girolami2017} (see also~\cite{Niknam2018}). In particular, in Appendix~\ref{sec:method} we describe the lower bound on the QFI in terms of curvature~\cite{Girolami2014b,Macri2016,Garttner2018} which can be measured by the multiple quantum coherence spectrum~\cite{Macri2016,Garttner2018}.

\subsection{Witnessing coherence as a proxy for witnessing entanglement} \label{sec:QFI}

We now explain how witnessing coherence via the QFI~\eqref{qfi} can be used to witness multipartite entanglement.\\

 A state $\rho$ of $N$ subsystems is \emph{multipartite-entangled} when it is not \emph{separable} $\rho\neq\rho_\text{sep}$ where
\begin{equation}\label{rho:sep}
\rho_{\rm sep} \equiv \sum_{j} p_{j}\, \varrho_{j}^{(1)} \otimes \cdots \otimes \varrho_{j}^{(N)}.
\end{equation}
with $\varrho^{(k)}$ describing the state of the $k$th subsystem.\\

 Multipartite entanglement is a resource for quantum metrology, when the phase to be estimated is encoded in a quantum state via a local observable~\cite{Giovannetti2004,Giovannetti2006,Giovannetti2011}
\begin{equation}\label{M}
M = \sum_{k=1}^{N} M^{(k)} ,
\end{equation}
where $M^{(k)}$ acts on the $k$th  subsystem. Indeed, in this case the maximum QFI
\begin{equation}
{\rm QFI}_{\rm max}(M) = \left( \sum_{k=1}^{N} \Delta M^{(k)} \right)^{2} ,
\end{equation} 
where $\Delta M^{(k)}$ is the difference between the extreme eigenvalues of $M^{(k)}$, is achieved by the superposition of the two extreme eigenvectors of $M$, which is a multipartite-entangled state. For an illustration consider an example of $M$ chosen as the $z$ magnetization of $N$ spin-$\frac{1}{2}$  particles $M_z\equiv\sum_{k=1}^N S_k^z$ (where $S_k^{x,y,z}$ are spin operators for the $k$th spin $1/2$). The maximum QFI is then achieved for the Greenberger-Horne-Zeilinger state $|\rm{GHZ}\rangle\equiv(|\!\!\downarrow \rangle^{\otimes N}+|\!\!\uparrow \rangle^{\otimes N})/\sqrt{2}$, which gives quadratic scaling of the QFI, ${\rm QFI}_{\rm max}(M) = N^{2}$, termed \emph{Heisenberg scaling}.

In contrast, for separable states, the QFI is not larger than the \emph{separability threshold}~\cite{Giovannetti2004,Giovannetti2006,Pezze2009,Hyllus2010,Hyllus2012,Toth2012,Strobel2014,Toth2014,Pezze2016,Hauke2016,Girolami2017}, 
\begin{equation}
{\rm QFI}_{\rm sep}(M) \equiv \max_{\rho_\text{sep}} {\rm QFI}(M,\rho_\text{sep}) = \sum_{k=1}^{N} (\Delta M^{(k)})^{2}, 
\label{thresh} 
\end{equation} 
which is achieved for a product state of the superposition of extreme eigenvectors of $M^{(k)}$ on each subsystem~\footnote{It is easy to see that this is the optimal product state, as classical correlations in~\eqref{rho:sep} do not increase the QFI due to its convexity.}. For the example  $N$ spin$-\frac{1}{2}$ particles with $M$ chosen as the $z$ magnetization, we have that the separability threshold ${\rm QFI}_{\rm sep}(M) = N$ is achieved for the product state $2^{{-N/2}} \, (|\!\!\downarrow \rangle+|\!\!\uparrow \rangle)^{\otimes N}$. The linear scaling of the QFI with system size is called \emph{standard scaling}.  Whenever the QFI~\eqref{qfi} or its lower bound is measured above the separability threshold~\eqref{thresh}, the presence of MPE is certified. 

\begin{figure}[ht!]
	\begin{center}
		\includegraphics[width=1.0\columnwidth]{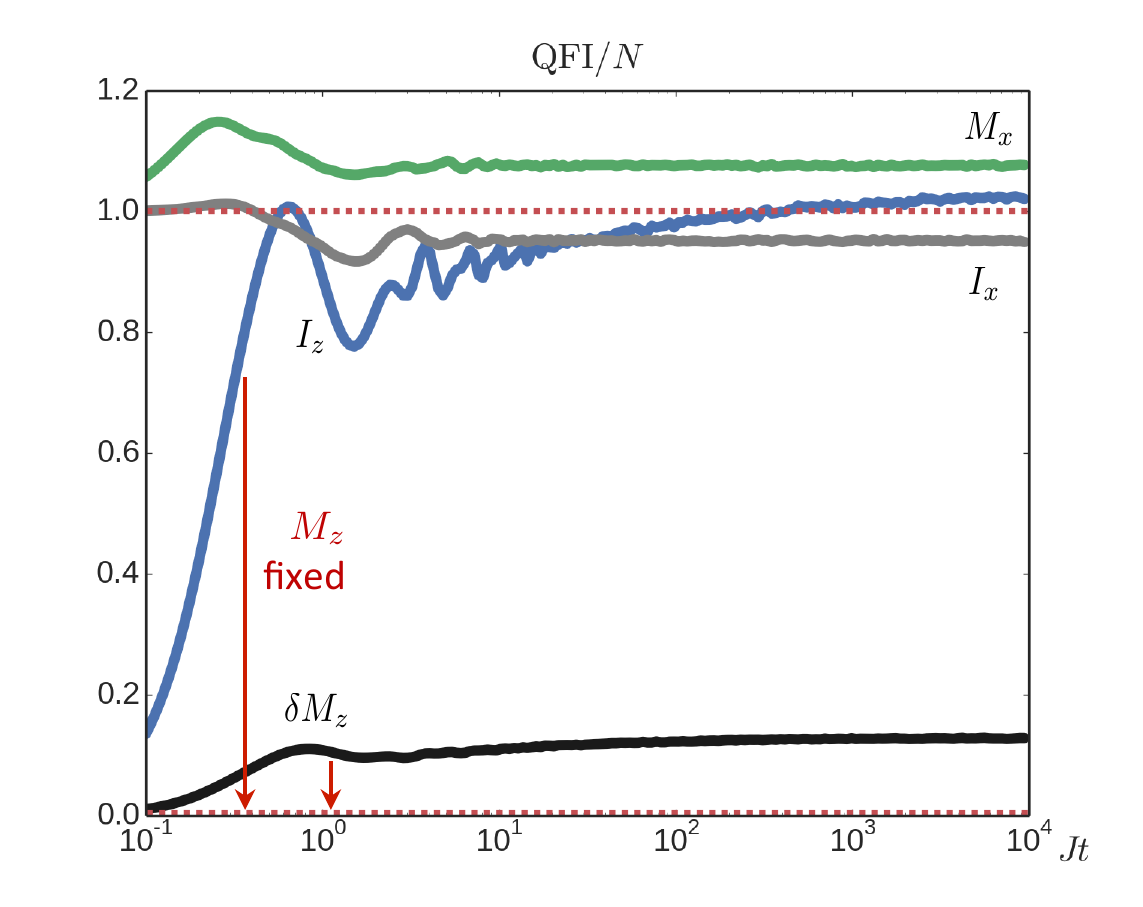}
		\vspace*{-4mm}
		\caption{
			{\bf QFI in an MBL system}. Throughout the paper we illustrate our results with the example of a many-body localized system, an $XXZ$ chain with strong disordered longitudinal field (see details in Sec.~\ref{sec:MBL}).
			The figure shows the evolution of the QFI per number of subsystems $N$. We consider four different phase encodings: $x$ magnetization $M_x\equiv\sum_{k=1}^N S^x_k$ [green (top) line], the difference of $z$ magnetization between two halves of the chain,  $\delta M_z\equiv\sum_{k=1}^{N/2} S^z_k- \sum_{k=N/2+1}^{N} S^z_k$  [black (bottom) line],  and $x$ and $z$ imbalance $I_{x,z}\equiv\sum_{k=1}^N (-1)^k S^{x,z}_k$ [blue and gray (at $Jt=10^{-1}$ lower and upper middle) lines, respectively]. The QFI witnesses entanglement only if the corresponding separability threshold $\text{QFI}_\text{sep}/N=1$ [red dashed (horizontal) line] is crossed, cf.~\er{thresh}. Due to the conservation of $M_z$ by the dynamics, for an initial state with a fixed $M_z$, the spin axes $x$ and $y$ are equivalent (and the QFI for $M_y$ and $I_y$ is equal to that the QFI for $M_x$ and $I_x$, respectively). For observables commuting with $M_z$ (here $\delta M_z$ and $I_z$) the separability threshold is reduced to zero, see Sec.~\ref{sec:witness}. The parameters of the dynamics [cf.~\era{XXZ}{master}] are $N=14$ spins, $V/J=2$, $h/J=5$, and $\gamma/J=0$.
		}\vspace*{-10mm}
		\label{fig:QFI}
	\end{center}
\end{figure}
Given that QFI witnesses MPE only if it goes beyond the threshold~\eqref{thresh}  that scales linearly with the system size, its usefulness as an entanglement witness is severely limited in many situations. This is what can occur, for example, when attempting to observe logarithmic growth of entanglement in MBL experiments~\cite{Smith2016} (see Fig.~\ref{fig:QFI}). Furthermore, since entanglement is not related to any specific basis, in general it is not obvious which local observables $M^{(k)}$ in~\eqref{M} are optimal, i.e., lead to the maximal value of the QFI (cf.~Fig.~\ref{fig:QFI}), and all local unitary transformations of the operators $M^{(k)}$ should be considered. Even with the optimal choice of the observable, crossing of the separability threshold is not guaranteed for entangled states~\cite{Hyllus2010}. Take the example of the Werner state of two qubits in~\eqref{Werner}. From the convexity of the QFI we have ${\rm QFI}(M,\rho_W)\leq p\ {\rm QFI}(M, |\Psi^-\rangle )\leq  4 p $, where the last inequality is achieved for the $M$ chosen as the imbalance in the $z$ direction (the staggered $z$ magnetization) $I_z=\sum_{k=1}^2 (-1)^k S^z_k$. Thus, for $p\leq 1/2$ the QFI cannot exceed the separability threshold ${\rm QFI}_{\rm sep}=N=2$, although the Werner state is entangled for $p>\frac{1}{3}$, as it is confirmed, e.g., by the concurrence~\cite{Hill1997}. %Interestingly, for pure separable states, i.e.,\ product states, the local unitary optimisation of the observables always leads to the QFI given by the separability threshold.
In Sec.~\ref{sec:witness} we will show that these issues are remedied in the presence of a fixed local charge, when the separability  threshold is reduced to zero for any observable commuting with the charge (see also Fig.~\ref{fig:QFI}).

We note that other methods to witness MPE with coherence include, among others, those in Refs.~\cite{Hofmann2003,Li2013,Toth2012,Marty2017} and the multiple quantum coherence (MQC) spectra~\cite{Garttner2018}, which we discuss in the next section. We also note that QFI can be considered a measure of quantum macroscopity~\cite{Yadin2016} (see also~\cite{Lee2011,Park2016}).

%Instead of searching for the optimal set of $M^{(k)}$, one can consider $M$ composed  from identical $M^{(k)}$ chosen as an element in an orthonormal basis of subsystem observables. The separability threshold for the average QFI (the sum of QFIs over all the elements of the local basis, which is basis-independent) scales both with the system size $N$ and the dimension $d$ of subsystems as $N(d-1)$, which is the value achieved by all pure separable states~\cite{Li2013}. The dual variance criterion~\cite{Hofmann2003} helps in a complementary way to detect the entanglement. Alternatively, for spin-$\frac{1}{2}$ systems the average QFI with respect to the total magnetization along $x$ , $y$  and $z$ axes, admits the separability threshold equal to $2N$, attained by any pure separable state~\cite{Toth2012}. The separability threshold additionally can be made dependent on the breaking of the permutation symmetry in a quantum state by using the generalized variance~\cite{Marty2017}.

\subsection{Coherence and asymmetry monotones}  \label{subsec:coherence}

 In Sec.~\ref{sec:measures} we will derive bounds on the amount of entanglement or quantum discord in systems with a fixed or conserved local charge, respectively, which depend on the amount of coherence and asymmetry present in the system and hold both for closed and open dynamics. Therefore, we now discuss how to quantify the amount of coherence present in a quantum state.

Below we review examples of coherence monotones, which are convex functions of a quantum state that attain a value of zero only for diagonal (incoherent) states, and are strongly nonincreasing under incoherent operations (operations which preserve the set of diagonal states both on average and probabilistically)~\cite{Baumgratz2014} (for a review see, e.g.,~\cite{Streltsov2017b}).

\subsubsection{Relative entropy of coherence} 

A coherence monotone can be defined as a \emph{bona fide} distance of a given state $\rho$ from the set of incoherent, i.e.,\ diagonal, states~\cite{Baumgratz2014}. For example, if we choose as the distance the relative entropy  $S(\rho || \sigma) \equiv\Tr \rho \log_2 \rho - \Tr\rho \log_2 \sigma \geq 0$,  we obtain the \emph{relative entropy of coherence}~\cite{Herbut2005,Aberg2006,Vaccaro2008,Gour2009,Baumgratz2014}
\begin{equation}
\C(\rho)\equiv\min_{\sigma_\text{diag}}S(\rho || \sigma_\text{diag})=-S(\rho)+S(\rho_\text{diag}),\label{Scoh}
\end{equation}
where $S(\rho)\equiv-\Tr \rho \log_2 \rho$ is the von Neumann entropy and $\rho_{\rm diag}\equiv\sum_i \rho_{ii}|i\rangle\!\langle i|$ is obtained from $\rho$ by removing all coherences. This is a consequence of the closest state from the incoherent set being $\rho_\text{diag}$, since for any other incoherent state $\sigma_\text{diag}$ we have $S(\rho || \sigma_\text{diag})=S(\rho || \rho_\text{diag})+S(\rho_\text{diag}||\sigma_\text{diag})\geq S(\rho || \rho_\text{diag})$, where the equality follows from the fact that $\rho_{\rm diag}$ and $\sigma_\text{diag}$ share the same eigenbasis. 

For the example of the Werner state,~\eqref{Werner}, we have that  $S(\rho_W)=-\frac{3 (1-p)}{4} \log_2 \frac{1-p}{4} -\frac{1+3p}{4} \log_2 \frac{1+3p}{4} $, while $S(\rho_\text{diag})=-\frac{ 1-p}{2} \log_2 \frac{1-p}{4} -\frac{1+p}{2} \log_2 \frac{1+p}{4}$, so $\C(\rho_W)=-\frac{1+p}{2} \log_2 \frac{1+p}{2} +\frac{1+p}{2}[1+ \tilde{p} \log_2 \tilde{p} +(1-\tilde{p})\log_2(1-\tilde{p})]$, where $\tilde{p}=\frac{1-p}{2(1+p)}$. Therefore, indeed $\C(\rho_W)>0$ for all $p>0$  when the state is coherent [we then have $\tilde{p}\neq 1/2$, so both terms in $C(\rho_W)$ are positive].

The relative entropy of coherence $\C(\rho)$ has direct operational interpretation as it corresponds to distillable coherence, i.e.,\ the rate at which maximally coherent qubits can be asymptotically distilled from many copies of $\rho$ by using an incoherent operation~\cite{Yuan2015,Winter2016}.  When the state $\rho$ is pure, the relative entropy of coherence equals  $S(\rho_\text{diag})$, cf.~\eqref{Scoh}, which can be accessed  by measuring occupation in the basis $\{|i\rangle\}_{i=1}^D$. For mixed states, however, it cannot be easily measured in experiment, as $S(\rho)$ generally requires full quantum state tomography (with a few exceptions such as noninteracting fermions~\cite{Song2012}).

In Secs.~\ref{subsec:measureMPE} and~\ref{subsec:measureMPD} we show that the relative entropy of coherence is a faithful upper bound on the amount of multipartite entanglement or multipartite discord in states with a fixed or conserved local charge, respectively.

\subsubsection{$l_1$ coherence}

Another coherence monotone that fulfills the axioms of resource theory of coherence with incoherent operations~\cite{Baumgratz2014} is the $L_1$-norm,
\begin{equation}
l_{1}(\rho) \equiv \sum_{i \neq j} | \rho_{ij} | .
\label{l1}
\end{equation}
For the example of the Werner state~\eqref{Werner}, we simply have $l_1(\rho_W)=p$, which is nonzero for all $p>0$. 

In this work we show that~\eqref{l1} is a faithful upper bound on multipartite discord measured by negativity of quantumness~\cite{Piani2011,Nakano2013,Adesso2016}. While $l_{1}(\rho)$ is a monotone of coherence, being nonpolynomial in $\rho$ means it cannot be directly related to observations, but requires full quantum tomography of the system state $\rho$. 

We will now introduce a new lower bound on~\eqref{l1}, which entails the so-called multiple quantum coherence spectrum~\cite{Baum1985,Baum1986,Munowitz1987} that can be experimentally accessed also in many-body systems~\cite{Baum1985,Macri2016,Garttner2017,Garttner2018,Wei2018} (see also Appendix~\ref{sec:method}). The MQC spectrum is defined for an observable $M$ diagonal in the computational basis $M=\sum_{i} m_i |i\rangle\!\langle i| $ as
\begin{equation}
I_m(\rho)\equiv \sum_{ij:\, m_i-m_j=m} |\rho_{ij}|^2,
\label{Im}
\end{equation}
where $\rho_{ij}\equiv \langle i|\rho| j\rangle$. 
We introduce 
\begin{equation}\label{l1blockM}
 l_{1,M}^\text{block}(\rho) \equiv \sum_{m\neq 0} \sqrt {I_m(\rho)}.
\end{equation}
From the inequality between $L_1$- and $L_2$-norms, we then obtain an experimentally accessible lower bound on $l_1$ coherence as, 
\begin{eqnarray}
l_{1}(\rho) &\geq & \!\! \sum_{ij:\,m_i\neq m_j} |\rho_{ij}|= \sum_{m\neq 0}\, \sum_{ij:\,m_i-m_j=m} |\rho_{ij}|\nonumber\\
 &\geq&  \sum_{m\neq 0} \sqrt{\sum_{ij:\,m_i-m_j=m} |\rho_{ij}|^2} = l_{1,M}^\text{block}(\rho).\label{l1blockM2}
\end{eqnarray}
In general, as $I_m(\rho)<1$, we have that the experimentally accessible lower bound fulfills $l_{1,M}^\text{block}(\rho)<d$, where $d$ is the number of different gaps in the $M$ spectrum, which for local observables, e.g., magnetization, makes it feasible to detect experimentally the growth of coherence with the system size. In general, the first inequality in  Eq.~\eqref{l1blockM2} is only saturated when $M$ is a nondegenerate observable, while the second inequality is saturated when all the gaps in $M$ spectrum are nondegenerate. Therefore, when $d=D(D-1)/2$, the bound in Eq.~\eqref{l1blockM2} is saturated. In particular, in the example of the Werner state, for the choice of the observable $M=I_z$ as the imbalance, we indeed have $l_{1,I_z}^\text{block}(\rho_W)=p=l_1(\rho_W)$ ($I_z$ features two different gaps, $m=\pm2$ in the subspace of $|\!\downarrow \!\uparrow \rangle$, and $|\!\uparrow \!\downarrow \rangle$, where the coherence is present).

\subsubsection{Asymmetry monotones}

In the case when not all basis states $|i\rangle$ in $\{ |i\rangle : i = 1,\ldots,D \}$ are distinguishable (e.g., due to degeneracy of a system Hamiltonian $H=\sum h_i |i\rangle\!\langle i|$), the amount of coherence between distinguishable  subspaces (e.g., energy eigenspaces) can be quantified with the resource theory of asymmetry as~\cite{Bartlett2007,Gour2008,Vaccaro2008,Gour2009,Marvian2012}  
\begin{equation}
\A(\rho)\equiv -S(\rho)+S(\rho_\text{block}),
 \label{Sas}
\end{equation}
where $\rho_\text{block}$ is obtained from $\rho$ by removing all coherences (i.e.,\ dephasing) between the different distinguishable subspaces (e.g., Hamiltonian eigenspaces $\rho_\text{block}=\sum_{i,j:\, h_i=h_j}\rho_{ij}\,|i\rangle\!\langle j|$). In the nondegenerate case~\cite{Marvian2016} we recover $\A(\rho)=\C(\rho)$ [cf.~\er{Scoh}]. For pure states,~\eqref{Sas} can be accessed experimentally by measuring occupation in the distinguishable subspaces (e.g., energies in the Hamiltonian), as the entropy of this distribution equals $S(\rho_\text{block})$. In Sec.~\ref{subsec:measureBP} we will consider systems with a fixed local charge and show that the bipartite entanglement is lower bounded by the asymmetry~\eqref{Sas} with respect to the charge of a subsystem in the partition. \\

In analogy to $l_1$ coherence,~\eqref{l1}, for $h$ indexing distinguishable subspaces (e.g., $h$ being an eigenvalue of the system Hamiltonian indexing energy eigenspaces), we now introduce
\begin{eqnarray}
l_{1}^\text{block}(\rho)&\equiv& \sum_{h\neq h'}\,  \sqrt{\sum_{i:\,h_i=h}\sum_{j:\,h_j=h'}  |\rho_{ij}|^2}.
\label{l1block}
\end{eqnarray} 
Note that $l_1^\text{block}$  is independent of the choice of basis elements inside a degenerate subspace.
It is not known whether it is an asymmetry monotone~\footnote{It may be an asymmetry monotone only with respect to a restricted subset of translationally invariant operations, in analogy to coherence monotones for the set of incoherent operations~\cite{Baumgratz2014} and the set of maximally incoherent operations~\cite{Aberg2006}.}. Nevertheless, it is bounded from below by experimentally accessible $l_{1,M}^\text{block}$ [Eq.~\eqref{l1blockM}], 
\begin{equation}
l_{1}^\text{block}(\rho)\geq l_{1,M}^\text{block}(\rho) \label{l1block2}
\end{equation}
when $M$ is diagonal in the computational basis, and chosen so  its eigenvalues for states $|i\rangle$ and $|j\rangle$ fulfill$m_i=m_j$ whenever $h_i=h_j$, cf. Eq.~\eqref{l1blockM2}. For example, when the observable is chosen as the Hamiltonian, $M=H$, the inequality \eqref{l1block2} is saturated for $H$ with nondegenerate gaps.

As we will show in Sec.~\ref{subsec:measureBP}, $l_{1}^\text{block}$ serves as a lower bound on bipartite entanglement in states with a fixed local charge and is experimentally accessible [cf.~\eqref{l1blockM} and~\eqref{l1block2}].

\section{Vanishing separability thresholds from fixed or conserved local charge}  \label{sec:witness}

We now derive the first set of our results. After defining a fixed and conserved charge in Sec.~\ref{subsec:charge}, in Sec.~\ref{subsec:witnessMPE}  we show how coherence implies MPE for states with a fixed local charge and how with an appropriate choice of observables the QFI faithfully witnesses MPE with the zero separability threshold (cf.~Sec.~\ref{sec:QFI}). In Sec.~\ref{subsec:witnessBPE} we further discuss how by appropriate choice of observables, also bipartite entanglement can be witnessed with zero threshold when a local charge is fixed. When a local charge is not fixed, but it is conserved, we demonstrate in Sec.~\ref{subsec:witnessMPD} that coherence is related to quantum discord.  We finish by discussing the relation to the superselection rules in Sec.~\ref{subsec:SSR}.

\subsection{Definition of fixed and conserved local charge} \label{subsec:charge}

We begin by defining a fixed and a conserved charge. A charge is understood to be an observable $Q$ on the system. When we consider a state of a multipartite quantum system consisting of $N$ subsystems, we refer to $Q$ as local when it is the sum of subsystem observables $Q=\sum_{k}Q^{(k)}$, e.g., a total magnetization of a spin-$\frac{1}{2}$ chain along the $z$ axis. 

When a quantum state $\rho$ is supported within only a single eigenspace of $Q$, 
\begin{equation}\label{fix}
Q \rho = q\, \rho,
\end{equation}
 we say that  this state is of a \emph{fixed charge} $q$ (see Fig.~\ref{fig:matrix}). For example, a general state of two qubits (two spins $\frac{1}{2}$) with total $z$ magnetization equal zero is given by (in the basis $|\!\!\downarrow \!\downarrow \rangle$, $|\!\!\downarrow \!\uparrow \rangle$, $|\!\!\uparrow \!\downarrow \rangle$, and $|\!\!\uparrow \!\uparrow \rangle$),
 \begin{eqnarray} \label{2Qfix}
 \rho_{2}^\text{fix}&\equiv&\left( \begin{array}{c c c c} 0&&&\\&p&c&\\&c^*&1-p&\\&&&0 \end{array}\right),
 %&\equiv& p |\!\downarrow \!\uparrow \rangle\!\langle 01| + (1-p) |\!\uparrow \!\downarrow \rangle\!\langle 10|\\\nonumber
 %&&+c\, |\!\downarrow \!\uparrow \rangle\!\langle 10|+ c^* |\!\uparrow \!\downarrow \rangle\!\langle 01|,
 \end{eqnarray}
 where $|c|^2\leq p(1-p)$ and $0\leq p\leq 1$. This state is coherent for all $|c|>0$.  %In particular, for $p=1/2$ and $c=\pm 1/2$ this is the Bell state, $|\Psi^\pm \rangle\equiv(|\!\downarrow \!\uparrow \rangle\pm|\!\uparrow \!\downarrow \rangle)/\sqrt{2}$. 

When the state $\rho$ features no coherences between different eigenspaces of $Q$, i.e.,\ no asymmetry with respect to $Q$, 
\begin{equation}\label{cons}
  [\rho,Q]= 0,
\end{equation}
we say that a quantum state is of a \emph{conserved charge} (see Fig.~\ref{fig:matrix}).  For example, the Werner state in Eq.~\eqref{Werner} is of conserved total $z$ magnetization (for all values $0\leq p\leq 1$), while the most general state of two qubits (two spins $\frac{1}{2}$) with conserved total $z$ magnetization is given by (in the basis $|\!\!\downarrow \!\downarrow \rangle$, $|\!\!\downarrow \!\uparrow \rangle$, $|\!\!\uparrow \!\downarrow \rangle$, and $|\!\!\uparrow \!\uparrow \rangle$)
 \begin{eqnarray} \label{2Qcons}
\rho_2^\text{cons}&\equiv&\left( \begin{array}{c c c c} p_{00}&&&\\&p_{01}&c&\\&c^*&p_{10}&\\&&&p_{11} \end{array}\right),
\end{eqnarray}
where $|c|^2\leq p_{01}p_{10}$, and $p_{ij}\geq 0$ with $\sum_{ij=0,1}p_{ij}=1$. The state $\rho_2^\text{cons}$ is coherent for all $|c|>0$.

\begin{figure*}[ht!]
	\begin{center}
		\includegraphics[width=\textwidth]{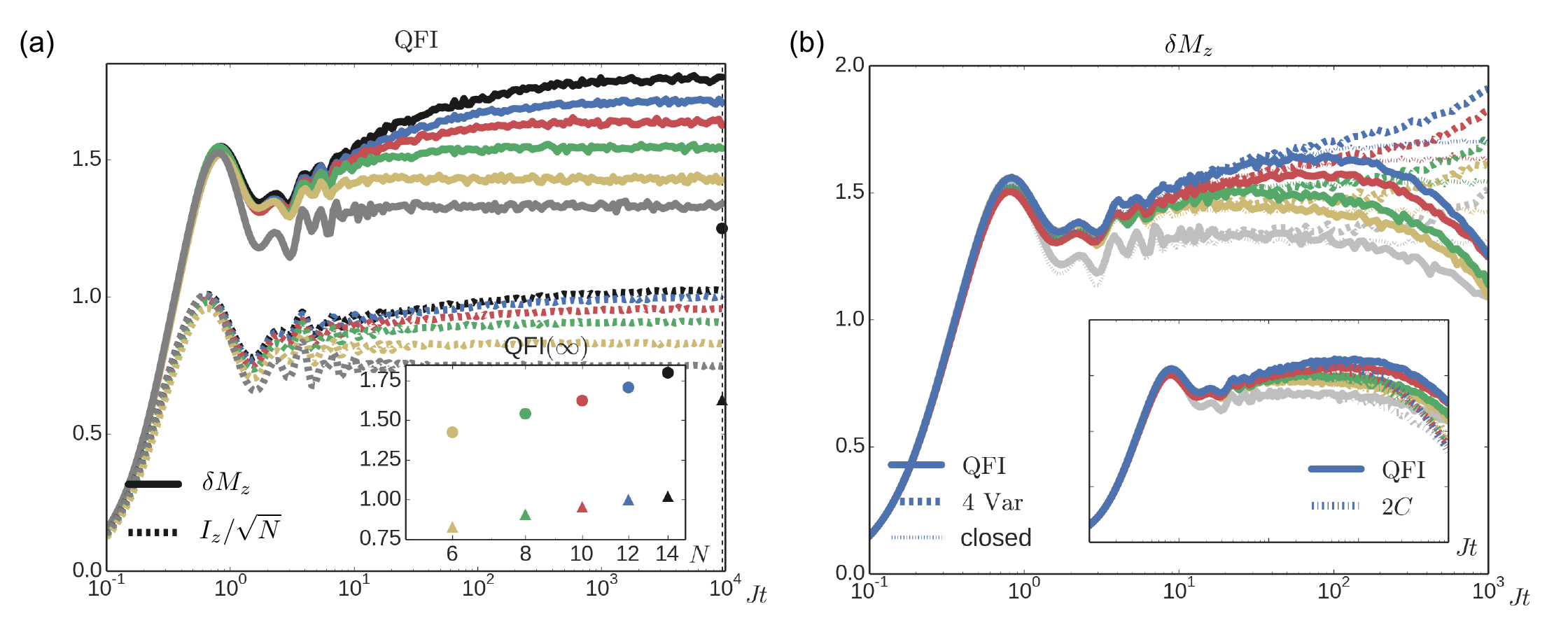} %{QFI_open_2.pdf}
		\vspace*{-7mm}
		\caption{
			{\bf Witnessing bipartite entanglement in an MBL system with conserved charge}. 
			We show for the $XXZ$  chain of Sec.~\ref{sec:MBL} that not only MPE can be witnessed, but also BPE, provided that the phase encoding observable commutes with subsystem charges $Q^{(A)}$ and $Q^{(B)}$ in the bipartition.  
			(a) For closed dynamics ($\gamma/J=0$) the QFI of the $z$-magnetization difference between two halves of the chain (solid lines) witnesses BPE at all times $t>0$. Similarly, the QFI of the $z$ imbalance (dashed lines) witnesses BPE in the staggered partition ($ABAB \cdots AB$ instead of $AA \cdots AB \cdots BB$) at all times $t$. The inset shows that asymptotic values (taken from $Jt=10^4$) of the QFI for the $z$-magnetization difference (circles) and the $z$ imbalance (triangles), as a function of size $N$, grow with system size (cf.\ Ref.~\cite{Bardarson2012}). For the $z$ imbalance the asymptotic value grows with $N$ even after rescaling by the system size. 
			%Interestingly, for $M_x$ and $I_x$ (not shown), $\mathrm{QFI}/N$ is independent of system size, see also Fig.~\ref{fig:QFI}. 
			(b) In the presence of local dephasing  ($\gamma/J=2\times 10^{-4}$) the QFI of the magnetization difference (solid lines) decays in comparison with the closed case (dotted lines), while the variance (dashed line) increases, thus overestimating the QFI [cf.~\er{C2}]. 
			The inset shows that both the QFI (solid lines) and its lower bound in terms of the curvature (dotted lines) (defined in Appendix~\ref{sec:method}) witness BPE for the observable chosen as the magnetization difference, although the curvature decays at faster rate dependent on the system size~\cite{Yadin2016,Garttner2017}. 
			The parameters of the dynamics [cf.~\era{XXZ}{master}] are 
			$N = 6, 8, 10, 12, 14$ [yellow, green, red, blue, and black (grayscale: light gray to black), respectively, open case only up to $N=12$],
			$V/J=2$, $h/J=5$, and $\gamma/J=2\times 10^{-4}$; gray (bottom) curves correspond to the noninteracting case $V/J=0$ with $N=12$ (closed dynamics) and $N=8$ (open dynamics). Here the QFI is independent of system size as entanglement obeys the  area law. 
		}\vspace*{-5mm}
		\label{fig:QFI_open}
	\end{center}
\end{figure*}

We now explain when, in a system undergoing closed or open dynamics, a fixed or a conserved charge are present at all times. In unitary dynamics, the state of the system at time $t$ is given by $\rho_t=e^{-it H} \rho_0 e^{it H}$, where $\rho_0$ is the initial state and $H$ the Hamiltonian of the system. When the Hamiltonian $H$ conserves the charge $Q$, $[Q,H]=0$, we have that $\rho_t$ is of a fixed (conserved) charge whenever the initial state $\rho_0$ is of a fixed (conserved) charge. This is the case for closed dynamics of the $XXZ$  spin chain with disorder, which we discuss in details in Sec.~\ref{sec:MBL} and show in Figs.~\ref{fig:QFI}--\ref{fig:thermal}. In this system the total magnetization in the $z$ direction is conserved. 

The same holds also in the case of open dynamics governed by a master equation~\cite{Lindblad1976,Gorini1976} when $Q$ is a generator of a strong symmetry of the dynamics~\cite{Buca2012,Albert2014}, i.e., both the Hamiltonian and jump operators in the master equation conserve the charge. For example, the disordered $XXZ$  spin chain in the presence of dephasing, which we also discuss in Sec.~\ref{sec:MBL}, fulfills this condition for the total magnetization in the $z$ direction. Finally, in the case when the local charge $Q$ is only a generator of weak symmetry of the dynamics~\cite{Buca2012,Albert2014}, i.e., the master equation commutes with the symmetry generated by $Q$ only as a whole, when $\rho_0$ is of a conserved charge, so is $\rho_t$. This is the case when the $XXZ$  spin chain exchanges excitations with the environment (see Sec.~\ref{sec:MBL}).\\

In Secs.~\ref{subsec:witnessMPE} and~\ref{subsec:witnessBPE} we show that entanglement in the presence of a fixed local charge is implied by the coherence with respect to subsystem charges. In the presence of a conserved local charge,  coherence instead implies quantum discord, as we argue in Sec.~\ref{subsec:witnessMPD}.

\subsection{Fixed local charge and witnessing multipartite entanglement} \label{subsec:witnessMPE}
Any additional information about a state $\rho$, in which the presence of multipartite entanglement is to be determined, can be used to refine the separability threshold,~\eqref{thresh}, as the threshold should be computed for separable states consistent with that information. In particular, we will now show that for a state with a fixed local charge, the separability threshold can be reduced to zero.

First, let us consider systems of a fixed charge $Q$ of a value $q$ [Eq.~\eqref{fix}].  For a mixed state, $\rho = \sum_{j} p_{j} \rho_{j}$ with probabilities $p_j$, a fixed charge implies that the charge of all $\rho_{j}$ is also fixed to $q$, i.e., $Q\rho_{j} = q\rho_{j}$, as follows. A fixed charge implies that $\var(Q,\rho) = 0$, but we also have 
\begin{equation}\label{Qvar0}
\var(Q,\rho) = \sum_j p_{j} \var(Q,\rho_j)+\sum_{j>j'} p_j p_{j'} (\langle Q\rangle_j-\langle Q\rangle_{j'})^2 , 
\end{equation}
which together give 
$\var(Q,\rho_{j}) = 0$ and $\langle Q\rangle_j=\langle Q\rangle_{j'}$ for all $j,j'$,
so  
\begin{equation}\label{Qvar1}
Q \rho_{j}  = q \,\rho_{j} .
\end{equation}

Second, for a separable state $\rho_{\rm sep}$ [Eq.~\eqref{rho:sep}], we have that $\rho_{j}= \varrho_{j}^{(1)} \otimes \cdots \otimes \varrho_{j}^{(N)}$ are product states. When the charge $Q$ is local, $Q = \sum_{k=1}^{N} Q^{(k)}$, we then have that 
\begin{equation}\label{Qvar2}
\var(Q, \rho_{j}) =
\sum_{k=1}^{N} p_{j} \var(Q^{(k)}, \varrho_{j}^{(k)}) .  
\end{equation}
This means that $\var(Q,\rho_{\rm sep}) = 0$ implies $\var(Q^{(k)}, \varrho_{j}^{(k)})=0$, that is,  the charge is fixed locally for each of the states $\varrho_{j}^{(k)}$.  In particular, when the operators $Q^{(k)}$ are nondegenerate, e.g., a single-spin magnetization, this implies that $\varrho_{j}^{(k)}$ are pure eigenstates of $Q^{(k)}$, so all $\rho_{j}$ are elements of one basis of the Hilbert space of the system. In other words, in the case of the fixed charge with nondegenerate operators $Q^{(k)}$, all separable states are always diagonal and with zero coherence in this basis. Moreover, as all diagonal states in the computation basis are separable, we conclude that:
\begin{quotation}
	\noindent
	{\emph{a quantum state with fixed local charge is \\
			coherent if and only if it is entangled}}.
\end{quotation} 
For example, the state of $N=2$ qubits with a fixed magnetization~\eqref{2Qfix} is entangled for all $|c|>0$.

Therefore, when the QFI,~\eqref{qfi} of any diagonal (i.e., commuting with local charges) observable $M$ is nonzero, the state is entangled, as {\em the separability threshold vanishes} [cf.\ \er{thresh} and Fig.~\ref{fig:QFI}]. Furthermore, when $M$ is nondegenerate, e.g., chosen as a linear combination of the local charges with appropriate fields $M= \sum_{k-1}^N h_{k} Q^{(k)}$,  the QFI becomes a {\em faithful} witness of MPE. We also note that the experimentally accessible lower bound on the QFI in terms of the curvature (see Appendix~\ref{sec:method}) is a faithful witness as well.

The case of degenerate $Q^{(k)}$ is discussed in the next section, where we consider bipartite entanglement.

\subsection{Fixed local charge and witnessing bipartite entanglement}\label{subsec:witnessBPE}

We now discuss how, in the presence of a fixed local charge, bipartite entanglement between two parts of a quantum system is related to the asymmetry with respect to the asymmetry of the charge difference.

A state $\rho$ is \emph{bipartite entangled}, if it is not \emph{bipartite separable} $\rho\neq\rho_\text{BP-sep}$, where
\begin{equation}
\label{rho:sepBP}
\rho_\text{BP-sep}\equiv\sum_{j} p_j \varrho_j^{(A)}\otimes\varrho_j^{(B)}
\end{equation}
with $\varrho_j^{(A)}$ ($\varrho_j^{(B)}$) being a state of subsystem $A$ ($B$) in the bipartition of the system [cf. Eq.~\eqref{rho:sep}]. In particular, in this work we consider a system composed of $N$ subsystems, which we divide into two groups labeled $A$ and $B$, e.g., two halves of a spin chain, and refer to them as subsystems $A$ and $B$ (see Fig.~\ref{fig:QFI_open}). \\

For a local observable $M=\sum_{k=1}^N M^{(k)}$ [Eq.~\eqref{M}], the $AB$-separability threshold corresponds to a tensor product of maximally entangled states inside $A$ and $B$ subsystems [cf.~\eqref{thresh}]
\begin{eqnarray}
{\rm QFI}_{\rm BP-sep}(M) &\equiv& \max_{\rho_\text{BP-sep}} {\rm QFI}(M,\rho_\text{BP-sep}) \\
&=& \Big(\sum_{k\in A} \Delta M^{(k)}\Big)^{2}+\Big(\sum_{k\in B} \Delta M^{(k)}\Big)^{2}, \nonumber
\label{threshAB} 
\end{eqnarray} 
where $\Delta M^{(k)}$ is the difference between the extreme eigenvalues of $M^{(k)}$.
This threshold in general scales quadratically in the subsystem size. For the case of  $N$ spin-$\frac{1}{2}$  particles with $M$ chosen as the staggered $z$ magnetization, $M=I_z\equiv\sum_{k=1}^N(-1)^k S_z^{(k)}$, we have that the separability threshold for the half chain partition (when $N$ is even) equals ${\rm QFI}_{\rm BP-sep}(I_z)=N^2/2$, and is achieved for the state $(|\!\downarrow \!\uparrow \!\downarrow\!\uparrow..\!\downarrow\!\uparrow \rangle+|\!\uparrow\!\downarrow\!\uparrow\!\downarrow..\!\uparrow\!\downarrow \rangle)\otimes(|\!\downarrow \!\uparrow\!\downarrow\!\uparrow..\!\downarrow\!\uparrow \rangle+|\!\uparrow \!\downarrow\!\uparrow\!\downarrow..\!\uparrow\!\downarrow \rangle)/2$. Therefore, in general, other methods to detect BPE are used~\cite{Daley2012,Islam2015}. We will now show, however, that in the presence of a fixed local charge, when an observable $M$ is chosen to commute with the subsystem charges, e.g., as the subsystem charge difference, the separability threshold~\eqref{threshAB} is again reduced to zero [cf.\ Fig.~\ref{fig:QFI_open}(a)].\\

Consider a quantum state of a fixed local charge,~\eqref{fix}, with the subsystem charges denoted by $Q^{(A)}$ and $Q^{(B)}$, $Q= Q^{(A)}+Q^{(B)}$, being in general degenerate (for systems with $N$ particles, $Q^{(A)}\equiv\sum_{k\in A} Q^{(k)}$ and $Q^{(B)}\equiv\sum_{k\in B} Q^{(k)}$, e.g., the magnetizations of the two parts of a spin chain). Although the charge is fixed~\eqref{fix}, the subsystem charges  $Q^{(A)}$ and $Q^{(B)}$ do not have to be in general. For a bipartite-separable state~\eqref{rho:sepBP}, however, we obtain from the fixed charge condition that
\begin{equation}
\var(Q^{(A)},\varrho_{j}^{(A)})=0=\var(Q^{(B)},\varrho_{j}^{(B)})
\end{equation}
[cf. Eqs.~\eqref{Qvar0}-\eqref{Qvar2}] and thus $\varrho_{j}^{(A)}$ and $\varrho_{j}^{(B)}$ are of a fixed subsystem charge $Q^{(A)}$ and $Q^{(B)}$, respectively. Therefore, the bipartite separable state $\rho_\text{BP-sep}$ is  block diagonal, 
\begin{equation}
0=[\rho^\text{BP}_\text{sep},Q^{(A)}]=[\rho^\text{BP}_\text{sep},Q^{(B)}],
\end{equation}
with respect to (degenerate) eigenspaces of $Q^{(A)}$ and $Q^{(B)}$, which are equivalent as the total charge $Q$ is fixed (cf.\ Fig.~\ref{fig:matrix}).

We conclude that the asymmetry (block coherence) with respect to the subsystem charge $Q^{(A)}$ or $Q^{(B)}$ (or equivalently $\delta Q=Q^{(A)}-Q^{(B)}$) implies the presence of BPE, i.e.
\begin{quotation}
	\noindent
	{\emph{a quantum state with fixed local charge is block-coherent
			 only if it is bipartite entangled}}.
\end{quotation} 
Furthermore, any block-diagonal observable $M$, i.e., an observable commuting with the subsystem charges $[M,Q^{(A)}]=0=[M,Q^{(B)}]$ such as the charge difference ($M=\delta Q$), e.g.,  a difference of the subsystem magnetizations for a fixed total magnetization, encodes phases only in coherences between different values of a subsystem charge. Therefore, the corresponding {\em QFI is a witness of BPE with zero separability threshold} (see Fig.~\ref{fig:QFI_open}). 

 We need to note, however, that a bipartite-entangled state does not need to feature asymmetry, but can be entangled within the individual blocks, e.g., $|\widetilde{\Psi}^-\rangle\equiv(|\!\!\uparrow \!\downarrow \rangle\otimes|\!\!\downarrow \!\uparrow \rangle-|\!\!\downarrow \!\uparrow \rangle\otimes|\!\!\uparrow \!\downarrow \rangle)/\sqrt{2}$ is a Bell $|\Psi^-\rangle$-like state, but without asymmetry $(q^{(A)}=0=q^{(B)})$. Therefore, the QFI for block-diagonal observables is in general \emph{not} faithful witness of BPE, as it cannot detect entanglement inside the blocks.

\subsection{Conserved local charge and witnessing quantum discord}\label{subsec:witnessMPD}

Let us stress that a fixed value of the charge $Q$ [Eq.~\eqref{fix}] is essential for the coherence to imply MPE and the separability threshold to disappear. In general, a separable state $\rho$ conserving the local charge [Eq.~\eqref{cons}] can be coherent. We now give two examples. 

When $Q$ is the total $z$ magnetization of $N$ 1/2-spins, the symmetrized product state $|\psi(p)\rangle=\sqrt{p}|\!\downarrow \rangle+\sqrt{1-p}|\!\uparrow \rangle$, given by
\begin{equation}
\bar{\rho}_N\equiv\int_{0}^{2\pi}\mathrm{d}\phi\frac{1}{2\pi} e^{-i\phi Q}[|\psi(p)\rangle\!\langle\psi(p)|]^{\otimes N} e^{i\phi Q} ,
\label{rn}
\end{equation}
is separable, as $e^{-i\phi Q}$ is a local unitary. Nevertheless, the state coherence is nonzero for $0<p<1$, and we have [cf. Eq.~\eqref{Scoh}]
\begin{equation}\label{rnC}
C(\bar{\rho}_N)=\sum_{k=1}^N p^k (1-p)^{N-k} {{N}\choose{k}} \log_2 {{N}\choose{k}},
\end{equation}
and [cf. Eq.~\eqref{l1}]
\begin{equation}\label{rnl1}
l_1(\bar{\rho}_N)=\sum_{k=1}^N p^k (1-p)^{N-k} {{N}\choose{k}} \left[ {{N}\choose{k}} -1 \right].
\end{equation}

Another good example is given by the Werner state $\rho_W$ [Eq.~\eqref{Werner}]. This state is invariant under all identical local transformations~\cite{Werner1989} $U\otimes U\rho_W U^\dagger\otimes U^\dagger=\rho_W$, thus implying conservation of all local charges. However, as we already discussed in Sec.~\ref{subsec:coherence}, the Werner state is coherent for all $p>0$, although it is entangled only for $p\geq1/3$. Similarly, a general state of two qubits with conserved magnetization $\rho_2^\text{cons}$ [Eq.~\eqref{2Qcons}] is coherent for all $|c|>0$, while it is entangled only when $|c|^2> p_{00}p_{11}$, as confirmed e.g., by the partial transpose~\cite{Horodecki1997}.

We will show that, although in the presence of local charge conservation [Eq.~\eqref{cons}] coherence and asymmetry no longer imply MPE and BPE, they instead imply \emph{multipartite and bipartite discord}, respectively (cf.\ Fig.~\ref{fig:matrix}). In particular, the states $\rho_W$ [Eq.~\eqref{Werner}], $\rho_2^\text{cons}$ [Eq.~\eqref{2Qcons}], and $\bar{\rho}_N$ [Eq.~\eqref{rn}], are discordant (for $p>0$, $|c|>0$, and $1>p>0$, respectively).

\subsubsection{Witnessing multipartite quantum discord}

The weakest type of quantum correlations that can be present even in separable quantum states, i.e., states without entanglement, corresponds to quantum discord~\cite{Ollivier2001,Zurek2000,Henderson2001}. For \emph{multipartite discord}~\cite{Modi2010,Okrasa2011}, also referred to as \emph{quantumness}, the classical states (i.e., those without MPD) can be characterized as diagonal in some orthogonal separable basis
\begin{equation}\label{rho:cl}
\rho_{\text{cl}}\equiv\sum_{i_1,...,i_N} \lambda_{i_1\cdots i_N} |e_{i_1}^{(1)}\rangle\!\langle e_{i_1}^{(1)}|\otimes\cdots\otimes  |e_{i_N}^{(N)}\rangle\!\langle e_{i_N}^{(N)}|,
\end{equation}
where $\{|e_{i_k}^{(k)}\rangle\}_{i_k}$  is an orthonormal basis in $k$th subsystem, $k=1,...,N$~\cite{Modi2010}.
Therefore, the minimal coherence with respect to all separable bases can be used as a witness of multipartite discord~\cite{Yao2015,Bromley2015,Adesso2016,Ma2016} (see also~\cite{Frerot2016,Malpetti2016} for thermal states). %This manifests the fact that for classical states all the contributions to the coherence are local, and therefore can be removed by an appropriate choice of local basis.  \\

In Appendix~\ref{app:discord} we show that in the presence of conserved local charge [Eq.~\eqref{cons}] for classical states~\eqref{rho:cl} we have that
\begin{equation}\label{Qc0}
[Q^{(k)},\rho_\text{cl}]=0 \quad \text{for}\quad k=1,...,N,
\end{equation}
so  $|e_{i_k}^{(k)}\rangle$ can be chosen as eigenstates of $Q^{(k)}$ charges. When $Q^{(k)}$ are nondegenerate, this basis is uniquely defined and all classical states are diagonal. Therefore, for a quantum state all the contributions to coherence correspond to multipartite discord and all diagonal states are classical. We thus conclude that
\begin{quotation}
	\noindent
	{\emph{a quantum state with conserved charge is \\
			coherent if and only if it is discordant.}}
\end{quotation}
Therefore, the QFI of any  diagonal, i.e., commuting with local charges nondegenerate observable becomes a {\emph{faithful witness of multipartite discord}}, and usual minimization over local basis is no longer necessary.  We note that the experimentally accessible lower bound on the QFI in terms of curvature is also faithful (see Appendix~\ref{sec:method}).  \\

\subsubsection{Witnessing bipartite quantum discord}

We now show that for a system with a conserved local charge divided into two parts, the asymmetry with respect to a subsystem charge, implies bipartite quantum discord (see Fig.~\ref{fig:matrix}), analogously to the case for BPE with a fixed local charge.\\

For a given bipartition of a system into subsystems $A$ and $B$, a system state $\rho$ features  \emph{symmetric bipartite discord} if it is not a bipartite classical-classical state~\cite{Modi2010} $\rho\neq \rho_\text{cl-cl}$,
\begin{equation}\label{clcl}
\rho_\text{cl-cl}\equiv\sum_{i_A,i_B}\lambda_{i_A,i_B} |e_{i_A}^{(A)}\rangle\!\langle e_{i_A}^{(A)}| \otimes |e_{i_B}^{(B)}\rangle\!\langle e_{i_B}^{(B)}|,
\end{equation}
with $|e_{i_A}^{(A)}\rangle$ ($|e_{i_A}^{(B)}\rangle$) being an element of orthonormal bases of $A$ ($B$).

A classical-classical state~\eqref{clcl} with a conserved local charge $[Q^{(A)}+Q^{(B)},\rho_\text{cl-cl}]=0$ is \emph{block diagonal} with respect to the subsystem charges $Q^{(A)}$ and $Q^{(B)}$,
 \begin{equation}\label{Qc4}
 [Q^{(A)},\rho_\text{cl-cl}]=0=  [Q^{(B)},\rho_\text{cl-cl}]
 \end{equation}
 [cf. Eq.~(\ref{Qc0})].
 In other words, when the charges for $A$ and $B$ parts are degenerate, classical-classical states can feature coherence inside the subsystem charge eigenspaces, but no asymmetry, i.e., coherence between different eigenspaces of $Q^{(A)}$ and $Q^{(B)}$ (see Fig.~\ref{fig:matrix}).  Similarly, in Appendix~\ref{app:discord} we show that for any bipartite classical-quantum state,
 \begin{equation} \label{rho:clq}
\rho_\text{cl-q}\equiv\sum_{i}\lambda_{i} |e_{i}^{(A)}\rangle\!\langle e_{i}^{(A)}| \otimes \rho_i^{(B)},
 \end{equation}
 the local charge conservation~\eqref{cons} again implies~\eqref{Qc4}. Therefore, also quantum-classical states are block diagonal with respect to $Q^{(A)}$ and $Q^{(B)}$.

We conclude that the asymmetry with respect to the charges $Q^{(A)}$ or $Q^{(B)}$, implies the presence of both symmetric and asymmetric BPD, i.e.
 \begin{quotation}
 	\noindent
 	{\emph{a quantum state with conserved local charge is block-coherent
 			only if it is bipartite discordant}}.
 \end{quotation} 
 Furthermore, the QFI for block-diagonal (i.e., commuting with the subsystem charges) observables is a \emph{witness of BPD}, analogously to the fixed charge case (cf.~Sec.~\ref{subsec:witnessBPE}). We note that this last result is also implied by Ref.~\cite{Cianciaruso2017}, which establishes the difference between the QFI for the total charge and the sum of the QFI for the subsystem charges, as a witness of bipartite discord (also without charge conservation). 
 
 Finally, note that the QFI for block-diagonal observables is not faithful in general. Consider, e.g., a Werner-like state $\tilde{\rho}_W=(1-p)\tilde{\mathds{1}}/4+p|\tilde{\Psi}^-\rangle\!\langle\tilde{\Psi}^-|$, where $\tilde{\mathds{1}}$ is the identity operator on the subspace of zero magnetization in both halves of the spin chain and $|\widetilde{\Psi}^-\rangle\equiv(|\!\!\uparrow \!\downarrow \rangle\otimes|\!\!\downarrow \!\uparrow \rangle-|\!\!\downarrow \!\uparrow \rangle\otimes|\!\!\uparrow \!\downarrow \rangle)/\sqrt{2}$ also belongs to this subspace. The state $\tilde{\rho}_W$ features no asymmetry, but it is known to be discordant for $p>0$. Moreover, note that even though  $\tilde{\rho}_W$ is of fixed charge (zero total magnetization), it is bipartite entangled only for $p>1/3$. This illustrates that, in contrast to the multipartite case, bipartite-discordant states with fixed charge are not necessarily bipartite entangled.

\subsection{Superselection rules, entanglement and QFI}~\label{subsec:SSR}

We now briefly discuss the closely related theory of entanglement in the presence of a superselection rule (SSR)~\cite{Verstraete2003,Schuch2004,Schuch2004b,Bartlett2007}, e.g., for the total number of particles.  

In the presence of a superselection rule given by a conservation of a local charge, a state $\rho$ with the conserved charge can, by definition, be constructed only as a probabilistic mixture of states with a fixed local charge (in general with different values). A probabilistic mixture of separable states with fixed charges (which are diagonal; cf. Sec.~\ref{subsec:witnessMPE}) is diagonal. Thus,  the \emph{coherence in a state with the conserved charge necessary implies multipartite entanglement}.  For example, the separable states $\rho_W$ [Eq.~\eqref{Werner}] and $\bar{\rho}_N$ [Eq.~\eqref{rn}] cannot be constructed from separable states of fixed charges and thus are SSR entangled~\cite{Verstraete2003}. In other words, in the presence of a SSR all multipartite-discordant states are SSR entangled, while the QFI  for nondegenerate observables commuting with the charge becomes a faithful witness of SSR MPE. In the bipartite case,  the QFI of the charge difference in a given bipartition not only becomes a witness of BPE entanglement for states with the conserved (rather than a fixed) charge, but it also quantifies the \emph{nonlocality} of the state   (another, beyond BPE, resource implied by SSR~\cite{Popescu1997,Schuch2004,Schuch2004b})  as a convex roof of \emph{superselection induced variance} (cf.~\cite{Toth2013,Yu2013}). 

%In Appendix~\ref{app:SSR} we show that the above relations also hold in an extended entanglement theory for states with conserved charge, where local operations and classical communication (LOCC) that obey SSR , i.e.,\ do not change the charge, are replaced by separable operations preserving the charge conservation in quantum states.

\section{Bounds on entanglement and discord monotones for systems with fixed or conserved charge}  \label{sec:measures}

In this section we show how the relation between coherence and quantum correlations, be it entanglement, or discord (quantumness) can be strengthened. By this we mean going from simply witnessing entanglement or discord to establishing a stronger quantitative relation between  monotones for entanglement and monotones for coherence or asymmetry. The bounds we present in \eraaa{ineqMPE}{ineqMPD}{ineqS}{ineqN} 
are the second set of central results for this paper.

\subsection{Faithful upper bound on multipartite entanglement} \label{subsec:measureMPE}

A monotone of multipartite entanglement of a state $\rho$ can be defined in terms of the relative entropy between the state and the closest separable state~\cite{Vedral1997,Vedral1998},
\begin{equation}\label{mpe}
 \E_\text{MP}(\rho) \equiv \min_{\sigma_{\rm sep}} S(\rho || \sigma_{\rm sep}),
 \end{equation}
 where the relative entropy $S(\rho || \sigma) \equiv \Tr\rho \log_2 \rho - \Tr\rho \log_2 \sigma$.  In general, this minimum is difficult to evaluate, in particular for mixed states of many-body systems~\cite{Yichen2014}.

\begin{figure}[ht!]
	\begin{center}
		\includegraphics[width=1.0\columnwidth]{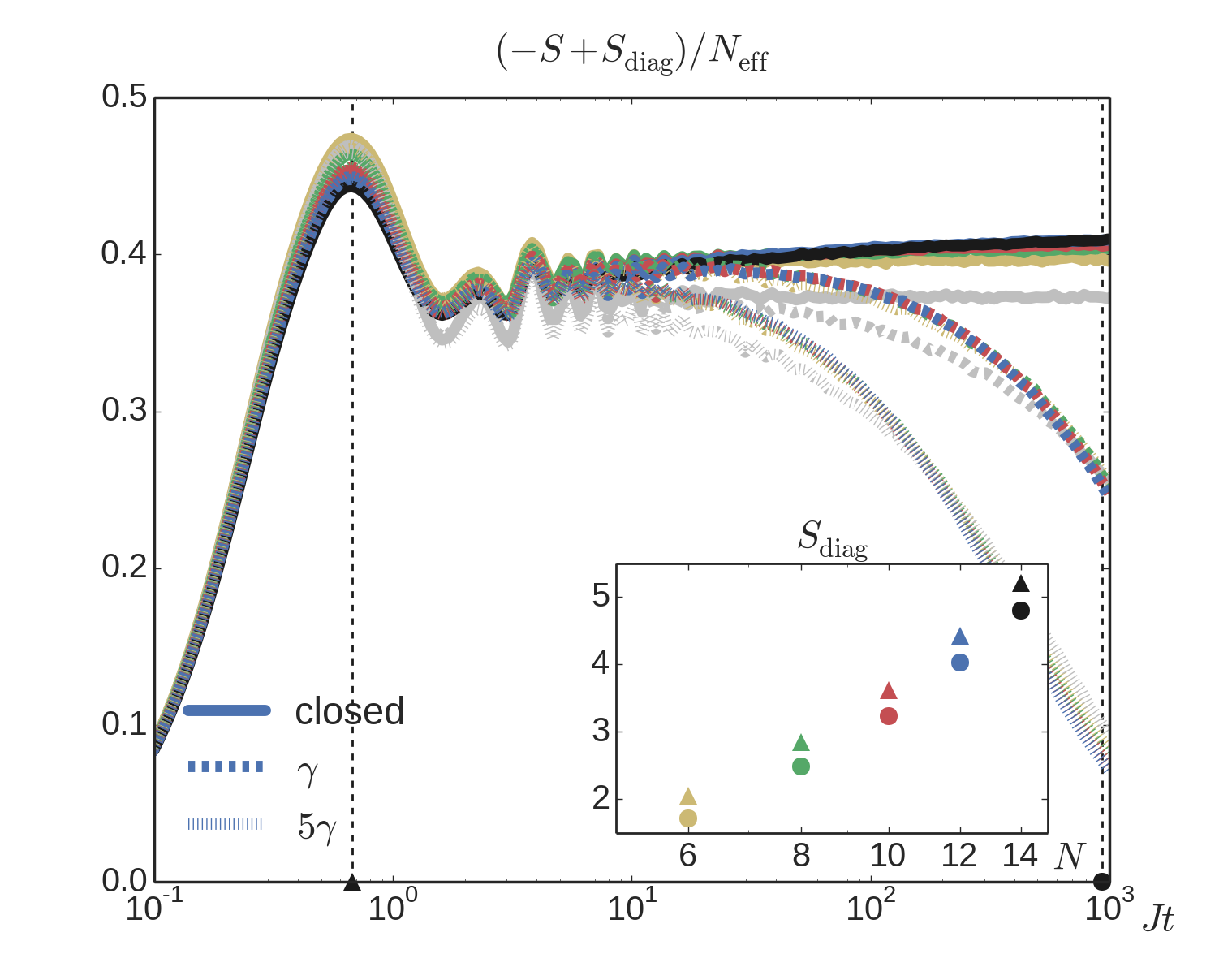}
		\vspace*{-4mm}
		\caption{
			{\bf Coherence as an upper bound on the MPE in an MBL system}. 
			We show coherence~\eqref{Scoh}, which is a faithful upper bound on MPE, cf.~\er{ineqMPE}, for the $XXZ$  chain of Sec.~\ref{sec:MBL}. The curves are rescaled by the effective system size $N_\text{eff}=\log_2{{N}\choose{N/2}}$ of the zero $z$ magnetization subspace. 
			In the presence of dephasing, coherence decays at a rate proportional to the dephasing strength $\gamma$ and is weakly dependent on the interaction strength, but not on the system size. 
			The inset shows that the coherence in the closed dynamics ($\gamma = 0$) at $t=1/J$ (triangles) and $t=10^3/J$ (circles) follows the same scaling with system size.
			The parameters of the dynamics [cf.~\era{XXZ}{master}] are $N=6, 8,10,12,14$ spins [yellow, green, blue, red, black (grayscale: light gray to black), respectively], $V/J=2$, $h/J=5$, and $\gamma/J=2\times 10^{-4}$; gray (bottom) curves are the noninteracting case $V/J=0$ with $N=8$ (here results are independent of system size as entanglement obeys the area law ). 
		}\vspace*{-5mm}
		\label{fig:MBL_MPE}
	\end{center}
\end{figure}

For a state $\rho$ with a fixed local charge, we now derive a { \emph{faithful upper bound for MPE in terms of coherence}} in the computational basis [cf. Eq.~\er{Scoh}]
\begin{equation}
\E_{\rm MP}(\rho) \leq - S(\rho)+ S(\rho_{\rm diag})\equiv\C(\rho), 
\label{ineqMPE}
\end{equation}
where $S(\rho)\equiv - \Tr\rho \log_2 \rho$ and $\rho_{\rm diag}\equiv\sum_i \rho_{ii}|i\rangle\!\langle i|$ is obtained from $\rho$ by removing all coherences, thus leaving only the diagonal (see~Fig.~\ref{fig:matrix}). Analogously to~\eqref{ineqMPE}, also the \emph{geometric entanglement}~\cite{Wei2003,wei2004} can be faithfully bounded from above in the presence of a fixed local charge by the geometric coherence~\cite{Streltsov2015} (see Appendix~\ref{app:infidelity}). Note that the bound~\eqref{ineqMPE} can be directly accessed by measuring occupations in the computational basis whenever $\rho$ is pure (see Fig.~\ref{fig:MBL_MPE}). \\

\emph{Derivation}. In order to arrive at~\eqref{ineqMPE}, for $\rho$ with a fixed local charge, instead of the distance from the set of all separable states in~\eqref{mpe}, consider the distance from the set of separable states with a fixed charge. Note that the latter set is smaller and thus the distance from it is greater, thus providing an upper bound on MPD. Furthermore, since separable states with a fixed charge are diagonal in the corresponding basis (cf.~Sec.~\ref{sec:witness}), the closest state from this set is given by $\rho_{\rm diag}$.  Finally, the bound~\eqref{ineqMPE} is faithful as it reaches zero whenever $\rho$ is separable. \\

We note that it is known that for $\rho$ obeying a local symmetry it is enough to consider the minimum over the separable states also obeying this symmetry~\cite{Vollbrecht2001}. For $\rho$ conserving a local charge $[Q,\rho]=0$, we thus have $\E_\text{MP}(\rho) = \min_{\sigma_{\rm sep}:\,[Q,\sigma_{\rm sep}]=0} S(\rho || \sigma_{\rm sep})$.  For example, a Bell diagonal state of two spins  (i.e., a state which diagonalizes in the Bell basis), with a fixed total magnetization [Eq.~\eqref{2Qfix} with $p=1/2$], the closest separable state is known to be given by $\rho_\text{diag}=(|01\rangle\!\langle01|+|10\rangle\!\langle10|)/2$, so $\E_{\rm MP}(\rho)=C(\rho)$~\cite{Vedral1998}. In general, however, for a state $\rho$ with a fixed local charge, the charge of the closest separable state, although conserved, is not fixed, as there might be a trade-off in spreading the support of $\rho_{\rm sep}$ to other eigenspaces of $Q$, where $\rho_{\rm sep}$ no longer needs to be diagonal (cf.~Sec.~\ref{subsec:witnessMPD}).  For example, consider a symmetric  $W$ state of $N=3$ spins,
\begin{equation} \label{W}
|W_3\rangle\equiv(|\!\uparrow \!\downarrow\!\downarrow \rangle+|\!\downarrow \!\uparrow\!\downarrow \rangle+|\!\downarrow\!\downarrow\!\uparrow \rangle)/\sqrt{3}. 
\end{equation}
It is known that the state $\bar{\rho}_3$ in~\er{rn} with $p=2/3$ is the closest separable state to  $|W_3\rangle$,  so  $\E_{\rm MP}(|W_3\rangle)=2 \log_2 (3/2)$~\cite{wei2004}. The bound~\eqref{ineqMPE} corresponds then to $ 2 \log_2 (3/2) \leq \log_2 3 $. \\

Entanglement monotones are required to be nonincreasing under local operations and
classical communications (LOCC)~\cite{Vedral1997,Vedral1998,Plenio2007}, which are considered free in the resource theory of entanglement. Actually, the relative entropy of entanglement is an entanglement monotone even under a larger set of separable operations, which transform separable states into separable states. Since LOCC and separable operations do not conserve a local charge, the bound~\eqref{ineqMPE} is in general not saturated, as illustrated by the example of~\eqref{W}. When only a restricted set of separable operations that preserve the charge conservation in a quantum state is considered, the coherence in~\eqref{ineqMPE} becomes an entanglement monotone (as such operations are incoherent and thus cannot increase the coherence), which we show in Appendix~\ref{app:SSR}.

\subsection{Faithful upper bound on multipartite discord} \label{subsec:measureMPD}
Multipartite discord can be quantified as the relative entropy to the closest classical state
\begin{equation}
\D_{\rm MP}(\rho) \equiv \min_{\rho_{\rm cl}} S(\rho || \rho_{\rm cl}),\label{DMP}
\end{equation}
where a classical state $\rho_{\rm cl}$ is diagonal is some local basis~\cite{Modi2010,Yao2015,Bromley2015,Adesso2016,Ma2016} [see Eq.~\eqref{rho:cl}].

For $\rho$ commuting with a local charge, i.e., the case of the conserved charge, we obtain a \emph{faithful upper bound} on MPD in~\eqref{DMP}  as follows. In analogy to~\eqref{ineqMPE}, instead of considering the distance to classical states, we now consider the smaller set of the classical states with the conserved charge, which are diagonal, as we showed in Sec.~\ref{subsec:witnessMPD}. This again gives a bound in terms of the coherence in the computational basis [cf. Eq.\eqref{Scoh}]
\begin{equation}
\D_{\rm MP}(\rho) \leq - S(\rho)+ S(\rho_{\rm diag})\equiv\C(\rho) .\label{ineqMPD}
\end{equation}
where $\rho_{\rm diag}$ is obtained from $\rho$ by removing all coherences, $\rho_{\rm diag}\equiv\sum_i \rho_{ii}|i\rangle\!\langle i|$  (cf.~Fig.~\ref{fig:matrix}). The bound~\eqref{ineqMPD} is faithful, as it is zero when the state $\rho$ is classical.

For the Werner state of $2$ qubits~\eqref{Werner} we have $\D_{\rm MP}(\rho_W)=C(\rho_W)$, cf. Sec.~\ref{subsec:coherence}, so the bound~\eqref{ineqMPD} is saturated. Similarly, for the example of the $W$ state of three spins~\eqref{W}, the MPD is exactly equal $\D_{\rm MP}(|W_3\rangle)= S(\rho_\text{diag})=\log_2 3$~\cite{Modi2010}. 
%~\footnote{This illustrates the fact that for pure states multipartite quantum correlations do not generally correspond only to MPE, as we have $\mathcal{E}_{\rm MP}(|\psi\rangle)=2 \log_2 (3/2)\leq\D_{\rm MP}(|\psi\rangle)=\log_2 3$~\cite{Modi2010}, in contrast to the bipartite correlations in pure  states~\cite{Vedral1998}.}. 
In general, however, Eq.~\eqref{ineqMPD} is only an upper bound for MPD, even when the state is pure, since the minimum in~\eqref{DMP} cannot be restricted to classical states with the conserved charge, as the set of classical states is not convex (cf.~\cite{Vollbrecht2001}). Indeed, it is instead known that the minimum in~\eqref{DMP} corresponds to dephasing of $\rho$ in some separable basis~\cite{Modi2010}, while Eq.~\eqref{ineqMPD} corresponds to dephasing in the computational basis. For example, take the separable state $\bar{\rho}_3$ given by Eq.~\eqref{rn} with $p=2/3$. In this case (for optimal dephasing in the $x$ basis~\cite{Modi2010}) we have $\D_{\rm MP}(\rho)= 0.942...\leq C(\bar{\rho}_3)=2/3 \log_2 3$ [cf. Eq.~\eqref{rnC}].

% Similarly, for $(|\!\uparrow 10\!\downarrow \rangle+|\!\downarrow 01\!\uparrow \rangle+|\!\uparrow 01\!\downarrow \rangle+|\!\downarrow 10\!\uparrow \rangle+|\!\downarrow 11\!\downarrow \rangle+|\!\uparrow 00\!\uparrow \rangle)/\sqrt{6}$, dephasing in $x$ basis gives $S(\rho^{x}_\text{diag})=\log_2 (8/\sqrt{3})<S(\rho_\text{diag})=\log_2 6$.

Analogously to~\eqref{ineqMPD}, in the presence of a conserved local charge, also the \emph{geometric multipartite discord/quantumness}~\cite{Streltsov2011} can be faithfully bounded from above by the geometric coherence~\cite{Streltsov2015},  while the negativity of quantumness~\cite{Piani2011,Nakano2013,Adesso2016} can be faithfully bounded by $l_1$ coherence (see Appendix~\ref{app:infidelity}). Those bounds are a demonstration of the fact that entanglement or quantum discord quantified geometrically by a \emph{bona fide} distance to a set of separable or classical states, e.g., by the relative entropy, is bounded from above by coherence in any separable basis~\cite{Yao2015,Bromley2015,Adesso2016}. Furthermore, it is known that the multipartite quantum discord quantified this way can be considered as minimum coherence in some separable basis~\cite{Modi2010,Yao2015,Bromley2015,Adesso2016,Ma2016}. A conserved local charge, however, guarantees that the bound~\eqref{ineqMPD} on MPD is faithful, while, when the charge is fixed, the analogous bound~\eqref{ineqMPE} holds also for MPE (cf.~Fig.~\ref{fig:matrix}). Interestingly, the QFI~\eqref{qfi} itself for a nondegenerate observable with a fixed spectrum which is minimized over a local choice of basis can also be viewed as a measure of multipartite discord (see~\cite{Girolami2013,Girolami2014} and cf.~\cite{Frerot2016,Malpetti2016,Cianciaruso2017}), and in this case the choice of computational basis again provides a faithful upper bound on MPD.

Finally, we note that there exist similar bounds from above on the amount of entanglement or quantum correlations created by incoherent operations between a system state and an ancilla, in terms of the corresponding coherence of the initial system state~\cite{Streltsov2015,Ma2016} (see also~\cite{Chitambar2016,Streltsov2017} in the context of bipartite settings).

\begin{figure*}[ht!]
\begin{center}
\includegraphics[width=\textwidth]{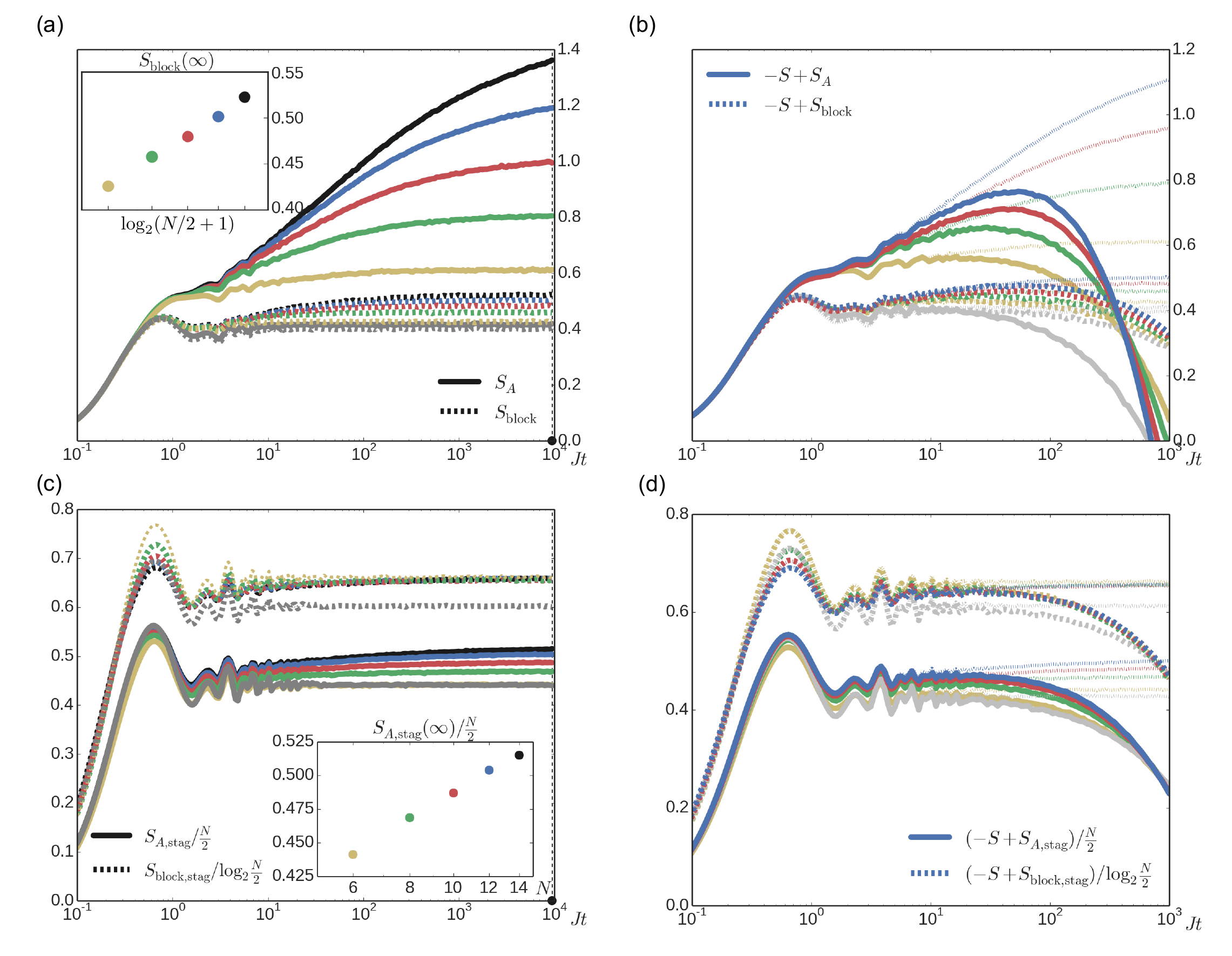}
\vspace*{-7mm}
\caption{
{\bf Bipartite entanglement and asymmetry in an MBL system}.
(a) Entanglement entropy $S(\rho_A)$ (solid lines) and asymmetry $S_{\rm block}$ of half-chain magnetization (dashed lines), [cf.~\eqref{ineqS}] 
for the $XXZ$  chain of Sec.~\ref{sec:MBL}, with closed dynamics. The asymmetry initially follows the area law ($Jt<1$), which is broken at later times, in analogy with the entanglement entropy. 
The inset shows 
the asymptotic value of the asymmetry (taken from $Jt=10^4$) scales as $\log_2(N/2+1)$ with system size (cf.\ Fig.~\ref{fig:MBL_block}). 
(That of the entanglement entropy, not shown, scales as $N$, as expected.)
(b) Similar, but with dephasing, so for  $-S(\rho)+S_{\rm block}$ (solid lines) and $-S(\rho)+S(\rho_A)$ (dashed lines).
Here the decay rate of the asymmetry is independent of the system size, but depends on the interaction strength (cf.\ Fig.~\ref{fig:MBL_block}).  
(c) Entanglement entropy (solid lines) for the staggered bipartition ($ABAB \cdots AB$ instead of $AA \cdots AB \cdots BB$) 
and corresponding asymmetry of the staggered magnetization. The curves are scaled by $N/2$ and $\log_2(N/2)$, respectively.
The inset shows that the asymptotic value of the entanglement entropy per site (taken from $Jt=10^4$) displays an additional weak dependence on $N$. 
(d) Same as (c) but for the dissipative case (subtracting the von Neumann entropy $S$). In the presence of dephasing, the decay of both sets of curves is independent of the system size, but depends on the interaction strength.
The parameters of the dynamics [cf.~\era{XXZ}{master}] are 
$N=6,8,10,12,14$ [yellow, green, red, blue, and black (grayscale: light gray to black), respectively, open case only up to $N=12$],  
$V/J=2$, $h/J=5$, and with $\gamma/J=2\times10^{-4}$ for the dissipative case [(b) and(d)]; gray (bottom) curves correspond to the noninteracting case $V/J=0$ and $N=12$ (closed dynamics) and $N=8$ (open dynamics), but in both cases the results follow area laws.
}\vspace*{-5mm}
\label{fig:MBL_BP}
\end{center}
\end{figure*}

\subsection{Lower bounds on bipartite entanglement} \label{subsec:measureBP}
Bipartite entanglement can be quantified by its relative entropy to the set of bipartite-separable states~\eqref{rho:sepBP}, so-called \emph{relative entropy of entanglement} or \emph{entanglement entropy}~\cite{Vedral1997,Vedral1998}, 
\begin{equation}
\E_{\rm BP}(\rho)\equiv \min_{\rho_{\rm BP-sep}} S(\rho || \rho_{\rm BP-sep}), \label{bpe}
\end{equation}
 which operationally corresponds to entanglement of distillation, i.e.,\ the asymptotic rate at which Bell states can be distilled from many copies of $\rho$ by LOCC. In particular, BPE of a pure state $|\psi\rangle$ is given simply by the von Neumann entropy of the reduced state to the subsystem $A$ or $B$ in the bipartition $\E_{\rm BPE}(|\psi\rangle)= S(\rho_A)=S(\rho_B)$, with $\rho_A=\Tr_{B}(|\psi\rangle\!\langle \psi|)$~\cite{Vedral1998}. For mixed states, however, a closed formula for~\eqref{bpe} is not known. 
 
 As the set of bipartite separable states~\eqref{rho:sepBP} for a given bipartition always contains separable states~\eqref{rho:sep}, we have that $\E_{\rm BP}(\rho) \leq \E_{\rm MP}(\rho)$ [see Eqs.~\eqref{mpe} and~\eqref{bpe}]. Thus, in the presence of a fixed local charge, the coherence~\eqref{Scoh} is also an upper bound for bipartite entanglement with respect to any partition of the system [cf. Eq.~\eqref{ineqMPE}]
 \begin{equation}\label{bpeC}
 \E_{\rm BP}(\rho)\leq -S(\rho)+S(\rho_\text{diag})\equiv C(\rho).
 \end{equation}
 This upper bound, however, is not in general faithful.  In this section, we remedy this situation by deriving new \emph{lower} bounds on bipartite entanglement monotones in Eqs.~\eqref{ineqS} and~\eqref{ineqN} in the case when a local charge is fixed.   
 
\noindent
\subsubsection{Lower bound on entanglement of formation} 

In Sec.~\ref{subsec:witnessBPE} we showed how bipartite separable states cannot feature coherences between subspaces with different values of the subsystem charge (cf.\ Fig.~\ref{fig:matrix}). Exploiting this structure, we obtain below a lower bound on the bipartite entanglement of formation in terms of the charge asymmetry, see Eq.~\eqref{ineqS}. 

\emph{Bipartite entanglement of formation}~\cite{Bennett1996} is defined as 
\begin{equation}
\E_{\rm BPE}^F\equiv \min_{\{p_j, |\psi_j\rangle\}} \sum_{j} p_j S(\rho_A^{(j)}) , \label{bpeF}
\end{equation}
where $\rho^{(j)}_A=\Tr_{B}(|\psi_j\rangle\!\langle \psi_j|)$ is the state of  $|\psi_j\rangle$ reduced to the subsystem $A$, while the minimization (so-called convex roof) is performed over all decompositions of $\rho$ into pure states $\rho=\sum_{j} p_j |\psi_j\rangle\!\langle \psi_j|$. The entanglement of formation  after regularization corresponds to the entanglement cost, i.e.,\ the asymptotic rate at which Bell states need to be supplied in order to prepare $\rho$ via LOCC~\cite{Hayden2001}. From the joint convexity of relative entropy, it is an upper bound on the relative-entropy bipartite entanglement \cite{Henderson2000} in Eq.~\eqref{bpe} 
\begin{equation}
\E_{\rm BPE}\leq\E_{\rm BPE}^F, 
\end{equation} 
and these measures coincide for pure $\rho$.\\

A \emph{new lower bound} on BPE of formation~\eqref{bpeF} in a state with a fixed local charge is given by the asymmetry of a subsystem charge~\eqref{Sas}
\begin{equation}
\E_{\rm BPE}^F(\rho)\geq -S(\rho)+S(\rho_{\rm block})\equiv\A(\rho) , \label{ineqS}
\end{equation}
where $\rho_{\rm block}\equiv \sum_{i,j:\, q_i^{(A)}=  q_j^{(A)} } \rho_{ij} |i\rangle\!\langle j|$ is obtained from $\rho$ by removing all coherences between different eigenspaces of the charge in the $A$ or $B$ subsystem (or equivalently of the charge difference between the subsystems) (cf. Fig.~\ref{fig:matrix}). Note that for pure states, the bound equals the entropy of the subsystem charge $Q^{(A)}$ statistics, which can be directly accessed by measuring $Q^{(A)}$. The derivation of~\eqref{ineqS} can be found in Appendix~\ref{app:measureBP}.  \\

Let us first consider Eq.~\eqref{ineqS} for the example of a Bell diagonal state of $N=2$ spins $\frac{1}{2}$ with a fixed total magnetization [Eq.~\eqref{2Qfix} with $p=\frac{1}{2}$]. Here the asymmetry equals the relative entropy of entanglement~\cite{Vedral1998}
\begin{equation}
 \A(\rho_2^\text{fix})=1-\mathcal{H} \Big(\frac{1}{2}+|c| \Big)=\E^\text{BP}(\rho_2^\text{fix}),
   \end{equation}
   where $\mathcal{H}(x)= - x\log_2 x +(1-x)\log_2 x$,  but the entanglement of formation is higher~\cite{Hill1997}, 
  \begin{equation}
  \E_{\rm BPE}^F(\rho_2^\text{fix})=\mathcal{H} \Big(\frac{1-\sqrt{1-4|c|^2}}{2} \Big),
   \end{equation}
   whenever the state is not pure ($|c|<\frac{1}{2}$)~\footnote{In Appendix~\ref{app:measureBP} we show that the bipartite entanglement is actually bounded by the asymmetry of formation. This bound is saturated for any Bell diagonal state with a fixed total magnetization.}.

Second, let us discuss the scaling of the bound~\eqref{ineqS} with the system size $N$ (see also Fig.~\ref{fig:MBL_BP}). For a system of $N$ spins $\frac{1}{2}$  (with $N$ even), the maximum entanglement in the bipartition into half chains is proportional to the number of spins  $S(\rho_A)= N/2$. When the total magnetization along one axis is fixed, the Hilbert space dimension $D = 2^N$ is reduced to ${{N}\choose{aN}}=\exp[ N \mu(a) + \mathcal{O}(\ln N)]$ (with $a$ the filling), where $\mu(a) = -a\log_2a-(1-a)\log_2(1-a)$,  which in the large-size limit is still exponential in size. In particular, for the biggest subspace of zero total magnetization ($a=1/2$), $S(\rho_A)= N/2$ is still achievable (by the uniform superposition of $2^{N/2}$ possible states of half a chain, each in the tensor product e.g., with the same state but with all spins flipped, which guarantees zero magnetization for an even number $N$ of spins) (cf.~Fig.~\ref{fig:MBL_BP}). 

In contrast, the maximal asymmetry scales logarithmically with the system size and we have $S(\rho_\text{block})\leq \log_2 (N/2+1)$ for the subspace of zero-magnetization [cf.~Figs.~\ref{fig:MBL_BP}(c) and~\ref{fig:MBL_BP}(d)], which is achieved for pure states uniformly distributed over the half-chain magnetization. It is, thus, still possible to certify breaking of the area law for one-dimensional systems, where it means constant scaling (independent of the system size). Indeed, for the example of the state achieving the maximal entanglement entropy $N/2$, and thus the volume law, we have $S(\rho_\text{block})\sim  \log_2(N/2)/2 $~\footnote{The distribution of the half-chain magnetization is binomial with probability $1/2$, which from the central limit theorem can be approximated as the Gaussian distribution with the variance $(N/2)\times 1/4$. In $S(\rho_\text{block})$ we neglected the constant contribution $\log_2(\pi e /2)/2$.}, so  the asymmetry displays a  logarithmic law  and thus breaking of the area law is detected. Furthermore, also for the ground state of the critical one-dimensional system of interacting spinless fermions, where the entanglement entropy  scales only logarithmically $S(\rho_A)\sim  \log_2(L)/6$ in the $A$-subsystem size $L$~\cite{Holzhey1994,Calabrese2004}, it is known that  $S(\rho_\text{block})\sim \log_2 (K \ln L )/2 $~\cite{Song2012,Goldstein2017}, where  $K$ is the Luttinger liquid parameter~\footnote{The distribution of the subsystem charge is Gaussian with the variance $K \ln L /\pi^2$ (see~\cite{Song2012}) and in $S(\rho_\text{block})$ we neglected the constant contribution $\log_2(2 e /\pi)/2$.}. Therefore, also in this case the bound in Eq.~\eqref{ineqS} detects breaking of the entanglement area law.  We observe an analogous behavior for an MBL system in Figs.~\ref{fig:MBL_BP}(a) and~\ref{fig:MBL_BP}(c).

Furthermore, for mixed states the bound~\eqref{ineqS} can be tighter than the known bound with the reduced state~\cite{Plenio2000}, 
\begin{equation}
-S(\rho)+S(\rho_A)\leq \E_{\rm BPE}(\rho)\leq \E_{\rm BPE}^F(\rho)  \label{ineqSA}
\end{equation}
[see Fig.~\ref{fig:MBL_BP}(b)]. In particular, the bound~\eqref{ineqS} is always positive,
\begin{equation}
-S(\rho)+S(\rho_\text{block})\geq 0 ,
\end{equation}
in contrast to what occurs with $-S(\rho)+S(\rho_A)$ [cf. Fig.~\ref{fig:MBL_BP}(b)].

In Sec.~\ref{sec:MBL} we discuss the bound~\eqref{ineqS} for the example of a many-body localized system as shown in Fig.~\ref{fig:MBL_BP}. In particular, for the case of closed dynamics we observe very good agreement between the asymmetry and the entanglement entropy in the Anderson-localized phase without interactions [cf. Fig.~\ref{fig:MBL_BP}(a)]. \\

Finally, we note that BPE of formation,~\eqref{bpeF}, is actually bounded by the convex roof of asymmetry (cf. the in Appendix~\ref{app:measureBP}). We can thus relate the bound~\eqref{ineqS} to the known equality~\cite{Winter2016} between the coherence of formation~\cite{Aberg2006,Yuan2015,Winter2016} for a state $\rho=\sum_{ij}\rho_{ij}|i\rangle\!\langle j|$, and the entanglement of formation~\cite{Bennett1996,Hill1997,Wootters1998,Henderson2000} for the corresponding maximally correlated state $\rho_\text{mc}\equiv\sum_{ij}\rho_{ij}|ii\rangle\!\langle jj|$, as follows. Namely, in the presence of a fixed local charge, $Q=Q^{(A)}+Q^{(B)}$, a given $A$-subsystem charge value $Q^{(A)}$ in the subsystem $A$ fully determines a value of the $B$-subsystem charge $Q^{(B)}$.  We also note that a relation similar  to that in~\cite{Winter2016} holds between the relative entropy of coherence and the relative entropy entanglement (cf.~\cite{Streltsov2015}). Thus, it is an interesting open question whether the asymmetry of charge difference is also a valid lower bound for the relative entropy entanglement in states with a fixed local charge.

\noindent
\subsubsection{Experimentally accessible lower bound on the convex roof of negativity}

When $\rho$ is mixed, the asymmetry of charge difference~\eqref{Sas} and~\eqref{ineqS} cannot be directly accessed experimentally, as $S(\rho)$  cannot be measured without full quantum state tomography with a few exceptions (such as noninteracting fermions~\cite{Song2012}; cf. also~\cite{Daley2012,Islam2015}).  Therefore, in Eq.~\eqref{ineqN} we derive an analogous lower bound to~\eqref{ineqS} on the convex roof of negativity of entanglement~\cite{Vidal2002,Lee2003}, which can be accessed experimentally by using multiple quantum coherence spectra, as explained in Appendix~\ref{sec:method}.

\begin{figure*}[ht!]
	\begin{center}
		\includegraphics[width=0.98\textwidth]{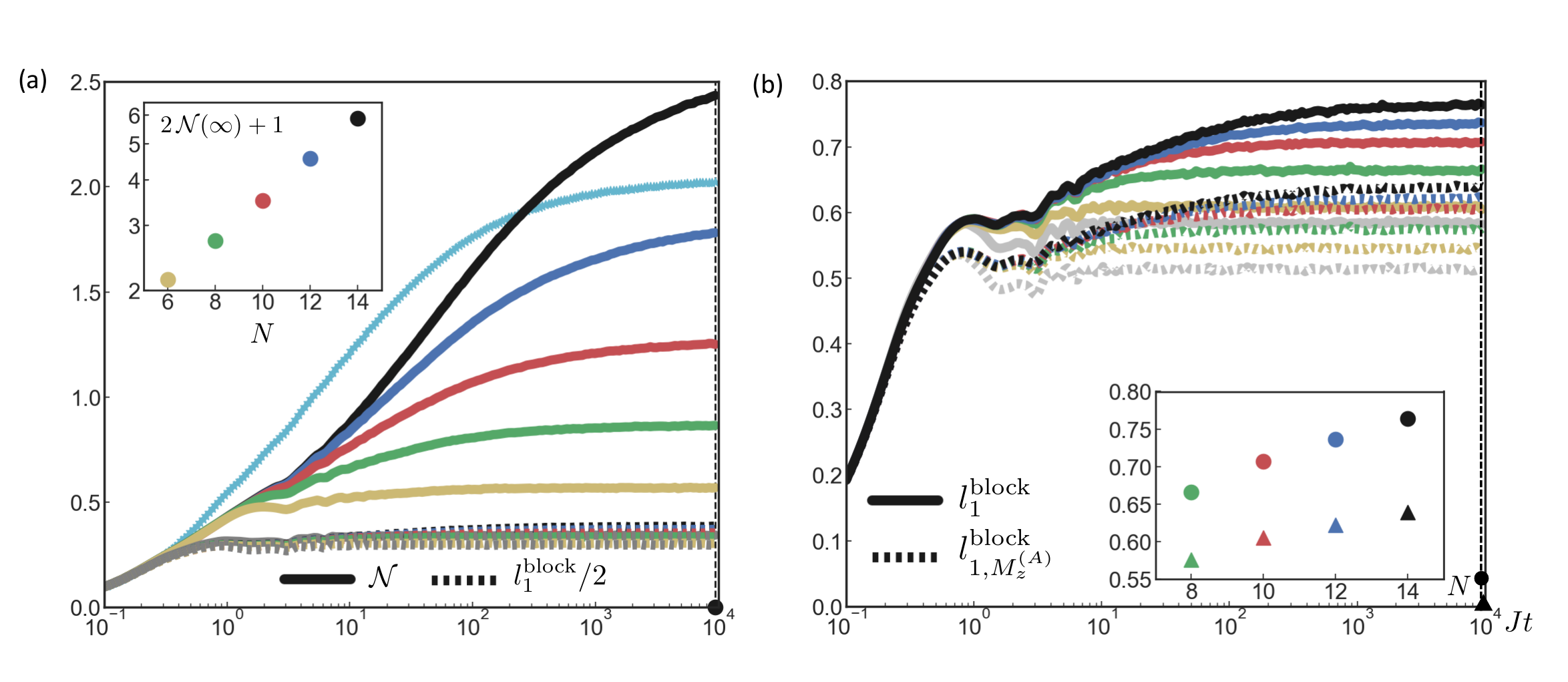}
		\vspace*{-2mm}
		\caption{
			{\bf Negativity and  experimentally accessible lower bound in a closed MBL system}.
			(a) We show the negativity of entanglement (~\er{N}) for the $XXZ$  chain of Sec.~\ref{sec:MBL}. The negativity (solid lines) initially follows an area law for times $Jt<2$ (with coefficient dependent on the interaction strength for $Jt>1$), but at later times becomes dependent on the system size. The lower bound by $l_1^\text{block}$ (Eq.~\eqref{ineqN}) is also shown (dashed lines) (cf.~Fig.~\ref{fig:MBL_l1}). 
		    The inset shows that the asymptotic values of $\mathcal{N}$ in the closed dynamics (at $Jt=10^4$) scales exponentially with the system size $N$ (note the logarithmic scale of the vertical axis).
			(b) We show the lower bounds $l_1^\text{block}\geq l_{1,M_z^{(A)}}^\text{block}$ on the convex roof of the negativity of entanglement (cf.~\er{ineqN}). Both $l_1^\text{block}$ (solid lines) and the experimentally accessible  $l_{1,M_z^{(A)}}^\text{block}$ (dashed lines) initially follow the area law for times $Jt<2$, and at later times become dependent on the system size. In particular, the asymptotic value of $l_1^\text{block}$ for the noninteracting system (gray) is crossed by $l_{1,M_z^{(A)}}^\text{block}$ for sizes above $N=8$. 
			The inset shows that the asymptotic values of $l_1^\text{block}$ (circles) and $l_{1,M_z^{(A)}}^\text{block}$ (triangles) in the closed dynamics (at $Jt=10^4$) increases with the system size.
			The parameters of the dynamics [cf.~\era{XXZ}{master}] are 
			$N=6,8,10,12,14$ spins [yellow, green, red, blue, and black (grayscale: light gray to black), respectively], $V/J=2$ and $h/J=5$; gray (bottom) curves correspond to the noninteracting case $V/J=0$ with (a) $N=12$  and (b) $N=8$. Light-blue stars in (a) correspond to a more strongly interacting system with $V/J=5$ and $N=12$. 
		}\vspace*{-7mm}
		\label{fig:negativity}
	\end{center}
\end{figure*}

\begin{figure}[ht!]
	\begin{center}
		\includegraphics[width=\columnwidth]{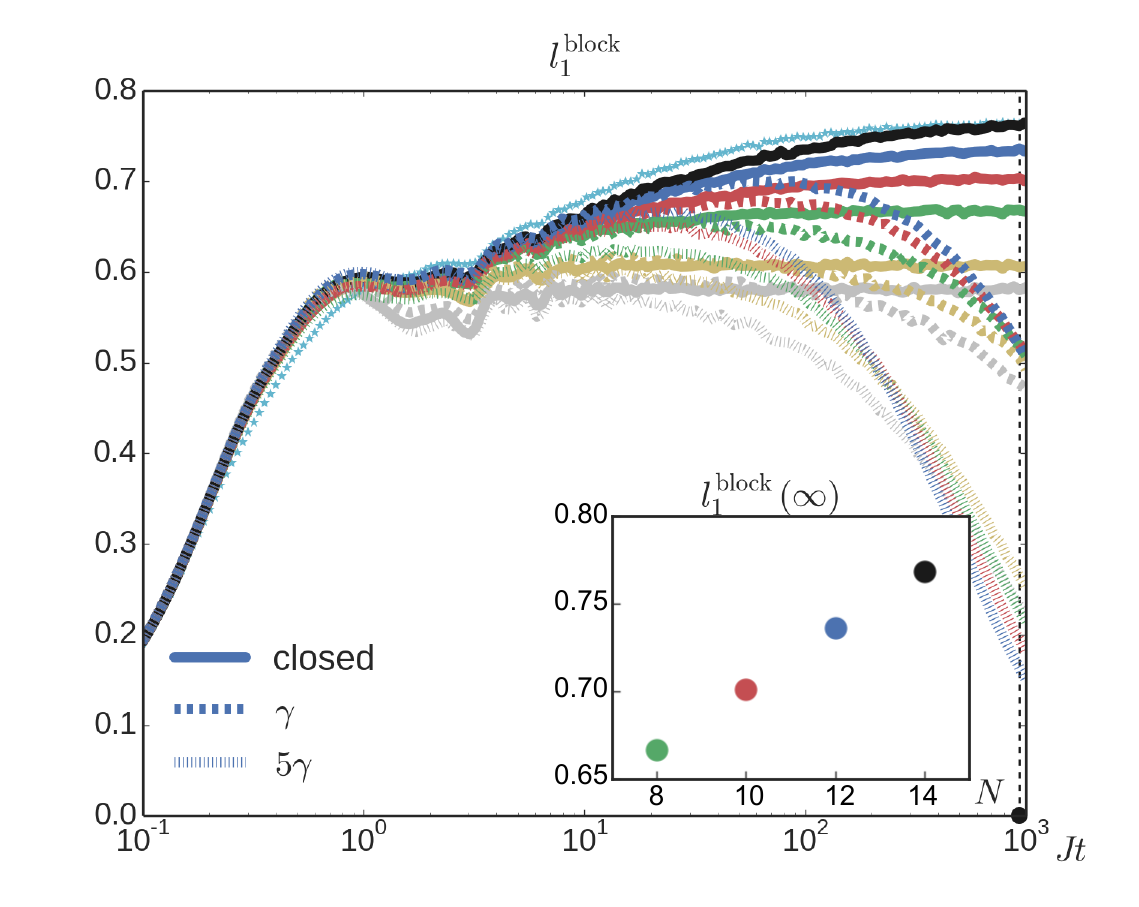}
		\vspace*{-7mm}
		\caption{
			{\bf Lower bound on BPE in an open MBL system}.
			We show the lower bound on the convex roof on the negativity of entanglement by $l_1^\text{block}$ [cf.~\er{ineqN}] for the $XXZ$  chain of Sec.~\ref{sec:MBL}.
			The solid curve is for the closed case and the dashed curves for the dissipative case. Decay in the presence of dephasing depends on both  the system size and interaction strength. 
			The inset shows that the asymptotic value of $l_1^\text{block}(\rho)$ in the closed dynamics at $t=10^3/J$ (circles) grows approximately linearly with $N$ (for $N=8,10,12,14$).
			The parameters of the dynamics [cf.~\era{XXZ}{master}] are 
			$N=6,8,10,12,14$ spins [yellow, green, red, blue, and black (grayscale: light gray to black), respectively, open case only up to $N=12$], $V/J=2$, $h/J=5$, and $\gamma/J=2\times 10^{-4}$; gray curves correspond to the noninteracting case $V/J=0$ with $N=12$ (closed dynamics) and with $N=8$ (open dynamics) (here results are independent of system size as the system follows an area law ). Light-blue stars in (a) correspond to a more strongly interacting system with $V/J=5$ and $N=12$. 
		}\vspace*{-7mm}
		\label{fig:MBL_l1}
	\end{center}
\end{figure}

The negativity of entanglement~\cite{Hill1997,Wootters1998,Vidal2002} is defined as 
\begin{equation}\label{N0}
\N(\rho)\equiv(\lVert\rho^{T_A}\rVert_1-1)/2,
\end{equation}
where $\lVert X \rVert_1=\Tr[X^\dagger X]^{1/2}$ denotes the trace norm and $\rho^{T_A}$ is the partial transpose of the subsystem $A$. This definition is independent of the choice of the separable basis and the subsystem $A$ or $B$, and for the case of two qubits ($D = 2 \times 2$) it corresponds to the  concurrence~\cite{Hill1997,Wootters1998}. The negativity is a witness of bipartite entanglement, as although the trace is conserved under partial transpose, the resulting matrix may no longer be positive if the state is entangled, giving the Peres criterion of separability~\cite{Peres1996}. For systems of dimensions $D = 2\times 2$  and $2\times 3$, the partial transpose is a faithful witness of bipartite entanglement~\cite{Horodecki1997}, but in general there exist entangled states, so-called positive-partial-transpose-bound (PPT-bound) states, which remain positive under the partial transpose~\cite{Horodecki1998}. Furthermore, negativity of entanglement is an entanglement monotone with respect to PPT operations~\cite{Vidal2002} and recently has been  proposed to be accessible experimentally~\cite{Gray2017} by measuring moments of the transposed state  $\Tr[(\rho^{T_A})^m]$ from which the efficient reconstruction of the negativity can be achieved using machine learning techniques. Finally, the positive partial transpose is a faithful witness of the bipartite entanglement for pure states,  and $\N(|\psi\rangle)=\sum_{i\neq i'} [\lambda_i\lambda_{i'}]^{1/2}/2$,  with $\lambda_i$ denoting the Schmidt coefficients for a pure $|\psi\rangle$. This allows for the negativity of entanglement~\eqref{N0} to be extended to a bipartite entanglement measure with respect to LOCC~\cite{Lee2003} via the convex roof  
\begin{equation}
\E_\text{BP}^\N(\rho)\equiv\inf_{p_j, |\psi_j\rangle} \sum_j p_j \N(|\psi_j\rangle),\label{N}
\end{equation}
where $\rho=\sum_{j}p_j|\psi_j\rangle\!\langle\psi_j|$ and $\E_\text{BP}^\N(\rho)=N(\rho)$ for a pure state $\rho$.\\

A \emph{lower bound} for the convex roof of negativity~\eqref{N} in states with a fixed local charge is given by
\begin{equation}
2\,\E_\text{BP}^\N(\rho)\,\geq\,  l_1^\text{block}(\rho)\, \geq  \,l_{1,Q^{(A)}}^\text{block}(\rho),
\label{ineqN}
\end{equation}
where $l_1^\text{block}$ is defined by Eq.~\eqref{l1block} with respect to the subsystem charge ($H=Q^{(A)}$), while $l_{1,Q_A}^\text{block}$ [see Eq.~\eqref{l1blockM}] can be accessed in experiment by measuring multiple quantum coherence spectra~\cite{Macri2016,Garttner2017,Garttner2018}, as we outlined in Appendix~\ref{sec:method}~\footnote{We also note that the negativity~\eqref{N0} itself also provides a bound $\E_\text{BP}^\N(\rho)\geq \N(\rho)$.}.

For the example of a state of $N=2$ spins $\frac{1}{2}$  with a fixed total magnetization Eq.~\eqref{2Qfix}  it is easy to verify that the lower bound equals the negativity $l_1^\text{block}(\rho_2^\text{fix})= l_{1,M_z^{(1)}}^\text{block}(\rho_2^\text{fix})= 2|c|= 2\N(\rho_2^\text{fix})$ (cf. Eq.~\eqref{N}). Furthermore, in this case, although the state is generally mixed, the convex roof of negativity and the negativity coincide (cf.~\cite{Wootters1998,Streltsov2017b}), so  the bound in Eq.~\eqref{ineqN} is saturated.

Let us now discuss the scaling of the bound~\eqref{ineqN} with the system size $N$. For a pure state of a system of $N$ spins $\frac{1}{2}$  with the total magnetization along one of the axes fixed,  the negativity obeys $\N(|\psi\rangle)\leq (2^{N/2} -1)/2$ and thus also its convex roof $\E_\text{BP}^{\N}(\rho)\leq (2^{N/2} -1)/2$ [with the bounds saturated exactly when the entanglement entropy is maximal $S(\rho_A)=N/2$] [cf. Fig.~\ref{fig:negativity}(a)]. Therefore, in this case the volume law corresponds to  $\log_2 [2\,\E_\text{BP}^{\N} (\rho)+1]$ scaling linearly with the subsystem size, while the area law for the one-dimensional system corresponds to a constant (see also the logarithmic negativity~\cite{Plenio2005} which is additive with respect to the tensor product).

In contrast, we have $l_1^\text{block}(\rho)\leq N/2$, which is saturated in the zero-magnetization subspace for the pure state uniformly distributed between $N/2+1$ values of the half-chain magnetization. Therefore, it is possible to detect breaking of the area law by the use of the bound~\eqref{ineqN} (see~Fig.~\ref{fig:MBL_l1}), similarly to  the case of the relative entropy of asymmetry~\eqref{ineqS}. Indeed, for the state with the maximal negativity, we have $l_1^\text{block}\sim \sqrt{\pi N} $~\cite{Note4}. Similarly, for a ground state of a critical one-dimensional system of interacting spinless fermions we have $l_1^\text{block}\sim 2\sqrt{2K \ln L /\pi} $, where $L$ is the subsystem-$A$ size and $K$ is the Luttinger liquid parameter~\cite{Note5}. 
 
Although, the experimentally accessible bound in~\eqref{ineqN} is necessarily smaller, it can also detect breaking of the area law. For example, when $l_1^\text{block}=N/2$ is maximal, we have $l_{1,M_z^{(A)}}^\text{block}\sim 2/3\sqrt{N/2}$, while for the state with the maximal negativity we have $l_{1,M_z^{(A)}}^\text{block}\sim 2 (\pi N/8)^{1/4}$~\cite{Note4}. Similarly, for the critical ground state $l_{1,Q^{(A)}}^\text{block}\sim 2 ( K \ln L /\pi)^{1/4}$~\cite{Note5}. In Figs.~\ref{fig:negativity} and~\ref{fig:MBL_l1} we also show the dynamics of the bounds for a closed and an open MBL system, which displays breaking of the area law. \\

The plethora of entanglement monotones~\cite{Plenio2007,Horodecki2009} seems to be in disagreement with the fact that asymptotically, for a large number of copies of a pure state, the bipartite entanglement is uniquely quantified by the relative entropy of entanglement~\cite{Popescu1997}. For a finite number of copies, however, bipartite entanglement needs to be characterized by a set of bipartite monotones determining the equivalence class for the Schmidt coefficients under LOCC~\cite{Vidal2000}. For mixed states, for an appropriate function quantifying the mixedness of the reduced state $\rho_A$, the entanglement monotones can be constructed via the convex roof~\cite{Vidal2000,Szilard2015}, as it is in the case of entanglement of formation~\cite{Bennett1996} and the convex roof of entanglement negativity~\cite{Lee2003}.  In Appendix~\ref{app:bounds}, we show that also for the mixedness of the reduced state quantified by the concurrence~\cite{Wootters1998,Hill1997} by Tsallis 2-entropy, the corresponding convex roof monotones~\cite{Rungta2001,Albeverio2001,Wootters2001,Szilard2015} can be bounded from below by the $L_2$-norm of block coherence, also accessible in experiments.

\subsubsection{Lower bounds on BPE as entanglement monotones in the presence of SSR}

We now explain that in the presence of a superselection rule related to a local charge, the asymmetry of a subsystem charge~\eqref{Sas} becomes a lower bound on the bipartite entanglement for the states with conserved charge (see also Sec.~\ref{subsec:SSR}).

In the presence of a SSR related to a local charge, a state of the conserved, but not fixed, charge can be created only as a probabilistic mixture of states with fixed values of the charge. This restriction is captured by the monotones of entanglement for mixed states, which are constructed from pure state monotones via the convex roof which obeys a given SSR~\cite{Schuch2004,Schuch2004b}. Therefore, in the presence of a SSR, the subsystem charge asymmetry~\eqref{Sas} becomes a lower bound on the bipartite entanglement for all states. This lower bound captures entanglement related to the resource of nonlocality induced by the SSR, also quantified by the QFI of the charge difference~\cite{Schuch2004,Schuch2004b,Toth2013,Yu2013}, as we now explain. 

First, the asymmetry disappears if and only if the QFI of the charge difference equals zero. Second, the asymmetry of a subsystem charge is nonincreasing under LOCC that obey the SSR, i.e.,\ conserve the charge (see Appendix~\ref{app:SSR}). Actually, it can even be shown that the asymmetry is nonincreasing with respect to a larger class of separable operations which preserve the charge conservation in a quantum state (see Appendix~\ref{app:SSR}).

\section{Application to a many-body localized system, without and with dissipation}\label{sec:MBL}

Throughout the paper up to now we have exemplified our results with the following model system, an $XXZ$  chain of spins $\frac{1}{2}$  in the presence of a disordered longitudinal magnetic field. This is a paradigmatic system widely believed to display a transition from a thermal phase at small disorder to an MBL phase at large disorder (for reviews on MBL see 
\cite{Nandkishore2015b,Altman2015,Abanin2017}). The Hamiltonian of this model is given by \cite{Pal2010}
\begin{eqnarray}
H_{XXZ}&\equiv & J \sum_{k=1}^{N-1} (S_k^{x}S_{k+1}^{x}+S_k^{y}S_{k+1}^{y})+ V \sum_{k=1}^{N-1} S_k^{z}S_{k+1}^{z} 
\nonumber\\
&&+ \sum_{k=1}^{N} h_k S_k^z, \label{XXZ}
\end{eqnarray}
where $S_k^{x,y,z}$ are the spin operators for the $k$th spin $1/2$, and the last term is a quenched random longitudinal field, with 
$h_k$ being random independent, identically distributed, and drawn uniformly from $[-h,h]$. Note that we consider open boundary conditions in order to remove residual symmetries. This Hamiltonian maps via a Jordan-Wigner transform to one of interacting spinless fermions in a random field, 
\begin{eqnarray}
H_{XXZ}^{(f)}&=& -\frac{J}{2} \sum_{k=1}^{N-1} (c_k^\dagger c_{k+1} +c_{k+1}^\dagger  c_k)+ V \sum_{k=1}^{N-1} n_k n_{k+1} \nonumber\\
&& -\sum_{k=1}^{N} h_k  n_k- V \sum_{k=1}^{N} n_k+\frac{V}{4}N+\sum_{k=1}^{N} h_k, 
\label{XXZf}
\end{eqnarray} 
where the fermion density on the $k$th site is given by $n_k\equiv c_k^\dagger c_{k}$, and we neglected a constant shift. We see that $J$ drives hopping of fermions, while $V$ is the strength of density-density interactions.

This is a convenient system to which apply the results of Secs.~\ref{sec:witness} and~\ref{sec:measures}, for two reasons. First,  Hamiltonian~\eqref{XXZf} \emph{conserves the total number of fermions} $\sum_{k=1}^{N} n_k$, which corresponds to the \emph{conservation of the total $z$ magnetization} $M_z\equiv\sum_{k=1}^{N} S_k^z$ in~\eqref{XXZ}.  Therefore, if the initial state has a fixed $z$ magnetization (or a fixed number of fermions) then unitary dynamics under $H$ will preserve it. 
[Note that while it is natural to consider a superselection rule for the number of fermions~\cite{Schuch2004,Schuch2004b} in the case of~\eqref{XXZf}, in the case of~\eqref{XXZ} no restrictions apply to possible operations on a quantum state.]  
For example, a state with fixed $z$ magnetization is the staggered state 
\begin{equation}
|\psi_{0} \rangle=|\!\uparrow \!\downarrow \!\uparrow \!\downarrow\cdots\rangle, 
\label{cdw}
\end{equation}
where $|\!\uparrow \rangle$ and $|\!\downarrow \rangle$ denote spin up and down, respectively, often used as an initial state, both in numerics \cite{Pal2010,Nandkishore2015b} and in experiments \cite{Schreiber2015}. Second, a key characteristic of the MBL state is how entanglement develops over time as the system evolves from an initial unentangled state~\cite{Bardarson2012,Serbyn2013}. 

\begin{figure}[t]
	\begin{center}
		\vspace*{-5mm}
		\includegraphics[width=1.0\columnwidth]{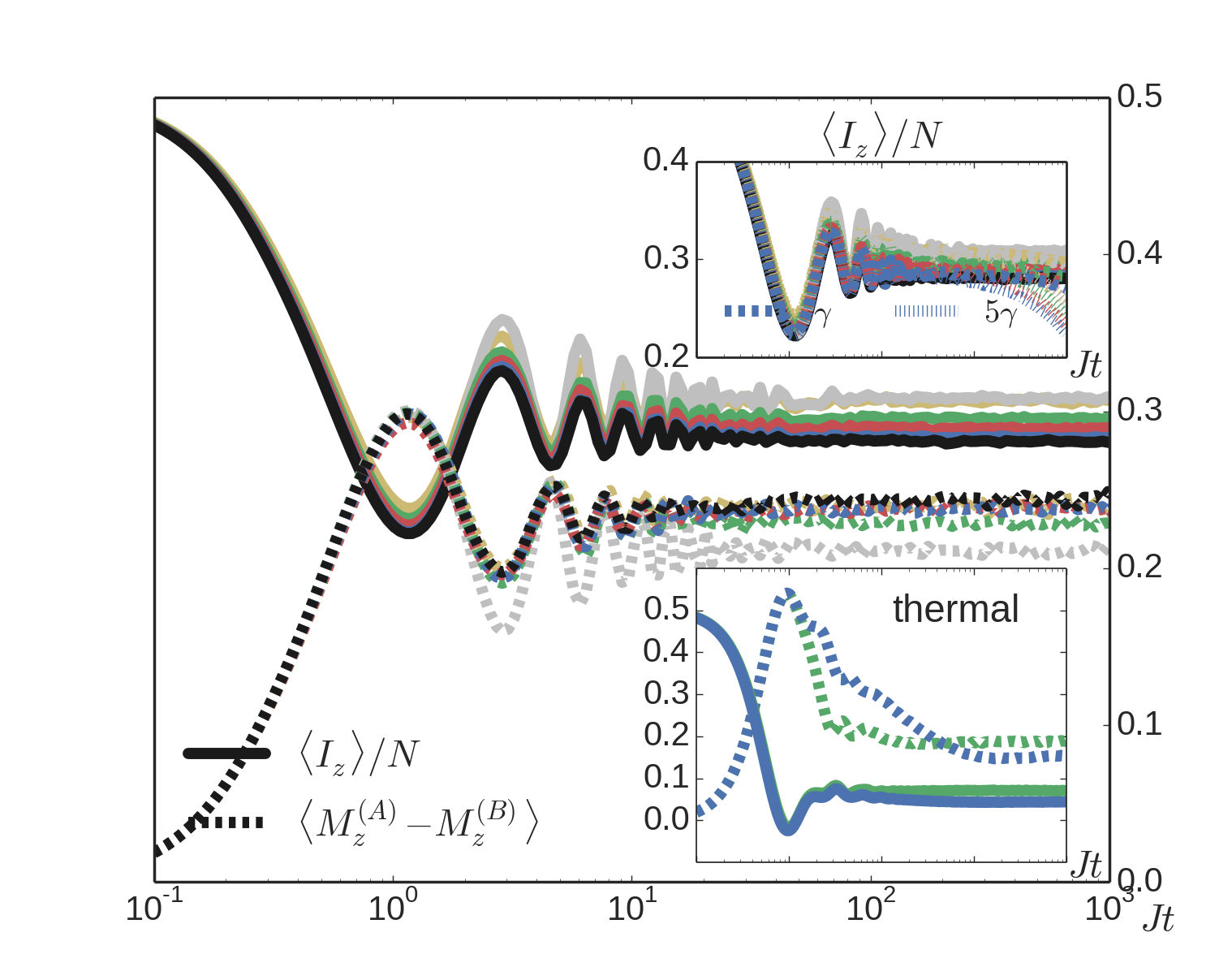}
		\vspace*{-7mm}
		\caption{
			{\bf Observable averages in MBL and thermal phases}:
			imbalance (solid lines) and average half-chain magnetizations difference (dashed lines) 
			for the $XXZ$  chain of Sec.~\ref{sec:MBL}. 
			They both show a quick approach to stationarity for, in contrast to the behavior of quantum correlations (cf.\ Fig.~\ref{fig:MBL_BP}). 
			In particular, there is no appreciable difference in timescales between the interacting and noninteracting cases.
			The parameters of the dynamics [cf.~\era{XXZ}{master}] are 
			$N=6,8,10,12,14$ [yellow, green, red, blue, and black (grayscale: light gray to black), respectively; open case only up to $N=12$], $V/J=2$ and $h/J=5$; gray (top solid and bottom dashed) curves are for $V/J=0$ and $N=8$. 
			The bottom inset shows that in the thermal phase ($V/J=2$ and $h/J=1$) the asymptotic values decrease with the system size [green (gray) $N=8$ and blue (darker gray) $N=12$]. The top inset shows that the average imbalance in the presence of dephasing $\gamma/J=2\times 10^{-4}$ (dashed curves) and $10^{-3}$ (dotted curves) decays to zero. 
		}\vspace*{-5mm}
		\label{fig:av}
	\end{center}
\end{figure}

For $h > h_c$ the system described by \eqref{XXZ}, or alternatively \eqref{XXZf}, is many-body localized. This is most immediately observed in the inability of the system to forget initial conditions, a key marker of non-ergodicity. An example is the behavior of the average {\em imbalance}  
$\langle I_z \rangle$ as a function of time, starting from the staggered initial state \er{cdw}, where the imbalance operator corresponds to a magnetization with the same stagger as the initial state, reading in spin language
\begin{equation}
I_z=\sum_{k=1}^N (-1)^k S^z_k ,
\label{Iz}
\end{equation}
so  for $|\psi_0 \rangle$ in Eq.~\eqref{cdw} it quantifies the degree of time correlation with the initial conditions. Figure \ref{fig:av} shows how for $|\psi_0 \rangle$ the imbalance becomes stationary at a value far from zero, contrary to what would occur if the system thermalized and became ergodic.

While in the thermal phase ($h<h_c$) the growth of entanglement is fast (see Appendix \ref{app:thermal}), in the MBL phase entanglement grows more slowly (cf.\ Fig.~\ref{fig:MBL_BP}), first towards an area law plateau similar to that of an Anderson-localized system ($V=0$ in $H$), only later growing logarithmically in time towards its asymptotic value (which obeys a volume law, but is nonetheless smaller than that of the thermal phase) \cite{Bardarson2012,Serbyn2013,Nandkishore2015b}. 

In order to apply our results both to pure and mixed states, we furthermore assume that the system can feature a \emph{local noise} that conserves the total $z$ magnetization (cf. Sec~\ref{subsec:charge}.) This condition is only fulfilled for \emph{dephasing}, when the system state $\rho$ evolves according to the master equation \cite{Levi2016}
\begin{equation}
\frac{\mathrm{d}}{\mathrm{d} t}\rho_t=\mathcal{L}(\rho_t)=-i[H_{XXZ},\rho_t]+\gamma\sum_{k=1}^N  S_k^z \rho_t\, S_k^z - N \gamma\rho_t,
\label{master}
\end{equation}
and $[S_k^z,M_z]=0$.
In particular, we consider weak dephasing where the MBL effects are expected to be robust~\cite{Nandkishore2014,Nandkishore2015,Levi2016}, in contrast to the limit of strong dephasing where the dynamics becomes classical~\cite{Medvedyeva2015,Fischer2016,Everest2017}. A more general local noise which preserves the conservation of $M_z$ in a quantum state can also feature thermal jumps $\sqrt{\kappa_{-}}{S_k^-}$ and $\sqrt{\kappa_{+}}{S_k^+}$ (cf.\ Sec.~\ref{subsec:charge} and Appendix~\ref{app:SSR}). We note that the dynamics in Eq.~\eqref{master} is allowed by the MQC protocol in Appendix~\ref{sec:method}, which delivers a faithful lower bound on the QFI.

Numerical simulations in Figs.~\ref{fig:QFI}-\ref{fig:thermal} are obtained (for each set of parameters) by averaging $10^4$ trajectories from exact diagonalization of the Hamiltonian \er{XXZ} in the closed case, and from numerical integration of the master equation \er{master} (using the BDF method), with an average over $5\times 10^3$ trajectories in the open case . The error bars are not shown, as they are smaller than the used widths of lines and symbols.

\subsection{Witnessing entanglement in MBL dynamics}

In Fig.~\ref{fig:QFI} we investigate how MPE in the chain of $N=14$ spins can be witnessed by measuring the QFI~\eqref{qfi} (or equivalently the variance, as dynamics is closed) of the local magnetization observables 
\begin{align}
%M_z^{(A)}-M_z^{(B)} 
\delta M_z&\equiv \sum_{k=1}^{N/2} S^z_k-\sum_{k=N/2+1}^{N} S^z_k ,\\
M_x &\equiv\sum_{k=1}^{N}(-1)^k S^z_k, \\
I_x &\equiv\sum_{k=1}^{N}(-1)^k S^x_k ,
\end{align}
and the $z$ imbalance $I_z$ [\er{Iz}] (cf.~\cite{Smith2016}). The QFI for $M_x$ is maximal and it certifies the presence of MPE at all times as it is larger than its separability threshold, which is equal to $N$ (see Fig.~\ref{fig:QFI}).

Since both $\delta M_z$ and $I_z$ commute with the $z$ magnetization $M_z$ [which is fixed to zero for the initial condition \eqref{cdw}], their separability threshold is reduced to zero (see Sec.~\ref{subsec:witnessMPE}). Therefore, exactly as $M_x$, they witness the MPE for all times as well (see Fig.~\ref{fig:QFI}). Interestingly, after the initial growth the QFI per spin for $M_x$ and $I_x$ saturates at times $t\sim J^{-1}$, while it continues to grow for $I_z$ and $\delta M_z$ in the interacting case [see Fig.~\ref{fig:QFI_open}(a)]. Moreover, even in the presence of interactions, the QFI per spin for $M_x$ and $I_x$ remains unchanged for different system sizes (not shown), while it increases for $I_z$ and $\delta M_z$ (the latter not rescaled by $N$) [see  Fig.~\ref{fig:QFI_open}(a)]. 

This growth of the QFI for $I_z$ and $\delta M_z$ is related to the fact that such a choice of observables also witnesses BPE for the staggered partition ($ABAB \cdots AB$) and the partition into half chains ($AA \cdots AB \cdots BB$), respectively. Indeed, the QFI for such observables detects the asymmetry with respect to the magnetization difference in the partitions $ABAB \cdots AB$ and $AA \cdots AB \cdots BB$, respectively, while the asymmetry can be present only in states with BPE  due to a fixed total magnetization $M_z$ (cf.\ Fig.~\ref{fig:matrix} and Sec.~\ref{subsec:witnessBPE}). Moreover, even in the presence of dephasing, when both the QFI and its experimentally accessible lower bound in terms of the curvature (see Appendix~\ref{sec:method}) are reduced due to mixedness of the system state, they continue to witness BPE, due to the bipartite-separability threshold being zero [see Fig.~\ref{fig:QFI_open}(b)]. As we discuss in the next section, it is possible not only to witness, but also to bound from below the bipartite entanglement present in the many-body localized system of Eqs.~\eqref{XXZ} and~\eqref{master}. 
 
We note that it is enough to measure the QFI of the magnetization difference $\delta M_z$ or the imbalance $I_z$ to distinguish the Anderson localized phase from the many-body localized phase (cf. Fig.~\ref{fig:QFI_open}), which is not possible by measuring simply the averages of those observables (cf. Fig.~\ref{fig:av}), whose saturation at nonzero value is used in experiments as markers of localization~\cite{Schreiber2015,Choi2016}. This observation was already made in Refs.~\cite{Bardarson2012,Smith2016}. As we will show in the next section, however, our bounds~\eqref{ineqS} and~\eqref{ineqN} not only distinguish the Anderson-localized phase from the many-body localized phase, but also bound from below the amount of bipartite entanglement.

\subsection{Measuring the growth of entanglement in MBL dynamics}

In Fig.~\ref{fig:MBL_MPE} we observe that the coherence in the basis of $z$ magnetization follows the volume law at all times [with respect to the effective size of the subspace with fixed zero-magnetization $N_\text{eff}=\log_2{{N}\choose{N/2}}$] both for closed dynamics and in the presence of the dephasing. The dephasing leads to exponential decay at a rate dependent only on the interactions, but not on the system size. As we derived in~\er{ineqMPE}, the coherence is a faithful upper bound on MPE quantified by relative entropy. MPE is also expected to follow the volume law, as, by definition, it is the entanglement across the partition of the system into $N$ subsystems (and thus the area of this partition is equal to the volume). Interestingly, the coherence is also an upper bound on the entanglement entropy [see Eq.~\eqref{bpeC}], which is known to be connected to the diagonal entropy in the so-called $l$-bit basis~\cite{Serbyn2013,Serbyn2013b,Huse2014}.

In order to investigate the bipartite entanglement, in Figs.~\ref{fig:MBL_BP}(a) and~\ref{fig:MBL_BP}(c) we show the growth in the entanglement entropy (solid lines) for two different bipartitions:  half chains ($AA \cdots AB \cdots BB$) and the staggered bipartition ($ABAB \cdots AB \cdots AB$). While in the presence of interactions the entanglement entropy between half chains shows pronounced \emph{logarithmic growth}~\cite{Bardarson2012,Serbyn2013} between the initial area law regime and the asymptotic saturation to the volume law, the entanglement between staggered partitions initially follows the volume law  [see Fig.~\ref{fig:MBL_BP}(c)]. This is due to the presence of $N$ boundaries faces between the $A$ and $B$ parts of the system in the staggered partitioning. The asymptotic value of the entanglement entropy per spin seems to follow a logarithmic dependence on the system size [see the inset in Fig.~\ref{fig:MBL_BP}(c)]. The behavior of entanglement entropy for the staggered bipartition is not directly captured by the usual MBL mechanism of dephasing in the basis of exponentially localized integrals of motions~\cite{Serbyn2013,Serbyn2013b,Huse2014,Ros2015}, as the minimum localization length is single site, i.e., of the partition size, and thus entanglement growth is necessary due to the neglected boundary effects and higher-order corrections to the logarithmic growth~\cite{Znidaric2018}.  
\begin{figure}[ht!]
	\begin{center}
		\includegraphics[width=\columnwidth]{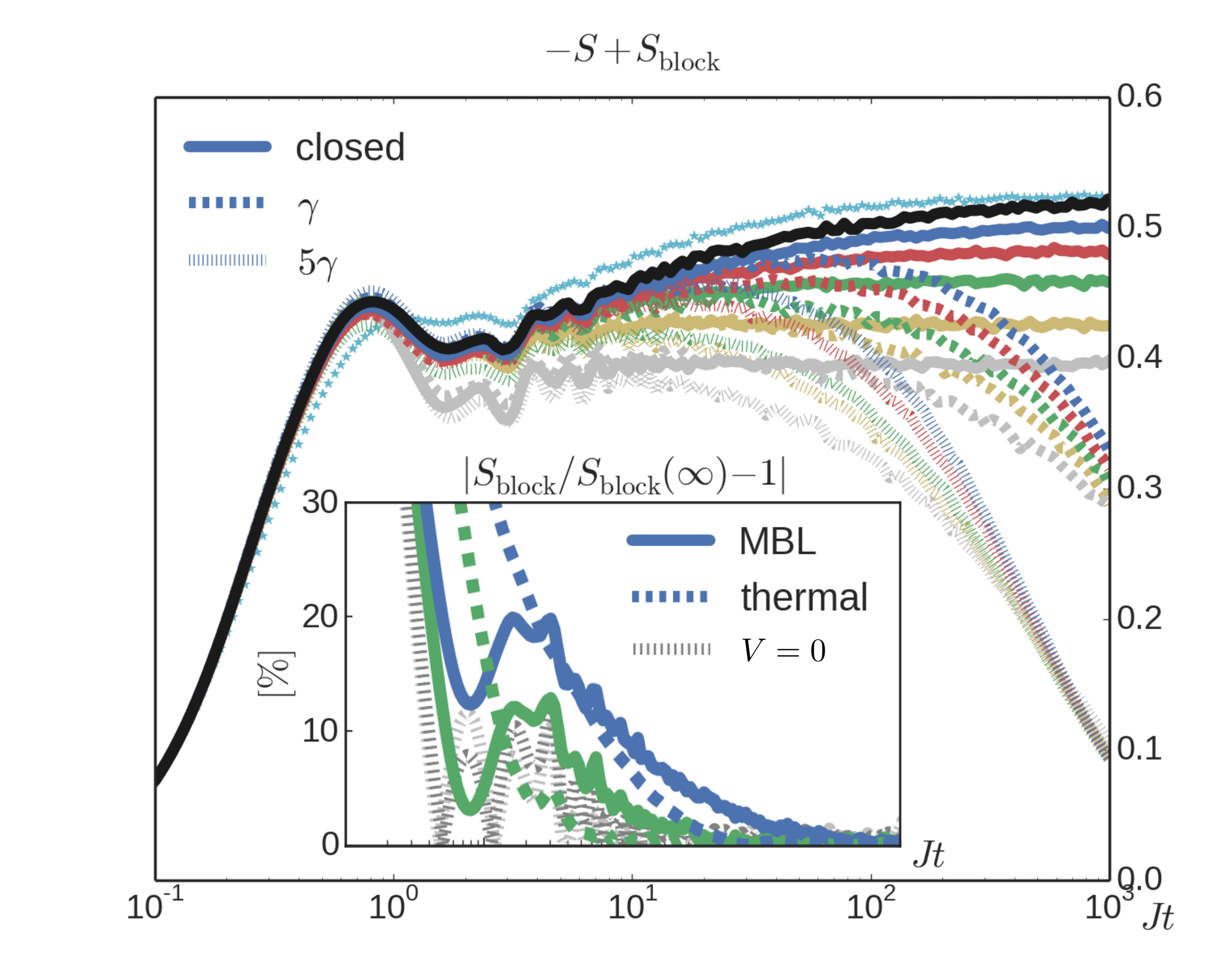}
		\vspace*{-7mm}
		\caption{
			{\bf Growth of asymmetry in an open MBL system}:
			asymmetry of half magnetization, cf.~\er{ineqS}, 
			for the $XXZ$  chain of Sec.~\ref{sec:MBL}, as also shown in Fig.~\ref{fig:MBL_BP}. Decay in the presence of dephasing ($\gamma/J=2\times 10^{-4}$) depends on the interaction strength, but not the system size. 
			The parameters of the dynamics [cf.~\era{XXZ}{master}] are 
			$N=6,8,10,12,14$ [yellow, green, red, blue, and black (grayscale: light gray to black), respectively, open case only up to $N=12$], $V/J=2$ and $h/J=5$; light blue stars are for interaction strength $V/J=5$ and $N=12$, and 
			gray curves correspond to the noninteracting case $V/J=0$ with $N=8$.
			The inset shows that although the asymmetry saturates one order of magnitude earlier that the entanglement entropy [cf.\ Fig.~\ref{fig:MBL_BP}(a)], its growth lasts for one order of magnitude longer than in the thermal phase, where asymptotic asymmetry is proportional to $\log_2(N/2)$ (cf.\ Appendix~\ref{app:thermal}).  
			The parameters in the inset are  $V/J=2$, $h/J=5$, and $N=8,12$ [green (gray) solid line and blue (darker gray) solid line], $V/J=2$, $h/J=1$, and $N=8,12$ [green (gray) dashed line and blue (darker gray) dashed line],  $V/J=0$, $h/J=5$, and $N=8,12$ [light gray dotted line and dark gray dotted line].
		}
		\vspace*{-5mm}
		\label{fig:MBL_block}
	\end{center}
\end{figure}

We now discuss the behavior of the asymmetry of the magnetization difference, which we derived as a new lower bound on the relative entropy of BPE in~\er{ineqS}. Note that in the closed case (dashed lines in Figs.~\ref{fig:MBL_BP}(a) and~\ref{fig:MBL_BP}(b)), the asymmetry simply equals the entropy of the magnetization difference distribution, which is created by (in the fermion language) fermions hopping across the boundary of the partition, and bounded by $\log_2(N/2\!+\!1)$ (cf.~Sec.~\ref{subsec:measureBP}). In particular, for the half chain bipartition, it initially follows the area law  which, in analogy to the entanglement entropy,  is broken at later times if the interactions are present in the system (see also Fig.~\ref{fig:MBL_block}). Although the saturation to the asymptotic value takes place an order of magnitude earlier than for the entanglement entropy, for the considered system size the growth is still longer than in the Anderson-localized phase (no interactions) or thermal phase (low disorder) (see the inset in Fig.~\ref{fig:MBL_block}, for quantum correlations in the thermal case see Appendix~\ref{app:thermal}). The asymptotic value for system sizes $N=8,10,12,14$ seems to follow the  logarithmic behavior, i.e., to be proportional to $\log_2(N/2\!+\!1)$. If this logarithmic scaling is present for all system sizes, asymmetry will detect breaking of the area law by the entanglement entropy.

Note that, similarly to the entanglement entropy in the staggered bipartition, the dynamics of asymmetry is not explained by  dephasing in the basis of exponentially localized integrals of motions~\cite{Serbyn2013,Serbyn2013b,Huse2014,Ros2015}, which is responsible for the logarithmic growth of entropy. The growth of asymmetry is instead contained within boundary effects (which are responsible for the area law of entanglement entropy in the noninteracting system) and higher-order corrections to the logarithmic growth in time of the entanglement entropy (at times later than $\sim J^{-1}$). In particular, in the closed case without interactions, i.e., in the Anderson-localized phase (gray curves in Fig.~\ref{fig:MBL_BP}), we observe good agreement between the asymmetry and entanglement entropy. We leave it as questions for future investigation what the exact mechanism of the asymmetry growth observed in Figs.~\ref{fig:MBL_BP} and~\ref{fig:MBL_block} is and whether this growth is unbounded. 

For the staggered bipartition, the asymmetry follows a logarithmic law  at all times due to fermions hopping through the $N$ faces of the partition boundary [see Fig.~\ref{fig:MBL_BP}(c,d)]. The dynamics of asymmetry is unchanged in the presence of weak enough dephasing [see Fig.~\ref{fig:MBL_BP}(b,d)], while the decay at long times is independent of the system size, but it changes with interactions (as it is also the case for coherence). 
% is it similarly for S_A?
In Figs.~\ref{fig:MBL_BP}(b) and~\ref{fig:MBL_BP}(d) we also show the lower bound on the relative entropy of BPE in Eq.~\eqref{ineqSA}~\footnote{We have  $S(\rho_A)=S(\rho_B)$ on average over trajectories, also in the dynamics with dephasing, as the  Hamiltonian \eqref{XXZ} is symmetric under flipping all spins, and the disorder distribution is also symmetric.}, which decays faster than the asymmetry, and thus the asymmetry is a tighter bound at long times.\\

As the asymmetry in not directly accessible in experiments with noise, in Figs.~\ref{fig:negativity} and~\ref{fig:MBL_l1}  we show the lower bounds on the convex roof of negativity of BPE,~\eqref{ineqN}, which demonstrate an analogous behavior to the asymmetry in Fig.~\ref{fig:MBL_BP}. 
This bounds are related to multiple quantum coherence spectra, [as shown in Eq.~\eqref{l1blockM} and explained in Appendix~\ref{sec:method}]. In particular, for the choice of an observable as the magnetization difference, the long-time dynamics of $l_{1,\delta M_z}^\text{block}$ is system-size dependent, thus again indicating breaking of the area law [cf. Fig.~\ref{fig:negativity}(b)].

Finally, we note that multiple coherence spectra featured in the bound~\eqref{ineqN} have been recently proposed as a witness of many-body localization in terms of the average correlation length~\cite{Wei2018}. This approach is successful even for highly mixed states, but many-body localization is considered in terms of the logarithmic growth of entropy of a subsystem (i.e., all bipartite correlations) rather than in terms of bipartite entanglement (i.e., only quantum bipartite correlations). Experimentally feasible witnesses of many-body localization have been discussed also in~\cite{Serbyn2014a} and~\cite{Goihl2016}.

\section{Conclusions} \label{sec:conclusions}

In this work we investigated the relation of entanglement and quantum discord to coherence and asymmetry in systems with a fixed or conserved local charge. Our results, which we summarize below, enable efficient witnessing of the presence of entanglement and discord in many-body systems, as well as the investigation of the entanglement scaling with the system size, the very property used to distinguish quantum phases.   

First, we showed that in the presence of a fixed local charge, coherence or asymmetry in the local charge basis is  related to the presence  in a quantum state of multipartite or bipartite entanglement, respectively. Therefore, the nonzero QFI of diagonal and block-diagonal observables becomes a witness of multipartite and bipartite entanglement. When the charge is conserved, but not fixed, we argued that coherence and asymmetry are instead connected to  multipartite and bipartite discord, and they can also be witnessed by the nonzero QFI.

Second, we found that the relation of entanglement to coherence and asymmetry is also quantitative. Namely, we showed that the amount of coherence serves as a faithful upper bound on the multipartite entanglement, and the amount of asymmetry as a lower bound on the bipartite entanglement. Importantly, the derived bounds are expressed as closed formulas for both  mixed and pure states, and in the bipartite case, where they can be accessed by multiple coherence spectra, breaking of the area law of entanglement can be detected in one-dimensional systems.  

We applied our results to the problem of many-body localization in a disordered $XXZ$  spin chain and demonstrated a slow growth of the asymmetry in the presence of interactions, which breaks the area law behavior obeyed by the system without interactions.

We leave the following questions open for further investigation: first, in the presence of a fixed local charge, whether the asymmetry of the subsystem charge,~\eqref{Sas} is also a lower bound on the entanglement entropy,~\eqref{bpe}, as it is the case in~\eqref{ineqS} for the bipartite entanglement of formation,~\eqref{bpeF}, second,  in the presence of a conserved local charge, whether the asymmetry of the subsystem charge is a lower bound on the relative entropy discord~\cite{Modi2010} (or whether $l_1^\text{block}$ is a lower bound on the negativity of quantumness~\cite{Piani2011,Nakano2013}). It is also not known whether the experimentally accessible $l_1^{\rm block}$ [Eq.~\eqref{l1block}], which appears as a lower bound on bipartite entanglement in~\eqref{ineqN}, is an asymmetry monotone. Finally, the mechanism of the asymmetry growth, which we observed numerically in the many-body localized phase, is not captured by the usual MBL mechanism of dephasing in exponentially localized basis and requires further explanation. \\

\emph{Note added}. Recently, we learned about a closely related work~\cite{Cornfeld2018} where a decomposition of negativity with respect to charge difference was discussed for systems with conserved charge. In particular, a scaling of negativity for ground states of one-dimensional critical systems was investigated, and a method to experimentally access thus decomposed negativity components was proposed based on~\cite{Gray2017}. An earlier work~\cite{Goldstein2017} discussed entanglement entropy for closed systems in the presence of charge conservation (and other symmetries), including the charge-resolved scaling of entanglement entropy within conformal theory. In our work we derived lower bounds on relative entropy of bipartite entanglement in both closed and open systems with a fixed charge in the context of asymmetry. We also proposed an experimentally accessible lower bound~in Eq.~\eqref{ineqN} to the convex roof of negativity. Furthermore, we discussed witnessing and quantifying multipartite entanglement (or discord when the charge is not fixed but conserved) in terms of coherence.

\acknowledgments

We  are thankful for access to the University of Nottingham High Performance Computing Facility. The research  leading  to  these  results received  funding  from the European Research Council under the European Union Seventh Framework Programme (FP/2007-2013) and ERC Grant Agreement No.\ 
335266 (ESCQUMA), the EPSRC Grant No. EP/M014266/1, and the H2020-FETPROACT-2014 Grant No. 640378 (RYSQ). I.L. gratefully acknowledges funding through the Royal Society Wolfson Research Merit Award. K.M. gratefully acknowledges support from Henslow Research Fellowship. We thank anonymous referees for their comments. \\

\begin{appendix}

%\section{Measuring MQC spectrum and curvature} \label{sec:method}
\Appendix{Measuring MQC spectrum and curvature} \label{sec:method}

We now briefly explain the methods of~\cite{Baum1985,Macri2016,Garttner2017,Garttner2018} to obtain the MQC spectrum~\eqref{Im}.%\\

%\subsection{Protocol}
\SubAppendix{Obtaining MQC spectrum}

Let the state of interest $\rho$ be a result of certain quantum dynamics from an initial pure state $|\psi_0\rangle$, i.e.,\ $\rho=\Lambda(|\psi_0 \rangle\!\langle\psi_0|)$ with a quantum channel (a completely positive and trace-preserving map) $\Lambda$. In the case of closed time-homogeneous dynamics, $\Lambda_t(\cdot)= e^{- i t H}(\cdot)e^{ i t H}$ corresponds to coherent evolution with the Hamiltonian $H$ in time $t$. 
In the case of open time-homogeneous dynamics, the evolution is given by $\Lambda_t=e^{t\mathcal{L}}$, where the superoperator $\mathcal{L}$ (often called the Lindbladian) is the generator in the master equation for $\rho$~\cite{Lindblad1976,Gorini1976}, 
\begin{equation}
\frac{\mathrm{d}}{\mathrm{d}t}\rho_t=\mathcal{L}(\rho_t) ,
\label{ME}
\end{equation}
with
\begin{equation}
\mathcal{L}(\cdot)=-i[ H, (\cdot)]+\sum_{j} L_j(\cdot) L_j^\dagger -\sum_{j} \frac{1}{2} \left\{ L_j^\dagger L_j , (\cdot) \right\} ,
\label{L}
\end{equation}
where $L_j$ are the jump operators. \\

\begin{figure}[t!]
	\begin{center}
		\includegraphics[width=1.0\columnwidth]{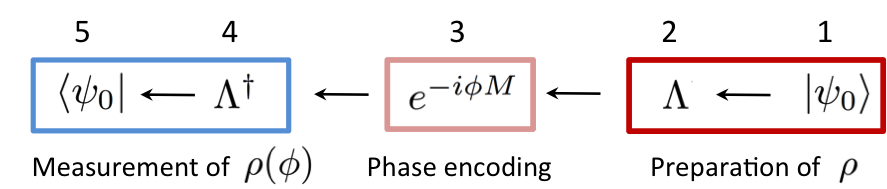}
		\vspace*{-4mm}
		\caption{
			{\bf Protocol for the curvature and the MQC spectrum}. For a system state $\rho$ being a result of dynamics from $|\psi_0\rangle$ (steps 1 and 2), its coherence with respect to the eigenbasis of an observable $M$ can be accessed by unitary perturbation encoding a phase $\phi$ (step 3), followed by a measurement (step 4 and 5) of the overlap between $\rho(\phi)$ and the unperturbed state $\rho$. This measurement scheme can also be used to estimate the encoded phase value $\phi$, and in the case of (noninteracting) closed dynamics it corresponds to Ramsey spectroscopy~\cite{Ramsey1956}.
		}\vspace*{-5mm}
		\label{fig:scheme}
	\end{center}
\end{figure}

The protocol to obtain the MQC spectrum consist of the following steps (see Fig.~\ref{fig:scheme}).
\begin{enumerate}
	\setlength{\itemsep}{0pt}
	\setlength{\parskip}{0pt}
	\item Preparation of the initial state $|\psi_0 \rangle$.
	\item Evolution of the initial state to  the state of interest $\rho=\Lambda(|\psi_0 \rangle\!\langle\psi_0|)$.
	\item Unitary phase $\phi$ encoding with an observable $M$, $\rho(\phi)=e^{-i\phi M} \rho e^{i\phi M}$.
	\item Conjugate evolution $\Lambda^\dagger[\rho(\phi)]$.
	\item Measurement of the overlap with the initial state $F(\phi)\equiv\langle \psi_0| \Lambda^\dagger[\rho(\phi)] |\psi_0 \rangle=\Tr[\rho \,\rho(\phi)] $.
\end{enumerate}

In the case when $\Lambda$ corresponds to time-homogeneous unitary dynamics~\cite{Macri2016},  step 4 corresponds to inverted system evolution, i.e.,\ evolution with Hamiltonian $-H$~\cite{Linnemann2016,Wei2018}. This is also the case for open dynamics with Hermitian jumps $L_j^\dagger=L_j$~\cite{Garttner2018}, e.g., dephasing, or with a set of jumps invariant under Hermitian conjugation $L_j^\dagger=L_{j'}$, e.g., infinite-temperature environments. The protocol can be  generalized to a mixed initial state preparation (discussed below). \\

Since $F(\phi)=\Tr[\rho\, \rho(\phi) ]$, we have from~\eqref{Im} that 
\begin{equation}
F(\phi)= \sum_{i j} e^{-i \phi (m_i-m_j)} |\rho_{ij}|^2 = \sum_{m} e^{-i \phi m}I_m(\rho).
\label{F}
\end{equation}
Therefore, the MQC spectrum for $M$ can be accessed by the Fourier transform of $F(\phi)$. 

The usefulness of the method for many-body systems relies on the fact that the initial pure state $|\psi_0\rangle$ is usually assumed classical, i.e.,\ a product of individual subsystem states, while the prepared state $\rho$ can be entangled. It then follows that the final measurement of the overlap can be implemented by measuring local subsystem observables~\cite{Macri2016,Garttner2017,Garttner2018,Wei2018}. For the example of a system of $N$ spins $1/2$ and the choice of $M$ as total spin magnetization~\cite{Garttner2017,Garttner2018,Wei2018}, the Fourier transform of $F(\phi)$ requires performing the protocol for $d=N$ values of $\phi= 2\pi\,k/N$,  $k=1,...,N$. For a classical initial state, each protocol simply requires measurement of $N$ individual spin magnetizations (so in total $d=N$ measurements).  In order to reconstruct $\rho_{ij}$ in the  quantum state tomography, $4^N$ measurements of individual spin magnetizations in all directions are required~\cite{Haffner2005}.

%
%
%\subsection{Curvature}
\SubAppendix{Curvature}
 
In Sec.~\ref{sec:witness} we discuss how the QFI,~\eqref{qfi}, can be used to witness multipartite entanglement. Here we consider the {\em curvature}~\cite{Lee2011,Girolami2014b,Park2016,Macri2016,Girolami2017,Garttner2018}
\begin{eqnarray}
C(M,\rho) &\equiv& \sum_{i \neq j} (m_{i} - m_{j})^{2}  | \rho_{ij} |^{2}=-\Tr( [M,\rho]^2)\nonumber\\
&=&\sum_{ij} ( \lambda_{i} -\lambda_{j})^{2} |\langle \lambda_{i} | M | \lambda_{j} \rangle|^{2},
\label{C}
\end{eqnarray}
where $\rho_{ij}\equiv\langle i|\rho|j\rangle$, and $| \lambda_{i} \rangle$ is an (orthonormal) eigenstate of $\rho$ that corresponds to an eigenvalue $\lambda_i$, i.e., $\rho=\sum_i \lambda_i | \lambda_{i} \rangle\!\langle\lambda_{i} |$. The curvature is a lower bound on the QFI~\cite{Girolami2017,Garttner2018},
\begin{equation}
{\rm QFI}(M,\rho) \geq 2\, C(M,\rho) ,
\label{qvc2}
\end{equation}
with the inequality saturated for pure states and mixed states of dimension $2$. Note that, analogously to the QFI, the curvature is also a faithful witness of coherence for nondegenerate $M$, as it is zero if and only if the QFI is zero [cf. Eq.~\eqref{C}]. Therefore, in the presence of a fixed (conserved) local charge for nondegenerate diagonal observables, the nonzero curvature is a faithful witness of multipartite entanglement (quantum discord) (cf. Sec.~\ref{sec:witness}). Similarly, for a system divided into two parts, for block-diagonal observables with respect to charges of the two parts, the nonzero curvature is the witness of bipartite entanglement (quantum discord), which performs as well as the QFI.

Most importantly, the curvature can be measured experimentally in several approaches, which we now review. 

First, the authors of~\cite{Garttner2018} note that the second moment of the MQC spectrum,~\eqref{Im}, the protocol for which we described above, yields the curvature,
\begin{equation}
\sum_m m^2 I_m (\rho) = C(M,\rho),
\end{equation}
so  the curvature can be calculated from the MQC spectrum. 
Second, the authors of~\cite{Macri2016,Garttner2018,Lee2011,Park2016} also note that it corresponds to the second derivative of $F(\phi)$ [Eq.~\eqref{F}], 
\begin{eqnarray}
-\partial^2_\phi F(\phi) |_{\phi=0}&=&\sum_{m} m^2 I_m(\rho) = C(M,\rho), \label{C2}
\end{eqnarray}
so  the curvature can be found from $F(\phi)$ by numerically calculating its second derivative around $\phi=0$.
In particular, for a pure state $\rho=|\psi\rangle\!\langle \psi|$ we have that $F(\phi)=|\langle \psi|\psi(\phi)\rangle|^2$ corresponds to the fidelity between $|\psi(\phi)\rangle$ and $|\psi\rangle$ (cf.~Fig~\ref{fig:scheme}). Therefore,  Eq.~\eqref{C2} describes the fact that the Bures metric equals the QFI~\eqref{qfi}. For a mixed state $\rho$, $F(\phi)=\Tr[\rho\,\rho(\phi)]$ is the probability of measuring $\rho$ on $\rho(\phi)$ and thus the curvature is the speed of the decay of that probability with $\phi$~\cite{Smerzi2012}.

 In case when $M$ is non-local, and thus the experimental implementation of the unitary rotation is challenging, it was proposed  in Refs.~\cite{Lee2011,Park2016}  to use dephasing along the $M$ observable, $\L_M(\rho)\equiv M\rho M - \frac{1}{2}(M^2\rho +\rho M^2 )$ [cf. Eq.~\eqref{L}] instead of the unitary rotation with $M$, in the step 3. of the protocol (cf.~Fig.~\ref{fig:scheme}), since
\begin{eqnarray}
C(M,\rho)&=& \Tr(\rho [M,[M,\rho]]) \\
&=& -\Tr[\rho ( 2 M\rho M - M^2\rho -\rho M^2)]\nonumber\\
&=& -\partial_\phi \Tr[\rho e^{2\phi \L_M}(\rho)] |_{\phi=0}.\nonumber
\end{eqnarray}
We note that, for the parity $P = (-1)^{M}$ we have $P^2=\mathds{1}$, so  
\begin{eqnarray} 
C(P,\rho)&=& \Tr( \rho [P,[P,\rho]]) =- 2 \Tr[\rho ( P \rho P )]+2 \Tr(\rho^2)\nonumber\\
&=& - 4 \Tr[\rho ( \Pi_{+} \rho \Pi_{+} +\Pi_{-} \rho  \Pi_{-})] + 4 \Tr(\rho^2),\quad \label{Cpar}
\end{eqnarray}
where $\Pi_{+}$ and $\Pi_{-}$ are the projections on the even and odd eigenspaces of $M$, so  $\Pi_{+}+\Pi_{-}=\mathds{1}$ and  $P=\Pi_{+}-\Pi_{-}$. Therefore, the curvature can also be measured in the protocol in~Fig.~\ref{fig:scheme} with the unitary rotation replaced by the non-demolition measurement of the system parity [while the second term in~\eqref{Cpar} corresponds to step 3 omitted].

Finally, we also note that a different method for accessing the curvature $C(M,\rho)$, by measuring the overlap of two copies of a state $\rho$ with one of them unitarily perturbed to $\rho(\phi)$, was proposed in Refs.~\cite{Girolami2014b,Girolami2017}.\\

%When $\rho$ is prepared via open dynamics, it may not be possible, due to dissipation, to implement unitary phase encoding of all $\phi$ values. The curvature, nevertheless, should still be accessible if small values $\phi$ can be realised, cf.~\er{C2}.

%\subsection{Extensions of the method}\label{app:method}
\SubAppendix{Extensions of the method}\label{app:method}

Here we propose extensions of the method~\cite{Baum1985,Macri2016,Garttner2017,Garttner2018}  to the case of a mixed initial state and in the presence of dissipation, which is not invariant under Hermitian conjugation.   \\

%\subsubsection{Mixed initial state}
\SubSubAppendix{Mixed initial state}

The method can be generalized to a mixed initial preparation of $\rho_0$ in the step 1, instead of a pure state $|\psi_0 \rangle$ (cf.~Fig.~\ref{fig:scheme}). In this case the final measurement in the step 5 needs to be replaced by measurements of the pure states being the eigenvectors of $\rho_0$  with nonzero eigenvalues~\cite{Garttner2017,Garttner2018}, i.e.,\ projections on $|\psi_0^{(i)}\rangle$  where $\rho_0=\sum_i\lambda_i |\psi_0^{(i)}\rangle\!\langle \psi_0^{(i)}|$.  The value of $F(\phi)$ can then be recovered as the weighted average $F(\phi)=\sum_i \lambda_i F_i(\rho)$ of the individual overlaps $F_i(\rho)\equiv \langle \psi_0^{(i)}| \rho(\phi) |\psi_0^{(i)}\rangle$.  When the initial state $\rho_0$ is classical, its eigenvectors can be chosen as product states, and thus the final measurement can be simply implemented by measuring the local observable with single-site resolution, while $\lambda_i$ can be extracted by measurement of that observable on $\rho_0$.\\

%\cite{what does Lukin do?}

%references to OTOC

%Moreover, for unitary dynamics, $F(\phi)$ is an out-of-time-order correlator, i.e.,\ is of the form $\langle W^\dagger_t V^\dagger W_t V \rangle$ with the Heisenberg evolved $W_t=e^{it H} W e^{-it H}$. The authors of~\cite{Garttner2017,Garttner2018} show that this relation holds for the choice $V=|\psi_0 \rangle \!\langle\psi_0|$ (or more generally $V$ s.t. $V\rho_0=\rho_0$) and $W(\phi)= e^{-i\phi M}$. Similarly as for initial mixed state in the protocol, $\langle W^\dagger_t(\phi) V^\dagger W_t(\phi) V \rangle$ can be actually obtained for any local observable $V$ and all initial (possibly mixed) states which are diagonal in the product basis corresponding to $V$. This requires a measurement of $V$ on $\rho_0$, before the evolution in the step 2., and the final measurement in the step 5. simply replaced by the second measurement of $V$. The resulting correlations of the two measurement outcomes yield exactly $\langle W^\dagger_t(\phi) V^\dagger W_t(\phi) V \rangle_{\rho_0}$.\\

%\subsubsection{Dissipation not invariant under Hermitian conjugation and a lower bound on QFI}
\SubSubAppendix{Dissipation not invariant under Hermitian conjugation and a lower bound on QFI}

We now consider the case when the dynamics $\mathcal{L}^\dagger$, required in step 4 of the protocol (cf.~Fig.~\ref{fig:scheme}), cannot be obtained from  $\mathcal{L}$ simply by a change in the Hamiltonian sign, and only the Hamiltonian in~\eqref{L} can be changed, while the dissipation is assumed to be given. We note that the case of the protocol with $\mathcal{L}^\dagger$ replaced by any other dynamics $\mathcal{L}'$, e.g., $\mathcal{L}'=\mathcal{L}+2i[H,(\cdot)]$ for the case of the reverted Hamiltonian, corresponds to a measurement of state $\rho_{\phi}$, which is composed of steps 4 and 5, as follows. This measurement has two possible outcomes $1$ and zero, corresponding to the projections $\Pi_0\equiv e^{t\mathcal{L}'^\dagger} (|\psi_0 \rangle\!\langle \psi_0|)$ and $\Pi_1\equiv \mathds{1}-e^{t\mathcal{L}'^\dagger} (|\psi_0 \rangle\!\langle \psi_0|)$. Therefore, we can consider the  Fisher information (FI) associated with this measurement, $\text{FI}(\rho_\phi,\{\Pi_x\}_x)\equiv \sum_{x=0,1} p_\phi(x)[\partial_\phi \log p_\phi (x) ]^2$, where the probability $p(x)\equiv\Tr(\Pi_x\rho_\phi)$, $x=0,1$. In the case of the closed dynamics, this corresponds to the Ramsey scheme~\cite{Ramsey1956} and the FI equals the QFI, while in general the FI is a lower bound on the QFI~\cite{Braunstein1994} (which equals the FI for the optimally chosen measurement).

%\section{Conservation of local charge in classical states} 
\Appendix{States without quantum discord in presence of conserved local charge} 
\label{app:discord} 
In this appendix we characterize states without multipartite or bipartite quantum discord in the presence of a conserved local charge. 

%\subsection{Classical states with conserved local charge}  
\SubAppendix{Classical states with conserved local charge} 

Here we prove that classical states~\eqref{rho:cl} with a conserved local charge, are block diagonal with respect to subsystem charges (see Eq.~\eqref{Qc0}). \\

\emph{Proof}. The charge conservation is defined as $[Q,\rho]=0$, while
\begin{equation}\label{Qc1}
-\mathrm{Tr} ([Q,\rho]^2)=\mathrm{Tr}(Q^2 \rho^2)-\mathrm{Tr}(Q \rho Q \rho)=C(Q,\rho)
\end{equation}
[cf.~\er{C}]. Moreover, when the conserved charge is local, $Q=\sum_{k=1}^N Q^{(k)}$,  for a classical state~\eqref{rho:cl} the curvature is additive in the charges,
\begin{eqnarray}
&&C(Q,\rho_{\text{cl}})=\sum_{\substack{i_1,...,i_N\\j_1,...,j_N}} (\lambda_{i_1,...,i_N}-\lambda_{j_1,...,j_N})^2 \nonumber \\
&&\qquad\qquad\qquad\times \,\left|\sum_{k=1}^N \langle e_{i_k}^{(k)} |Q^{(k)}|e_{j_k}^{(k)}\rangle \,\,  \prod_{\substack{l=1\\l\neq k}}^N \delta_{i_l,j_l}\right|^2 \nonumber\\
&&=\sum_{k=1}^N  \sum_{\substack{i_1,...,i_N\\j_1,...,j_N}} (\lambda_{i_1,...,i_N}-\lambda_{j_1,...,j_N})^2 |\langle e_{i_k}^{(k)} |Q^{(k)}|e_{j_k}^{(k)}\rangle|^2  \nonumber\\
&&=\sum_{k=1}^NC(Q^{(k)},\rho_\text{cl}),\label{Qc2}
\end{eqnarray}
where in the first equality we exploited orthogonality of the basis in $\rho_\text{cl}$ [cf.~\er{C}], while the second equality follows from noting that the difference $(\lambda_{i_1,...,i_N}-\lambda_{j_1,...,j_N})$ requires the indices to differ. Therefore, for a conserved charge, from the positivity of curvature in~\eqref{Qc2} together with~\eqref{Qc1} we arrive at
\begin{equation}\label{Qc3}
[Q^{(k)},\rho_\text{cl}]=0 \quad \text{for}\quad k=1,...,N.
\end{equation}
%Therefore the product eigenbasis of classical states, $|e_{i_k}^{(k)}\rangle$, can be chosen as $Q^{(k)}$ charges.When $Q^{(k)}$ are nondegenerate, the basis defined by the tensor product of $Q^{(k)}$ eigenstates is unique.  We thus conclude that when this basis is chosen as the computational basis, all classical states that conserve $Q$ are diagonal, and thus feature no coherence. We finish the proof by noting that all diagonal states are classical, cf.~Eq.~\eqref{rho:cl}. 

%\subsection{Classical-quantum states with conserved local charge} 
\SubAppendix{Classical-quantum states with conserved local charge} 
Here we prove that classical-quantum states~\eqref{rho:clq}, in the presence of local charge conservation, are block diagonal in the eigenspaces of $Q^{(A)}$ and $Q^{(B)}$.            \\                

\emph{Proof}. We have 
\begin{eqnarray}\label{qcl1}
-\mathrm{Tr}([Q,\rho_\text{cl-q}]^2)&=&C(Q,\rho_\text{cl-q})\\\nonumber
&=&\sum_{ij}\sum_{kl} (\lambda_{i}\lambda_k^{(i)}-\lambda_{j}\lambda_l^{(j)})^2 |\langle \lambda_{ik}| Q|\lambda_{jl}\rangle|^2
\end{eqnarray}
[cf.~\er{C}], where we have introduced the eigendecomposition of $\rho_i^{(B)}=\sum_{k}\lambda_k^{(i)}| e_k^{(B,i)}\rangle\!\langle e_k^{(B,i)}|$ and have defined $|\lambda_{ik}\rangle\equiv|  e_i^{(A)}\rangle\otimes |e_k^{(B,i)}\rangle$. When the charge is local $Q=Q^{(A)}+Q^{(B)}$, we furthermore have that 
 \begin{eqnarray}\label{qcl2}
 \langle \lambda_{ik}| Q|\lambda_{jl}\rangle&=&  \langle e_i^{(A)} |Q^{(A)}|  e_j^{(A)}\rangle \,\langle e_k^{(B,i)} |e_l^{(B,j)}\rangle\\\nonumber
 &&+ \,\delta_{i,j}\, \langle e_k^{(B,i)}|Q^{(B)} |e_l^{(B,j)}\rangle
 \end{eqnarray}
 [cf.~\eqref{Qc2}]. Note that when $i\neq j$, we have only the first term in~\eqref{qcl2}, while for $i=j$ we have that the $Q^{(A)}$-term is multiplied by $\langle e_k^{(B,i)}|e_l^{(B,i)}\rangle=\delta_{k,l}$, imposing the multiplicative term to vanish, $\lambda_{i}\lambda_k^{(i)}-\lambda_{j}\lambda_l^{(j)}=0$. Therefore, the only contribution comes from the first term, and thus the curvature is additive in the charge,
 \begin{eqnarray}\label{qcl3}
C(Q,\rho_\text{cl-q})
 &=&C(Q^{(A)},\rho_\text{cl-q})+C(Q^{(B)},\rho_\text{cl-q}).
 \end{eqnarray}
  We thus conclude, using positivity of the curvature, that any classical-quantum state with a conserved local charge $[Q^{(A)}+Q^{(B)},\rho_\text{cl-q}]=0$, fulfills
  \begin{eqnarray}\label{qcl3}
 [Q^{(A)},\rho_\text{cl-q}]=0=[Q^{(B)},\rho_\text{cl-q}].
  \end{eqnarray}

%\section{Separable operations conserving local charge or preserving local charge conservation} \label{app:SSR}
\Appendix{Separable operations conserving local charge or preserving local charge conservation} \label{app:SSR}

In this appendix we derive conditions on separable operations conserving a local charge or preserving its conservation in a quantum state. We show that under such operations, the asymmetry of subsystem charge and the coherence are nonincreasing, in the bipartite and the multipartite case, respectively. We also discuss their relation to LOCC with ancillas in the presence of a SSR~\cite{Schuch2004b}.

%\subsection{Separable operations conserving a local charge}
\SubAppendix{Separable operations conserving a local charge}

We consider the bipartite case, where a local charge $Q=Q^{(A)}   +Q^{(B)}$ is conserved by separable operations both on average and probabilistically, i.e.,\ $\Lambda(\rho)=\sum_j p_j K_j\rho\, K_j^\dagger$. By definition, this requires Kraus operators to be separable, $K_j= K_j^{(A)}\otimes K_j^{(B)}$ (here also LOCC are included), and to conserve the charge, $[K_j,Q]=0$ for all $j$. This property is also known as strong symmetry with respect to $Q$ (cf.~\cite{Albert2014}).

Let $K_j^{(A)}(q'_A,q_A)$ be a restriction on $K_j^{(A)}=\sum_{q_A,q'_A} K_j^{(A)}(q'_A,q_A)$ to mapping from states of a fixed value $q_A$ to $q_A'$. We have
\begin{eqnarray}\label{SSR:occ0}
0=[K_j,Q]&=&\sum_{q_A q'_A q_B q'_B} (q_A-q'_A+q_B-q'_B) \\\nonumber
&&\qquad \quad\times \quad K_j^{(A)}(q'_A;q_A)\otimes K_j^{(B)}(q'_B;q_B),
\end{eqnarray}
which implies $q_A-q'_A=q_B-q'_B$ for all $q_A$, $q'_A$, $q_B$, and $q'_B$. As the Kraus operator is separable, the shift in charge must be constant, $q_A-q'_A=\alpha_j$, 
\begin{equation}\label{SSR:occ1}
[K_j,Q^{(A)}]= \alpha_j K_j,
\end{equation}
and compensated  by the rest of the system, $q_B-q'_B=\beta_j=-\alpha_j$ (similarly, in the multipartite case, in which a fixed subsystem charge is mapped to a fixed subsystem charge by a uniform shift which cancels over the whole system,  $\alpha_j+\beta_j+...=0$ is required). In particular, when the shift $\alpha_j=\alpha$ is the same for all Kraus operators, they correspond to block-diagonal operators on a subsystem and a local ancilla, i.e., conserving the total charge of a subsystem and an ancilla~\cite{Schuch2004b}. \\

%The QFI for the difference of subsystem charges, $\delta Q=Q^{(A)}-Q^{(B)}$, is known to be a nonlocality monotone~\cite{Schuch2004,Schuch2004b}, i.e., it is nonincreasing under LOCC operations which conserve the charge, cf.~\cite{Toth2013,Yu2013}. As the QFI is unchanged by a constant shift in an observable, cf.~\er{qfi}, it remains the nonlocality monotone~\cite{Schuch2004b} also with respect to the separable operations conserving the charge.
%THIS NEEDS TO BE CHECKED!

We now argue that the relative entropy of asymmetry~\eqref{Sas} is a nonlocality monotone with respect to a subsystem charge, as follows. The separable operations conserving the charge belong to the set of free operations in the asymmetry theory~\cite{Bartlett2007,Gour2008,Vaccaro2008,Gour2009,Marvian2012}, since they preserve the symmetric states, by causing only a shift by a constant in the subsystem charge [cf.~\eqref{SSR:occ1}]. Therefore, they cannot increase any asymmetry monotone, including the relative entropy of asymmetry~\eqref{Sas}.

%\subsection{Separable operations preserving local charge conservation}
\SubAppendix{Separable operations preserving local charge conservation}

Here we consider separable operations that transform a state with a  conserved charge into another state with a conserved charge, both on average and probabilistically, $[K_j \rho \,K_j^\dagger, Q]=0$, for all $j$, whenever $[\rho,Q]=0$.

In the bipartite case the separable operation is given by $K_j= K_j^{(A)}\otimes K_j^{(B)}$. For a state $\rho$ of fixed subsystem charges of values $q_A$ and $q_B$, we thus have 
\begin{eqnarray}
&&0=[K_j \rho \,K_j^\dagger, Q] \label{SSR:occ}
\\\nonumber
&&=\sum_{q'_A q''_A, q'_B q''_B} (q''_A+q''_B-q'_A-q'_B)  \times
\\\nonumber
&&  K_j^{(A)}(q'_A;q_A)\!\otimes\! K_j^{(B)}(q'_B;q_B) \, \rho \, \left[K_j^{(A)}(q''_A;q_A)\!\otimes\!  K_j^{(B)}(q''_B;q_B)\right]^\dagger
\end{eqnarray}
and the condition $q''_A+q''_B=q'_A+q'_B$ follows. Since $K_j= K_j^{(A)}\otimes K_j^{(B)}$, by considering a separable state $\rho=\rho_A\otimes\rho_B$, we see that this condition can only be fulfilled when a given charge $q_A$  is mapped into a another single charge $q'_A= q_A+\alpha_j(q_A)$ and analogously $q_B$ is mapped into $q_B'=q_B+\beta_j(q_B)$. Therefore, the set of separable operations preserving the local charge conservation transforms states of a fixed subsystem charge into states of another fixed subsystem charge.\\

From this observation we can already conclude that the asymmetry of the subsystem charge,~\eqref{Sas}, in the states with conserved total charge again remains strictly nonincreasing  with respect to those operations (i.e., it remains a nonlocality monotone). This follows from the fact that such operations preserve the set of symmetric states with respect to the subsystem charge (also compare the proof for relative entropy in Supplemental Material of~\cite{Baumgratz2014}). 

Similarly, in the multipartite case, the separable operations which preserve charge conservation in the state require that a fixed charge is transformed into a fixed charge. This leads to the coherence~\eqref{Scoh} being not only a faithful upper bound~\eqref{ineqMPE}, but also an entanglement monotone.  In contrast to the degenerate bipartite case, this entanglement monotone is faithful. \\

In order to finish the characterisation of the separable operations preserving local charge conservation, we now investigate mapping of coherences in $\rho$. For the bipartite case, consider the coherence between different fixed subsystem charges, denoted by $q_A$ and $q_B$, and by $q'_A$ and $q'_B$. First, from the charge conservation in $\rho$, we have $q_A+q_B=q'_A+q'_B$, i.e.,\ $q_A=q_A'-\delta q$ and $q_B=q'_B+\delta q$ for some $\delta q$. Second, from the conservation of the charge in $K_j\rho K_j^\dagger$ [cf. Eq.~\eqref{SSR:occ}] 
 we have that $q_A$ ($q_B$) is mapped to a single value $\alpha_j(q_A)$ [$\beta_j(q_B)$], as well as 
\begin{eqnarray}\label{aabb}
\alpha_j(q_A')-\alpha_j(q_A)=-[\beta_j(q_B')-\beta_j(q_B)]
\end{eqnarray}
for all $q_A$, $q_A'$, $q_B$, and $q_B'$ such that $K_j^{(A)}$ ($K_j^{(B)}$) acts nontrivially (is nonzero) both on states with charges $q_A$ and $q_A'$ ($q_B$ and $q_B'$). Therefore, for such $q_A$ and $q_A'$ ($q_B$ and $q_B'$) we have that~\eqref{aabb} corresponds to a function of the subsystem charge difference $\gamma_j( \delta q)$, i.e.,
\begin{eqnarray}
\alpha_j(q_A+\delta q)&=&\gamma_j( \delta q)+\alpha_j(q_A),\\
\beta_j(q_B-\delta q)&=&\beta_j(q_B)-\gamma_j( \delta q ).
\end{eqnarray}
In particular, if $K_j^{(A)}$ connects nontrivially all charges, this function  is additive in the subsystem charge difference $\delta q$, i.e., $\gamma_j(\delta q)=c_j \,\delta q $ [note that for the Kraus operators that conserve charge, we have $c_j=0$, i.e., the shift is constant; cf.~\eqref{SSR:occ1}], which corresponds to weak symmetry. This is not possible to implement by local ancillas~\cite{Schuch2004b}.\\

%\section{Faithful upper bounds on geometric MPE and MPD, and on negativity of quantumness}\label{app:infidelity}
\Appendix{Faithful upper bounds on geometric MPE, geometric MPD, and negativity of quantumness}\label{app:infidelity}

%\subsection{Upper bounds on geometric MPE and MPD quantified with infidelity}
\SubAppendix{Upper bounds on geometric MPE and MPD quantified with infidelity}

Here we discuss how the geometric entanglement~\cite{Wei2003,wei2004} and the geometric multipartite quantum discord~\cite{Streltsov2011} can be faithfully bounded from above by the geometric coherence~\cite{Streltsov2015}.\\

The geometric entanglement~\cite{Wei2003,wei2004}  is defined as the convex roof of \emph{infidelity}, $1-F(|\psi\rangle,|\phi\rangle)\equiv 1-|\langle\phi|\psi\rangle|^2$, to the set of pure separable states 
\begin{equation}
\E^G_{\rm MP}(\rho)\equiv\min_{\{p_j, |\psi_j\rangle\}} [1- \sum_{j}p_j \max_{|\phi_\text{sep}\rangle} F(|\psi^{(j)}\rangle,|\phi_\text{sep}\rangle)].
\end{equation}
 Therefore, for $\rho$ with a fixed local charge, by considering the infidelity to the smaller set of pure separable states with a fixed local charge (i.e., the elements of the computational basis), we again obtain a faithful upper bound in terms of the geometric coherence $\C^G(\rho)$~\cite{Streltsov2015} whenever $\rho$ also is of the fixed local charge, 
\begin{eqnarray}
\E_{\rm MP}^G(\rho) &\leq& \min_{\{p_j, |\psi_j\rangle\}}  [1- \sum_{j} p_j \max_{|\phi_\text{diag}\rangle} F(|\psi^{(j)}\rangle,|\phi_\text{diag}\rangle)]\nonumber\\
&=& 1- \max_{\{p_j, |\psi_j\rangle\}}  \sum_{j} p_j \max_{\sigma_\text{diag}} F(|\psi^{(j)}\rangle\!\langle\psi^{(j)}| ,\sigma_\text{diag})\nonumber\\
&=& 1- \max_{\sigma_\text{diag}} F(\rho,\sigma_\text{diag}) \equiv\C^G(\rho)   
\label{ineqMPEg}
\end{eqnarray}
with $F(\rho,\sigma)\equiv \left[\Tr\left(\rho^{1/2}\,\sigma\rho^{1/2}\right)^{1/2}\right]^2$. In particular, for the pure state $F(|\psi\rangle\!\langle\psi| ,\sigma)=\langle \psi|\sigma|\psi\rangle $, so    $C^G(|\psi\rangle\!\langle\psi|)= \max_{\sigma_\text{diag}} F(|\psi\rangle\!\langle\psi| ,\sigma_\text{diag})=\max_{|\phi_\text{diag}\rangle} F(|\psi\rangle,|\phi_\text{diag}\rangle)$~\cite{Streltsov2017b}, which we used in the second line. The third line follows from the fact that the geometric coherence is equal to its convex roof (cf.~\cite{Streltsov2017b}).

Similarly, the geometric multipartite quantum discord~\cite{Streltsov2011} is quantified by infidelity to the classical states, 
\begin{equation}
\D^G_{\rm MP}(\rho)\equiv 1-\max_{\sigma_\text{cl}}  F(\rho,\sigma_\text{cl}).
\end{equation}
Therefore, for $\rho$ with a conserved local charge, by considering the infidelity to the smaller set of classical states with a conserved local charge (i.e., diagonal states), we again obtain a faithful upper bound by the geometric coherence
\begin{eqnarray}
\D_{\rm MP}^G(\rho) &\leq& 1- \max_{\sigma_\text{diag}} F(\rho,\sigma_\text{diag}) \equiv\C^G(\rho). \qquad
\label{ineqMPDg}
\end{eqnarray}
We note, however, that in contrast to the relative entropy coherence in Eq.~\eqref{Scoh}, the geometric coherence $\C^G(\rho)$ is usually difficult to evaluate.  

%\subsection{Upper bounds on negativity of quantumness}
\SubAppendix{Upper bounds on negativity of quantumness}

Multipartite discord can also be measured by  $l_1$ coherence minimized over the choice of a separable basis, which yields so-called \emph{negativity of quantumness} $\D_{\rm MP}^N(\rho)$~\cite{Piani2011,Nakano2013,Adesso2016}. Therefore, also in this case $l_1$ coherence in the computational basis~\eqref{l1} becomes a faithful upper bound on negativity of quantumness
\begin{eqnarray}
\D_{\rm MP}^N(\rho) &\leq& l_1(\rho).
\label{ineqMPDn}
\end{eqnarray}

%\section{Derivation of lower bounds on bipartite entanglement}\label{app:measureBP}
\Appendix{Lower bounds on bipartite entanglement}\label{app:measureBP}

Here we derive the lower bounds on bipartite entanglement given in Eqs.~\eqref{ineqS} and~\eqref{ineqN}. We also derive analogous lower bounds in the case of bipartite entanglement quantified with concurrence and Tsallis 2-entropy.  %\\

%\subsection{Lower bound on the bipartite entanglement of formation}
\SubAppendix{Lower bound on the bipartite entanglement of formation}

We now derive Eq.~\eqref{ineqS}.

First, recall that a pure state $|\psi\rangle$ can be represented in Schmidt decomposition~\cite{Schmidt1907} as $|\psi\rangle=\sum_i\sqrt{\lambda_i} |e_{i}^{(A)}\rangle\otimes|e_{i}^{(B)}\rangle$, where the Schmidt vectors $\{|e_{i}^{(A,B)}\rangle\}_i$ form orthonormal bases in the subsystems $A$ and $B$, while the Schmidt coefficients $\lambda_i\geq 0$. In particular, the von Neumann entropy of the reduced state $\rho_A$ equals the Shannon entropy of the Schmidt coefficients,  $S(\rho_A)= -\sum_i\lambda_i \log_2 \lambda_i$. It is known (see e.g., ~\cite{Schuch2004,Goldstein2017}) that when the local charge $Q=Q^{(A)}+Q^{(B)}$ is fixed, the subsystem charges are also fixed  (but not necessarily the same) in all Schmidt vectors, i.e.,\ $|e_{i}^{(A)}\rangle$ is of a fixed charge $Q^{(A)}$ (and analogously for $B$), as follows. We have 
\begin{equation}
q|\psi\rangle=Q^{(A)}|\psi\rangle+Q^{(B)}|\psi\rangle ,\label{Schmidt0}
\end{equation}
and by grouping orthogonal terms we have 
\begin{align}
q |e_{i}^{(A)}\rangle\otimes |e_{i}^{(B)}\rangle= & (Q^{(A)}|e_{i}^{(A)}\rangle)\otimes |e_{i}^{(B)}\rangle
\nonumber \\
& +|e_{i}^{(A)}\rangle\otimes ( Q^{(B)}|e_{i}^{(B)}\rangle) ,\label{Schmidt1}
\end{align}
for each $i$. Thus, we obtain the proportionality $Q^{(A)}|e_{i}^{(A)}\rangle=q_i^{(A)} |e_{i}^{(A)}\rangle$ , and analogously for $B$, with $q_i^{(A)}+q_i^{(B)}=q$ for all $i$.

Therefore, we can write, for the von Neumann entropy of the reduced state,
\begin{eqnarray}
S(\rho_A) &=&-\sum_{i}\lambda_i \log_2 \lambda_i\label{ineqS1}\\
&\geq&-\sum_{q^{(A)}} \sum_{i:\,q_i^{(A)}=q^{(A)}}\lambda_i\, \log_2\!\Big(\!\!\!\!\sum_{i:\,q_i^{(A)}=q^{(A)}}\!\!\!\! \!\lambda_i \Big) 
=S(\rho_{\rm block}), \nonumber
\end{eqnarray}
where  $q_i^{(A)}$ denotes the value of the subsystem charge $Q^{(A)}$ for the $i$th Schmidt vector (analogously for $B$). We now explain the last  equality in Eq.~\eqref{ineqS1}. For a pure $\rho=|\psi\rangle\!\langle\psi|$ we have $\rho_{\rm block}= \sum_{q^{(A)}} p_{q^{(A)}} |\psi_{ q^{(A)}}\rangle\!\langle\psi_{ q^{(A)}}|$, where  $ p_{q^{(A)}} \equiv \sum_{i:\,q_i^{(A)}=q^{(A)}}\lambda_i$ and $|\psi_{ q^{(A)}}\rangle\equiv \sum_{i:\, q_i^{(A)}=q^{(A)}}   \sqrt{\lambda_i} |e_{i}^{(A)}\rangle  \otimes |e_{i}^{(B)}\rangle / \sqrt {p_{q^{(A)}}} $ (cf. Fig.~\ref{fig:matrix}). Therefore,  $S(\rho_{\rm block})= -\sum_{q^{(A)}} p_{q^{(A)}} \log_2 p_{q^{(A)}}$, and Eq.~\eqref{ineqS1} follows. Note that, since $Q^{(A)}+Q^{(B)}=Q$ is fixed, we can also equivalently consider the eigenspaces and eigenvalues of the charge difference $\delta Q\equiv Q^{(A)}-Q^{(B)}$. 

Second, when a state $\rho$ is of a fixed local charge $Q$, we have that all pure states $|\psi_j\rangle\!\langle \psi_j|$ in $\rho=\sum_{j} p_j |\psi_j\rangle\!\langle \psi_j|$  are of the same fixed charge as $\rho$ (cf.~Sec.~\ref{subsec:witnessMPE}). Therefore, from~\eqref{ineqS1} we have
\begin{eqnarray}
\sum_j p_j S(\rho^j_A)&\geq & \sum_j p_j S(\rho^j_{\rm block})=\sum_j  p_j S( |\psi_j\rangle\!\langle \psi_j|\, ||\,\rho^j_{\rm block})\nonumber 
\\
&\geq& S\Big(\sum_j  p_j |\psi_j\rangle\!\langle \psi_j| \,|| \sum_j  p_j  \rho^j_{\rm block}\Big), \label{ineqS2}
\end{eqnarray}
where the second inequality is the joint convexity of the relative entropy. By observing that $\sum_j p_j\rho^j_{\rm block}=\rho_{\rm block}$, we finally arrive at the new lower bound in Eq.~\eqref{ineqS} in terms of the asymmetry~\eqref{Sas} of a subsystem charge (or equivalently the charge difference).\\

%\subsection{Lower bound on the convex roof of negativity}
\SubAppendix{Lower bound on the bipartite entanglement of formation}

We now derive Eq.~\eqref{ineqN}.

First let us recall from Eqs.~\eqref{Schmidt0} and~\eqref{Schmidt1} that for a pure state $|\psi\rangle$ with a fixed charge $Q$, the Schmidt vectors are always of a fixed subsystem charge. It then follows that 
\begin{eqnarray}
2 \,\N(|\psi\rangle)&=& \sum_{i\neq i'} [\lambda_i\lambda_{i'}]^{1/2}  \nonumber\\
&\geq& \sum_{q^{(A)}\neq q'^{(A)}}  \Big[\sum_{i:\,q^{(A)}_{i}=q^{(A)} } \lambda_i \sum_{i':\,q^{(A)}_{i'}=q^{(A)} }  \lambda_{i'}\Big]^{1/2}  \nonumber\\
&=& \sum_{q^{(A)}\neq q'^{(A)}} \Big[ \sum_{i:\,q^{(A)}_{i}=q^{(A)} }  |\psi_{i}|^2 \sum_{i':\,q^{(A)}_{i'}=q^{(A)} } |\psi_{i'}|^2 \Big]^{1/2} \nonumber\\
&=& l_1^\text{block}(|\psi\rangle\!\langle\psi|), 
\label{ineqN0}
\end{eqnarray}
where $\lambda_i$ are Schmidt coefficients, $\psi_{i}\equiv\langle i|\psi\rangle$ are coordinates of the state $|\psi\rangle$ in the computational basis, and  $l_1^\text{block}$ is calculated with respect to the subsystem-$A$ charge [see~\eqref{l1block}]. In the second line we used the inequality between $L_1$- and $L_2$-norms. The third line is  independent of the choice of the basis, as long as it features both a fixed system and subsystem charges (in particular it can be chosen with Schmidt vectors as elements, which gives the equality in the second line). The last equality follows by noting that $\rho_{ii'}= \psi_{i} \psi_{i'}^*$ for the pure $\rho=|\psi\rangle\!\langle\psi|$ [cf.~\eqref{l1block}].

We note that a related result for pure states connects the entanglement negativity to the minimal $l_1$ coherence in an AB-separable basis~\cite{Piani2011,Nakano2013} $\N(|\psi\rangle)=\min_{\text{AB-sep basis}} l_1(|\psi\rangle\!\langle\psi|)/2$. 

Second, when a state $\rho$ is of a fixed local charge $Q$, we have that all pure states $|\psi_j\rangle\!\langle \psi_j|$ in $\rho=\sum_{j} p_j |\psi_j\rangle\!\langle \psi_j|$  are of the same fixed charge as $\rho$ (cf.~Sec.~\ref{subsec:witnessMPE}). Therefore, from~\eqref{ineqN0} we have [cf.~\eqref{N}]
\begin{eqnarray}\label{ineqN1}
2 \,\E_\text{BP}^\N(\rho)&\geq& \min_{p_j,|\psi_j\rangle} \sum_j p_j \,l_1^\text{block}(|\psi_j\rangle\!\langle\psi_j|)  \\
&=& \min_{p_j,|\psi_j\rangle}  \sum_{q^{(A)}\neq q'^{(A)}} \sum_j p_j\Big[\sum_{\substack{i:\,q^{(A)}_{i}=q^{(A)}\\i':\,q^{(A)}_{i'}=q'^{(A)}}} |\psi_{i}^{(j)} \psi_{i'}^{(j)*}|^2\Big]^{1/2} \nonumber\\
&\geq & \min_{p_j,|\psi_j\rangle} \sum_{q^{(A)}\neq q'^{(A)}} \Big[\sum_{\substack{i:\,q^{(A)}_{i}=q^{(A)}\\i':\,q^{(A)}_{i'}=q'^{(A)}}}  |\sum_j p_j\psi_{i}^{(j)} \psi_{i'}^{(j)*} |^2\Big]^{1/2}
\nonumber
\end{eqnarray}
where the second inequality follows from the convexity of $L_2$-norm. By noting that $ \sum_j p_j\psi_{i}^{(j)} \psi_{i'}^{(j)*}=\rho_{i i'}$  as $\rho=\sum_{j}p_j|\psi_j\rangle\!\langle\psi_j|$, we finally arrive at a new lower bound~\eqref{ineqN} on the convex roof of the negativity in Eq.~\eqref{N}.\\

%\subsection{Lower bounds on bipartite entanglement quantified with concurrence and Tsallis 2-entropy}\label{app:bounds}
\SubAppendix{Lower bounds on bipartite entanglement quantified with concurrence and Tsallis 2-entropy}\label{app:bounds}
%\subsubection{$L_2$-norm for coherence and asymmetry}
\SubSubAppendix{$L_2$-norm for coherence and asymmetry}

In a given computational basis $\{|i\rangle\}_{i=1}^D$, in analogy to the $L_1$-norm coherence monotone~\eqref{l1}, one can define the $L_2$-norm
\begin{equation}
l_{2}(\rho) \equiv \sqrt{\sum_{i \neq j} | \rho_{ij} |^2 }.
\label{l2}
\end{equation}
It is known that the square of the norm $l_{2}^2(\rho)$ can increase under incoherent operations and thus is not a coherence monotone~\cite{Baumgratz2014}. For pure states $l_{2}(\rho)$ is related the state {\em purity}~\cite{Horodecki2003,Singh2015,Cheng2015} of the diagonal state, $\rho_{\rm diag}\equiv\sum_i \rho_{ii}|i\rangle\!\langle i|$,
\begin{equation}
l_{2}(\rho) = \sqrt{1 - \Tr(\rho^{2}_{\rm diag})} ,
\end{equation}
 while for mixed states we have 
\begin{equation}
l_{2}(\rho) = \sqrt{\Tr(\rho^2) - \Tr(\rho^{2}_{\rm diag})} .
\end{equation}

For the case of asymmetry, in analogy to~\eqref{l1block} we now introduce
\begin{eqnarray}\label{l2block}
l_{2}^\text{block}(\rho)&\equiv& \sqrt{ \sum_{i,j:\,h_i\neq h_j}  |\rho_{ij}|^2} \\\nonumber
&=& \sqrt{\Tr(\rho^2) - \Tr(\rho^{2}_{\rm block})} ,
\end{eqnarray} 
where $\rho_\text{block}$ is obtained from $\rho$ by removing all coherences (i.e.,\ dephasing) between distinguishable subspaces, $\rho_\text{block}=\sum_{i,j:\, h_i=h_j}\rho_{ij}\,|i\rangle\!\langle j|$, e.g., eigenspaces of a Hamiltonian $H=\sum_{i} h_i |i\rangle\!\langle i|$.

\begin{figure}[ht!]
	\begin{center}
		\vspace*{-5mm}
		\includegraphics[width=\columnwidth]{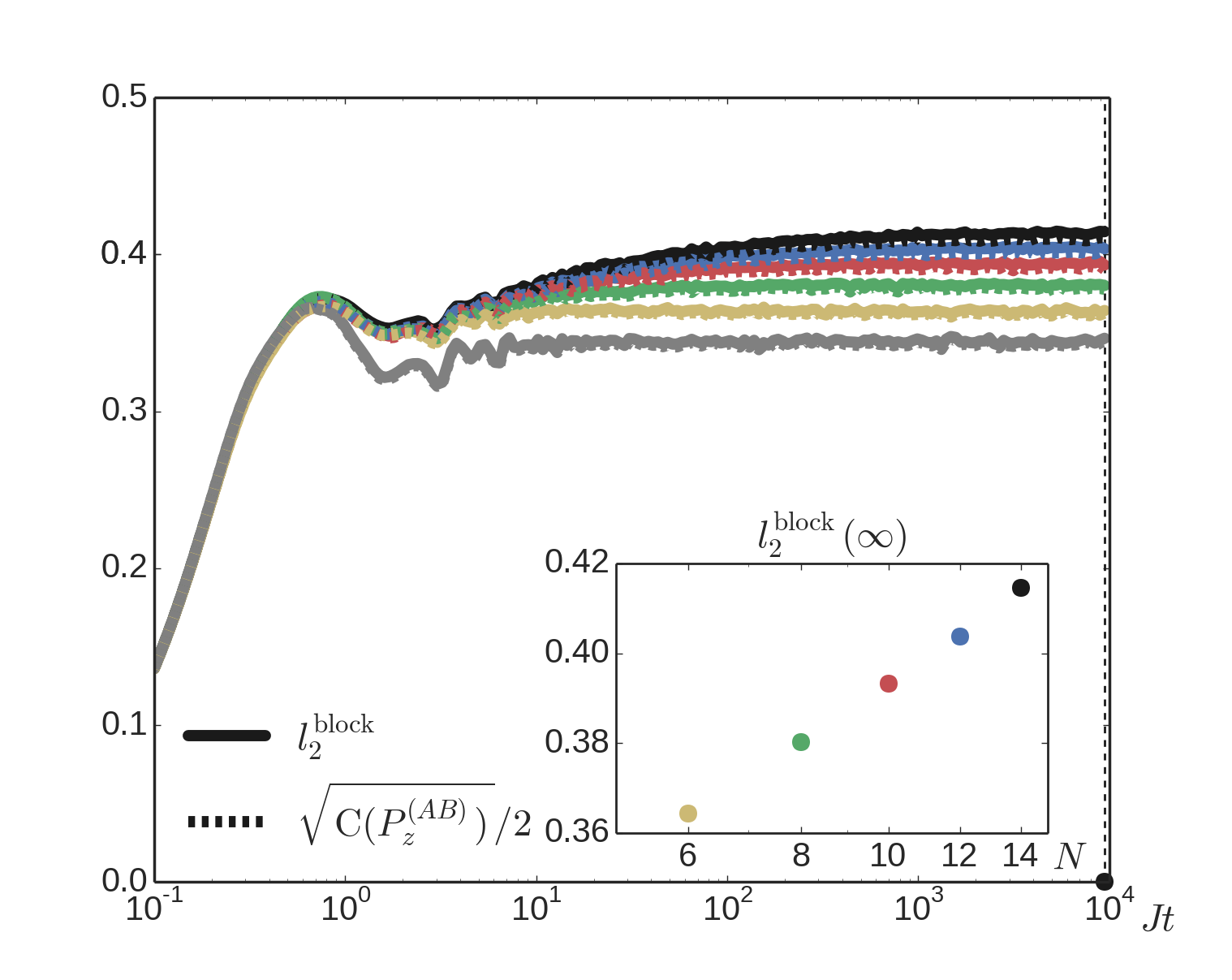} %{QFI_open.pdf}
		\vspace*{-7mm}
		\caption{
			{\bf Experimentally accessible lower bounds on BPE in an MBL system}. We show the dynamics of $l_2^\mathrm{block}(\rho)$ (solid lines) and $C(P_z^{(AB)},\rho)$, where the parity $P_z^{(AB)}\equiv (-1)^{\delta M/2}$  for the $XXZ$  discorded chain of $N=6,8,10,12,14$ spins [yellow, green, red, blue, and black (grayscale: light gray to black), respectively]. The inset shows asymptotic values of $l_2^\mathrm{block}(\rho)$ at $Jt=10^4$. Parameters of the dynamics~\eqref{XXZ} and~\eqref{master} are $V/J=2$, $h/J=5$ and $\gamma/J=0$, while the gray (bottom) curves corresponds to the noninteracting case $V/J=0$ and $N=12$. 
		}\vspace*{-5mm}
		\label{fig:l2_block}
	\end{center}
\end{figure}

Similarly to the case of $l_1^\text{block}$, it is not known whether $l_2^{\rm block}$ corresponds to a measure of asymmetry. Nevertheless, as we show below, $l_{2}^\text{block}$ serves as a lower bound on bipartite entanglement in states with a fixed local charge. It is also a lower bound on~\eqref{l2},
\begin{equation}
l_{2}(\rho)\geq l_{2}^\text{block}(\rho)\label{l2ineq},
\end{equation}
which is saturated for nondegenerate Hamiltonians. As $l_2(\rho) \leq \sqrt{(D-1)/D}<1$ is bounded, we need to adjust the notion of the volume law in the thermodynamic limit, but for moderate sizes $N$, breaking of the area law can still be detected (cf. Fig.~\ref{fig:l2_block}).

Importantly, Eq.~\eqref{l2block} can be obtained experimentally from the MQC spectrum~\eqref{Im} [cf.~\eqref{l1block2}],
\begin{eqnarray}\nonumber
l_2^\text{block}(\rho) &=& \sqrt{\sum_{m\neq 0}\sum_{i,j:\,m_i-m_j=m}  |\rho_{ij}|^2} \\ &=&\sqrt{\sum_{m\neq 0} I_m(\rho)},\label{l2block2},
\end{eqnarray} 
for diagonal observable $M=\sum_{i} m_i |i\rangle\!\langle i|$, which distinguishes the subspaces, i.e., fulfills $m_i=m_j$ if and only if $h_i=h_j$, e.g., $M=H$. Otherwise, the right-hand side of Eq.~\eqref{l2block2} provides a lower bound. Alternatively, $l_2^\text{block}(\rho)$ can be also obtained from the MQC protocol by replacing the unitary rotation with the generator $M$ by strong collective dephasing with $M$, which takes $\rho$ into $\rho_\text{block}$, and thus the outcome measured in the protocol is given by $F=\Tr(\rho_\text{block}^2)$ [while $F=\Tr(\rho^2)$ without dephasing] (cf.~Appendix~\ref{sec:method}). 

Finally, Eq.~\eqref{l2block} can also be bounded from below by the curvature ,~\eqref{C},
\begin{equation}\label{l2block3}
 l_2^\text{block}(\rho) \geq  \frac{\sqrt{C(M,\rho)}}{\Delta M} ,
\end{equation}
where $\Delta M$  is the difference between the extreme eigenvalues of a diagonal observable $M=\sum_{i} m_i |i\rangle\!\langle i|$, which fulfills $m_i=m_j$ when $h_i=h_j$.

In Fig.~\ref{fig:l2_block} we show $l_2^\mathrm{block}$ and the curvature of the parity with respect to the half-chain magnetization,  in the dynamics of a disordered $XXZ$  chain in Eq.~\eqref{XXZ} initially in the staggered state (see~Sec.~\ref{sec:MBL}). We observe very good agreement between both quantities. Indeed, the parity is a good choice for using the curvature as a lower bound of $l_2^\text{block}$ [Eq.~\eqref{l2block3}], as its spectrum is bounded, $\Delta P = 2$, as well as $C(P,\rho)/4  = {\sum_{ij}}' | \rho_{ij} |^{2}$, where $\sum'_{ij}$ is restricted to differences in magnetization being odd.  In particular, the good agreement between two approaches in Fig.~\ref{fig:l2_block} is a consequence of the Hamiltonian eigenstates being localized in the Hilbert space, so  only the states with magnetization differing by $-\frac{1}{2}$ and $\frac{1}{2}$ from the initial value contribute significantly to the dynamics. 
%
%

%\subsubsection{Bounds on bipartite entanglement quantified with concurrence and Tsallis 2-entropy}
\SubSubAppendix{Bounds on bipartite entanglement quantified with concurrence and Tsallis 2-entropy}

For the mixedness of the reduced state quantified by the concurrence~\cite{Wootters1998,Hill1997}, $ C_2(\rho)\equiv [1-\Tr(\rho^2)]^{1/2}/\sqrt{2}$~\footnote{We note that the entanglement negativity $\N(\rho)$ and concurrence $C_2(\rho)$ coincide for the pure system of two qubits~\cite{Wootters1998,Hill1997}, and can be considered as two nonequivalent extensions of that case.} or by Tsallis 2-entropy, $S_2^\text{Ts}(\rho)\equiv 1-\Tr(\rho^2)$, the corresponding convex roofs~\cite{Rungta2001,Albeverio2001,Wootters2001,Szilard2015}  
\begin{equation}
\E_{\rm BP}^{C_2}\equiv \min_{\{p_j, |\psi_j\rangle\}} \sum_{j} p_j C_2(\rho^j_A)
\end{equation}
and 
\begin{equation}
\E_{\rm BP}^\text{Ts}\equiv \min_{\{p_j, |\psi_j\rangle\}} \sum_{j} p_j S_2^\text{Ts}(\rho^j_A)
\end{equation}
are entanglement monotones.

Below we show that for $\rho$ with a fixed local charge $Q=Q^{(A)}+Q^{(B)}$ can be bounded from below by the introduced above $l_2^{\rm block}$ [Eq.~\eqref{l2block}],
\begin{eqnarray}
\E_{\rm BP}^{C_2}(\rho) &\geq& \sqrt{2}\,l_2^{\rm block}(\rho),\label{ineqC}\\
\E_{\rm BP}^\text{Ts}(\rho)&\geq& [l_2^{\rm block}(\rho)]^2. \label{ineqTs}
\end{eqnarray}
These bounds are accessible in experiments for both  pure and mixed states.

We also note that when a state $\rho$ is pure, from concavity of the logarithm, we have $\E_{\rm BP}(\rho)=S(\rho_A)=-\Tr\rho_A\log\rho_A \geq -\log \Tr(\rho_A^2)=-\log (1-C_2^2(\rho))$, where $-\log \Tr(\rho_A^2)$ is Renyi's 2-entropy~\cite{Vidal2000}, and analogously for the lower bounds~\eqref{ineqS} and~\eqref{ineqC} we have $S(\rho_\text{block})\geq-\log \Tr(\rho_\text{block}^2)$ (cf.~\cite{wei2004}).\\

\emph{Derivation}. We now derive the bounds in Eqs.~\eqref{ineqC} and~\eqref{ineqTs}.

First, for pure $|\psi_j\rangle$ with Schmidt  coefficients $\lambda^{(j)}_i\geq 0$, we have
\begin{equation}
\Tr(\rho_A^j)^2=\sum_i |\lambda^{(j)}_i|^2\leq \sum_q\! \Big(\!\!\!\sum_{i:\, q_i^{(A)}=q}\!\! \lambda_i^{(j)}\Big)^2=\Tr(\rho_{\rm block}^j)^2, \label{ineqC1}
\end{equation} 
where  $q_i^{(A)}$ denotes the value of the subsystem charge $Q^{(A)}$ for the $i$th Schmidt vector. 
Equation~\eqref{ineqC1} follows from Eqs.~\eqref{Schmidt0} and~\eqref{Schmidt1}, which imply that for a pure state $|\psi\rangle$ with a fixed charge $Q$, the Schmidt vectors are always of a fixed subsystem charge. Observing that $\Tr(|\phi\rangle\! \langle \phi|\,\rho_{\rm block})=\Tr(\rho_{\rm block}^2)$, we further have from the absolute homogeneity $\lVert x X \rVert_{2}=|x|\lVert  X \rVert_{2}$ of the Hilbert-Schmidt norm $\lVert X \rVert_{2}\equiv[\Tr(X^\dagger X)]^{1/2}$ and the triangle inequality  that
\begin{eqnarray}
&&\sum_j p_j\left[1-\Tr[(\rho_{\rm block}^j)^2\right]^{1/2} = \sum_j p_j  \lVert  |\phi_j\rangle\! \langle \phi_j|-\rho_{\rm block}^j\rVert_{2} \qquad\,\,\,\label{ineqC2} \\\nonumber
&&
\geq \Big\lVert \sum_j \left(p_j |\phi_j\rangle\! \langle \phi_j|-p_j\rho_{\rm block}^j\right)\Big\rVert_{2}=\left[\Tr(\rho^2)-  \Tr(\rho_{\rm block}^2)\right]^{1/2}\!\!\!\!,
\end{eqnarray}
 while from the operator convexity of the square function
\begin{eqnarray}
&& 1-\sum_j p_j\Tr(\rho_{\rm block}^j)^2= \sum_j p_j\Tr(|\phi_j\rangle\! \langle \phi_j|-\rho_{\rm block}^j)^2\qquad\quad\label{ineqTs2}
\\\nonumber
&&\geq \Tr\Big(\!\sum_j p_j |\phi_j\rangle\! \langle \phi_j|-p_j\rho_{\rm block}^j\Big)^2=\Tr(\rho^2)-  \Tr(\rho_{\rm block}^2).
\end{eqnarray}
Bringing together Eqs.~(\ref{ineqC1}) and~(\ref{ineqC2}), or  Eqs.~(\ref{ineqC1}) and~(\ref{ineqTs2}), we arrive at the bounds in Eqs.~\eqref{ineqC} and~\eqref{ineqTs}, respectively.

\newpage
\vspace*{-10mm}
%\section{Thermal phase in a disordered system}\label{app:thermal}
\Appendix{Thermal phase in a disordered system}\label{app:thermal}
\vspace*{100mm}
\begin{figure}[htb!]
\begin{center}
	\vspace*{-105mm}
\includegraphics[width=1.0\columnwidth]{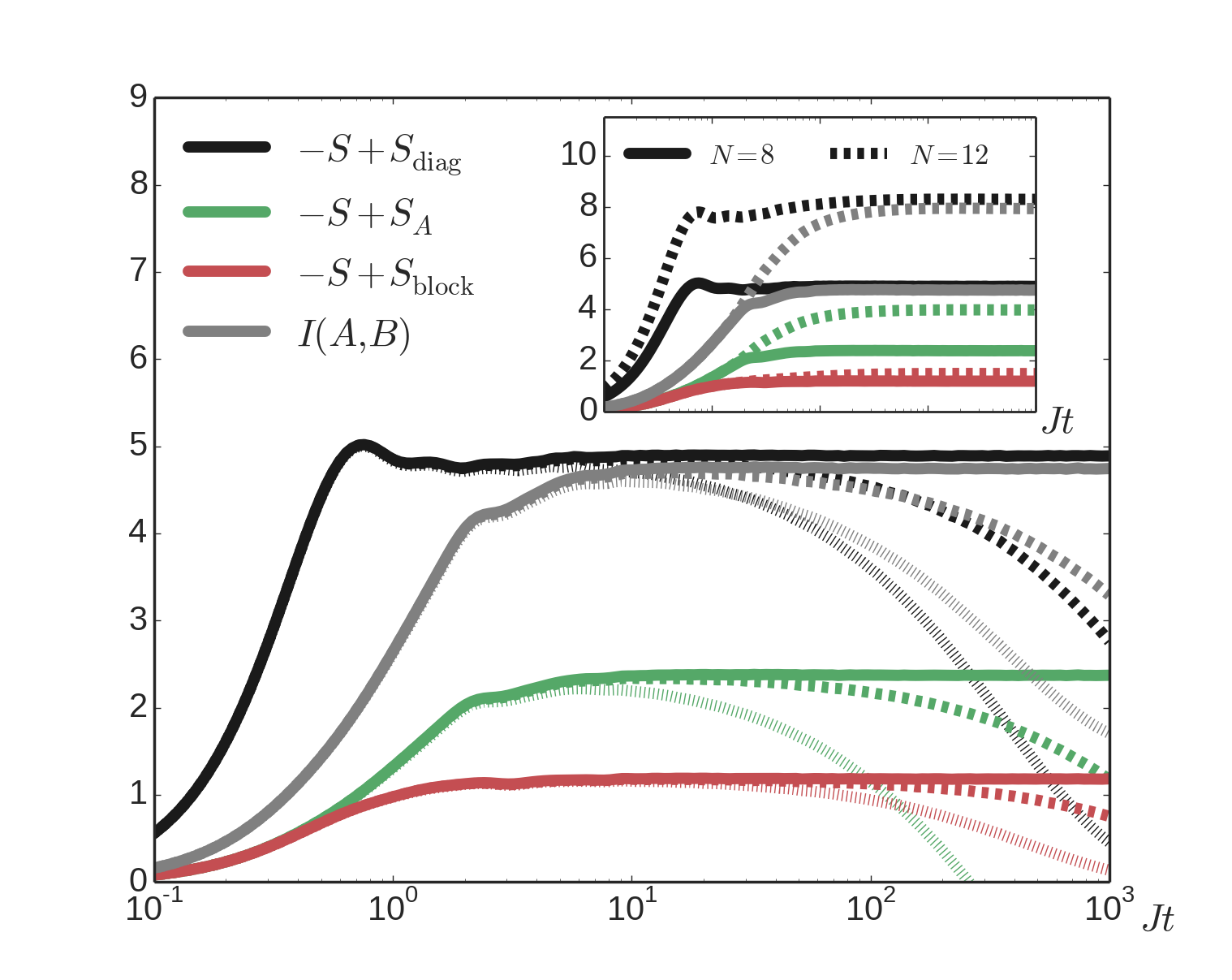}
\vspace*{-8mm}
\caption{
{\bf Correlations in the thermal phase of an MBL system}: relative entropy coherence~\eqref{Scoh}  [black (top) curve], a lower bound on the relative entropy of entanglement~\cite{Plenio2000} $-S(\rho)+S(\rho_A)$ [green (lower middle) curve], half-chain magnetization asymmetry~\eqref{Sas} [red (bottom) curve], and mutual information (total classical and quantum correlations) $I=(A,B)\equiv S(\rho_A)+S(\rho_B)-S(\rho)$ [gray (upper middle) curve] in thermal phase of MBL system ($V/J=2$, $h/J=1$, and $N=8$). Solid lines corresponds to the closed case, while the open case with dephasing is illustrated by dashed ($\gamma/J=2\times 10^{-4}$) and dotted ($\gamma/J=10^{-3}$) lines. The inset shows the closed case for $N=8$ (solid lines) and $N=12$ (dashed lines) spins. 
}\vspace*{-10mm}
\label{fig:thermal}
\end{center}
\end{figure}

\end{appendix}

\bibliography{mblqfi.bib}

\end{document}